# Photo-Assisted Pd–Nb$_2$O$_5$/Carbon Nanocomposites for Enhanced Ethanol Electro-Oxidation Kinetics and CO Tolerance in Alkaline Media

João V. T. Neves [a,b], Stephanie S. Aristides-Barros [a,b], Aline B. Trench [c], Ivani M. Costa [d], Mauro C. Santos [c], Giancarlo R. Salazar-Banda [a,b,*], Katlin I. B. Eguiluz [a,b,*]

[a] *Tiradentes University (UNIT), Graduate Program in Process Engineering (PEP), Av. Murilo Dantas, 300, Aracaju-SE, Brazil, ZIP code 49032-490.*

[b] *Laboratory of Electrochemistry and Nanotechnology (LEN), Institute of Technology and Research (ITP), 49.032-490, Aracaju, Sergipe, Brazil.*

[c] *Center for Natural and Human Sciences (CCNH), Federal University of ABC (UFABC). Rua Santa Adélia 166, Bairro Bangu, 09210-170, Santo André - SP, Brazil.*

[d] *Sertão Campus, Federal University of Alagoas, 57480-000, Delmiro Gouveia, AL, Brazil.*

*Corresponding author:* gianrsb@gmail.com (Giancarlo Richard Salazar Banda)

*Corresponding author:* katlinbarrios@gmail.com (Katlin Ivon Barrios Eguiluz)

# Graphical Abstract

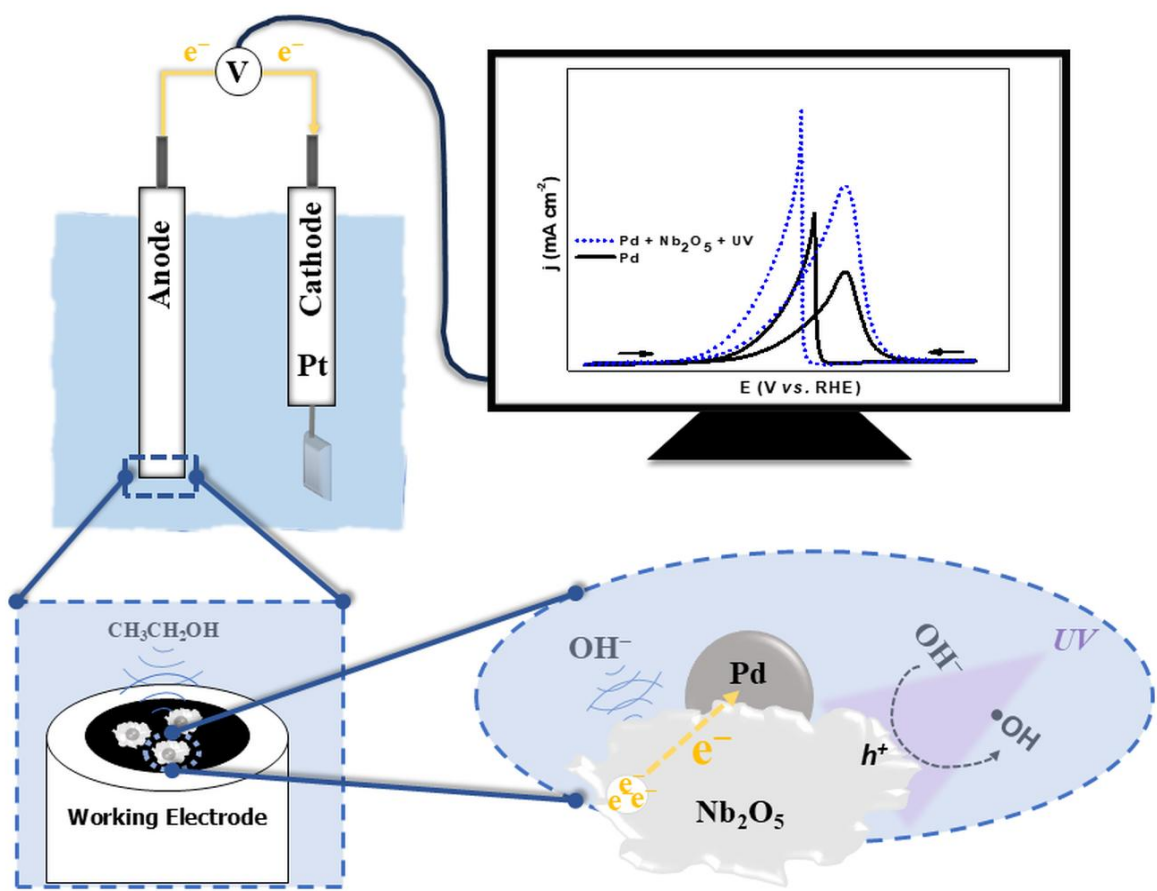

# Highlights

Nb$_2$O$_5$ incorporation lowers CO oxidation potential, mitigating Pd surface poisoning

Pd–Nb$_2$O$_5$ sites boots ethanol oxidation via bifunctional and electronic effects

Optimized Pd$_{(0.5)}$Nb$_2$O$_{5(0.5)}$/C shows the highest activity due to strong metal-support interaction

Nb$_2$O$_5$ enhances long-term Pd durability by limiting poisoning and structural degradation

Pd–Nb$_2$O$_5$/C photoelectrocatalysts exhibit high efficiency for ethanol electrooxidation

**Abstract**

Pd–based anodes for alcohol oxidation suffer from surface poisoning and sluggish kinetics. Here, we developed Pd–$Nb_2O_5$/C nanocomposites to improve ethanol electrooxidation kinetics and CO tolerance in alkaline media. Orthorhombic $Nb_2O_5$ prepared by the Pechini route was combined with fcc Pd nanoparticles via polyol reduction, yielding $Pd_{(x)}$-$Nb_2O_{5(y)}$/C nanocomposites with $x$:$y$ = 100:0, 70:30, 50:50, 30:70, 0:100. Rietveld-refined X-ray diffraction confirmed phase purity and showed similar Pd crystallite sizes (4.46 nm for Pd/C and 4.92–5.08 nm for $Nb_2O_5$-containing catalysts). Transmission and scanning electron microscopies coupled with energy-dispersive X-ray spectroscopy reveal uniformly dispersed Pd nanoparticles on $Nb_2O_5$ and carbon. UV–Vis diffuse reflectance indicated a band gap of 3.10 eV, and chopped-light photocurrent measurements confirm the strong ultraviolet responsiveness of $Nb_2O_5$. X-ray photoelectron spectroscopy reveals that $Pd_{(0.5)}Nb_2O_{5(0.5)}$/C had the highest $Pd^0$ content (58.99%). Electrochemical testing demonstrates that, relative to Pd/C, optimized $Pd_{(0.5)}Nb_2O_{5(0.5)}$/C reduces the ethanol oxidation onset potential by up to 160 mV, increases poisoning tolerance by a factor of five at a fixed potential, and raises the current density from 1.59 to 1.76 mA cm$^{-2}$. Under light irradiation, the current density increases from 1.07 to 2.10 mA cm$^{-2}$, accompanied by improved stability and extended durability, attributed to light-induced electron–hole generation and enhanced $OH^-$ adsorption. These results highlight the synergistic contribution of oxide–metal interactions and photoactivation to ethanol oxidation and provide insights for designing efficient catalysts for alkaline fuel cells.

# 1. Introduction

The sustained increase in global energy demand is a pressing concern [1], with studies attributing the trend to factors such as urban population growth [2,3] and the intensified production of goods and services [4,5]. Although fossil fuels still dominate the global energy portfolio, their adverse environmental impacts and price volatility are accelerating the search for cleaner, more efficient, and economically viable alternatives [6].

In this context, hydrogen-powered fuel cells are a promising option, benefiting from hydrogen's abundance, diverse production pathways, minimal greenhouse gas emissions at the point of use, and high-energy conversion efficiency [7]. However, widespread adoption is hindered by the high cost of hydrogen production, the energy demands of its conversion to electricity, and the infrastructure required for its storage and transport, given its low volumetric density and high diffusivity [8].

Liquid alcohols are another viable class of fuels for fuel cells, with methanol ($CH_3OH$) and ethanol ($C_2H_5OH$) being the most studied. Methanol, despite its well-established electrochemical oxidation pathways, has significant drawbacks, including toxicity, high flammability, reliance on fossil-derived feedstocks, and lower performance than hydrogen. These limitations have limited the commercialization of direct methanol fuel cells [9]. Ethanol, in contrast, offers several advantages, including large-scale renewable production, compatibility with existing fuel infrastructure, and favorable electrochemical properties [10]. Its theoretical cell potential (1.17 V), electrical efficiency (97%), and gravimetric energy density (8 kWh $kg^{-1}$) are comparable to or higher than those of methanol (1.21 V, 97%, and 6.09 kWh $kg^{-1}$) [11].

Efficient electrocatalysts are critical for ethanol oxidation in fuel cells. In alkaline media, the reaction is thermodynamically favored by the abundance of hydroxide ions ($OH^-$), which facilitate the oxidative pathway [12]. Pd/C catalysts have demonstrated higher ethanol oxidation activity than Pt/C under these conditions and benefit from greater elemental abundance and availability [13,14]. Nevertheless, the widespread application of palladium is constrained by its high price and susceptibility to deactivation via poisoning by reaction intermediates, both reducing catalytic efficiency [15,16].

To overcome these limitations, Alvarenga and Villullas [17] reviewed the use of transition metal oxides in the electrocatalytic oxidation of methanol and ethanol on noble metal nanoparticles. They reported that such oxides, when employed as supports or co-catalysts, enhance electrocatalytic activity due to their intrinsic oxophilic character and their ability to modify the electronic structure of the active metal. Furthermore, transition metal oxides exhibit high resistance to dissolution, thereby improving catalyst durability. In particular, niobium pentoxide ($Nb_2O_5$) has emerged as a highly promising material for efficient electrocatalysis, owing to its low toxicity, favorable physicochemical properties, and relatively low cost [18].

For instance, Rocha *et al*. [19] synthesized a series of catalysts, $Pt_{(1)}Nb_2O_{5(1)(amorphous)}/C$, $Pt_{(1)}T\text{-}Nb_2O_{5(1)(commercial)}/C$, $Pt_{(1)}Nb_2O_{5(1)(amorphous\text{-}commercial)}/C$, and Pt/C to study ethanol electrooxidation in acidic media. They found that $Pt_{(1)}Nb_2O_{5(1)(amorphous)}/C$ exhibited the lowest onset potential for CO removal, which was 0.10 V lower (versus the reversible hydrogen electrode, RHE) than that of Pt/C. Furthermore, the $Pt_{(1)}Nb_2O_{5(1)(amorphous\text{-}commercial)}/C$ showed the lowest onset potential for ethanol electrooxidation, with a 0.16 V reduction compared to Pt/C. In contrast, $Pt_{(1)}T\text{-}Nb_2O_{5(1)(commercial)}/C$ generated the highest current peak among the tested catalysts.

$Nb_2O_5$ has been extensively investigated in diverse applications, particularly as a photocatalyst, due to its n-type semiconducting nature and band gap in the range of 3.1–4.0 eV [20,21]. Reported applications include pollutant photodegradation [22], hydrogen production [23], $CO_2$ photoreduction [24], and selective transformation of organic molecules [25]. In a comprehensive review, Su *et al*. [26] explored the correlation between $Nb_2O_5$ crystalline structure and photocatalytic performance, highlighting that pseudo-hexagonal, orthorhombic, and monoclinic phases can achieve high activity. Among these polymorphs, the orthorhombic and pseudo-hexagonal phases are the most extensively investigated [27,28].

To date, no studies have reported the use of $Nb_2O_5$ in photo-assisted fuel cells for ethanol oxidation. In such systems, photocatalysis and electrocatalysis operate cooperatively, with light activation of the photocatalyst facilitating ethanol oxidation and enhancing overall reaction efficiency [29]. This mechanism differs from that of photoelectrochemical cells (PECs), where light directly drives the main reaction and is indispensable for its occurrence [30]. Hu, Zhai, and Zhu [31] reviewed metal/semiconductor-based photoresponsive catalysts such as PtNiRu/$TiO_2$, Au/ZnO, Au/CuI, Pt/$Fe_2O_3$/C, and Au/$TiO_2$, demonstrating ethanol electrooxidation enhancements ranging from 1.1- to 16.9-fold under illumination compared with dark conditions.

Given the early stage of photoassisted fuel cell development, continued theoretical and experimental efforts are required to guide material design and system optimization. Therefore, this study explores the synthesis of binary photoelectrocatalysts composed of Pd and orthorhombic $Nb_2O_5$ supported on carbon, aiming to correlate the $Nb_2O_5$ incorporation and UV illumination with the ethanol oxidation pathway. The findings highlight the synergistic role of metal–support interactions and light-induced processes in improving activity, stability, and resistance to catalyst poisoning, offering new perspectives for the design of photoassisted electrocatalysts.

## 2. Experimental

Details on the reagents and chemicals employed the electrochemical setup (Fig. S1), and the procedure for crystallite size determination are available in the Supporting Information.

## 2.1 Synthesis of Niobium Pentoxide (Nb$_2$O$_5$)

Nb$_2$O$_5$ was synthesized using the Pechini method, as schematically illustrated in Fig. S2. Ethylene glycol, citric acid, and niobium(V) chloride were combined in a molar ratio of 2:1:0.5 mol L$^{-1}$, with a total solution volume of 0.002 L [32,33]. Ethylene glycol was heated to 60 °C on a hot plate to reduce viscosity, followed by the addition of citric acid, which was dissolved by manual stirring with a glass rod. Next, niobium(V) chloride was added, and the temperature was raised to 90 °C. After homogenization, a polymeric resin was formed, cooled to room temperature, and subsequently pre-calcined at 400 °C for 2 h (10 °C min$^{-1}$) in a muffle furnace (model FDG 3P-S). After pre-calcination, the resulting solid was ground and subjected to a second calcination at 600 °C for 2 h and using the same heating rate. The final Nb$_2$O$_5$ powder was ground once more and stored for subsequent characterization and catalyst preparation.

## 2.2 Synthesis of Pd$_{(x)}$Nb$_2$O$_{5(y)}$/C Catalysts

The Pd/C catalytic composite (20 wt% Pd supported on Vulcan XC-72) and the series of Pd$_{(x)}$Nb$_2$O$_{5(y)}$/C catalysts (x:y = 70:30, 50:50, 30:70, 0:100), all with a total active phase loading of 20 wt% (metals and/or oxide) and 80 wt% carbon, were synthesized using the polyol method adapted from Kowal *et al*. [34] and Jiang *et al*. [35]. Fig. S3 schematically illustrates the main synthesis steps.

An aqueous solution of tetrachloropalladic acid (H$_2$PdCl$_4$, 10 mmol L$^{-1}$) was prepared by dissolving PdCl$_2$ in 20 mmol L$^{-1}$ HCl, followed by sonication to ensure complete dissolution. Ethylene glycol (25 mL) was then added, serving as both solvent and reducing agent. Separately, Nb$_2$O$_5$ was dispersed in 5 mL of ethylene glycol, and the mixture was sonicated for 30 minutes in an ultrasonic bath to ensure complete dispersion. This suspension was then transferred to the Erlenmeyer flask containing the H$_2$PdCl$_4$ solution and magnetically stirred for 30 min. The pH of the solution was adjusted to 12 using 1 mol L$^{-1}$ NaOH to promote the formation of small nanoparticles. Subsequently, the mixture was heated to 130 °C under nitrogen bubbling for 2 hours to remove oxygen, and then cooled to room temperature.

Independently, 80 mg of Vulcan XC-72 carbon was dispersed in 3 mL of ethylene glycol, sonicated, and then added to the Pd–$Nb_2O_5$ precursor solution. The resulting suspension was stirred for 30 min, filtered, and washed with ultrapure water. The solids were dried in an oven at 70 °C for 2.5 h and ground prior to characterization. The nanoparticulated catalysts were designated according to composition: Pd/C, $Nb_2O_5$/C, $Pd_{(0.7)}Nb_2O_{5(0.3)}$/C, $Pd_{(0.5)}Nb_2O_{5(0.5)}$/C, and $Pd_{(0.3)}Nb_2O_{5(0.7)}$/C.

## *2.3 Physicochemical Characterization*

Thermogravimetric analysis (TGA) of $Nb_2O_5$ was carried out on a Hitachi STA7200RV under a nitrogen flow (20 mL $min^{-1}$), heating the pre-calcined oxide to 900 °C at 10 °C $min^{-1}$. Structural and quantitative phase analyses of the catalysts were performed by X-ray diffraction (XRD) combined with the Rietveld refinement [36]. The refinements were carried out using the GSAS-EXPGUI software [37,38]. XRD patterns were recorded in the 15° < 2θ < 80° range at a scan rate of 0.5° $min^{-1}$ using a Shimadzu XRD-7000 diffractometer with CuKα radiation (λ = 0.15416 nm).

Morphological characterization of $Nb_2O_5$ was performed by scanning electron microscopy (SEM) using a JSM6510LV instrument, with energy-dispersive X-ray spectroscopy (EDS) mapping performed simultaneously. Transmission electron microscopy (TEM) images were obtained using a JEM1400plus microscope. Diffuse reflectance UV–vis spectroscopy (DRS) was carried out at room temperature on a Shimadzu UV-2600 spectrophotometer over the 200–800 nm range, using $BaSO_4$ as the reference. The optical band gap was estimated using the Tauc equation by extrapolating the linear portion of the $(αhν)^n$ versus hν curve to the x-axis [39,40].

X-ray photoelectron spectroscopy (XPS) was performed using a Scienta Omicron ESCA + spectrometer (Germany) with monochromatic Al Kα radiation (1486.7 eV) as the radiation source. The inelastic background of high-resolution center-level C 1s spectra was subtracted using the Shirley method. Spectral fitting was performed without imposing restrictions using multiple Voigt profiles. Data analysis was carried out using CasaXPS software.

*2.4 Electrochemical Experiments*

A 1 mol L$^{-1}$ KOH solution was used as the supporting electrolyte. For experiments without UV irradiation, nitrogen was bubbled through the solution for 10 min before measurements. In UV-assisted experiments, N$_2$ purging was omitted to (i) retain dissolved oxygen, which acts as an electron acceptor and promotes reactive species generation under UV light [41], and (ii) prevent optical interference from bubble scattering and reflection of UV light [42]. Electrochemical measurements were first performed in KOH alone, followed by tests in 1 mol L$^{-1}$ ethanol-containing electrolyte. All experiments were conducted at room temperature.

The electrochemical activation was carried out by cycling the potential 25 times in a 1 mol L$^{-1}$ KOH solution between 0.05 and 1.20 V vs. RHE at 200 mV s$^{-1}$. Subsequently, two additional cycles were collected at 20 mV s$^{-1}$ over the same potential range. Ethanol electrooxidation was studied in 1 mol L$^{-1}$ KOH + 1 mol L$^{-1}$ ethanol. Two voltammetric cycles were recorded at 20 mV s$^{-1}$. The electrocatalyst activities toward ethanol oxidation were evaluated through polarization curves and Tafel plots. Quasi-steady-state polarization curves were obtained in the potential range from 0.30 to 0.65 V vs. RHE at 1 mV s$^{-1}$. Chronoamperometric analysis was then performed by polarizing each catalyst at 0.55 V vs. RHE for 7200 s. The value of 0.55 V was determined from polarization curves as the onset region for significant ethanol oxidation.

Electrochemical impedance spectroscopy (EIS) was performed in 1 mol $L^{-1}$ KOH + 1 mol $L^{-1}$ ethanol after $N_2$ deaeration. Spectra were collected using ten points per decade from 1000 to 0.1 Hz at 0.77 V vs. RHE with a 5 mV AC amplitude. For the best-performing catalyst, EIS was recorded sequentially in the dark and under UV illumination under otherwise identical conditions.

UV-assisted electrochemical measurements followed the same general procedure as the dark experiments, without $N_2$ purging. We performed cyclic voltammetry, polarization curves, and chronoamperometry (600 s) in an ethanol-containing electrolyte. Each experiment was performed sequentially in the dark and under UV illumination. For the transient photoresponse evaluation, photocurrent on/off cycles were recorded during chronoamperometry at a fixed potential (as specified in Fig. S12 and Fig. 7C) in 1 mol $L^{-1}$ KOH + 1 mol $L^{-1}$ ethanol. Stability tests were carried out in 1 mol $L^{-1}$ KOH + 1 mol $L^{-1}$ ethanol by cyclic voltammetry between 0.05 and 1.20 V vs. RHE at 100 mV $s^{-1}$, applying 2500 consecutive cycles in the dark and 600 cycles under UV light.

CO oxidative desorption analyses were performed after electrode activation, previously described. Two baseline cyclic voltammograms (0.05–1.20 V vs. RHE, 20 mV $s^{-1}$) were recorded without CO. Next, the electrode was polarized at 0.05 V for 1500 s, with CO bubbling during the first 300 s to allow adsorption, followed by $N_2$ purging for 1200 s to remove dissolved CO. Subsequently, two potential sweeps were run between 0.05 and 1.20 V vs. RHE at 20 mV $s^{-1}$: the first cycle corresponding to oxidation of adsorbed CO on the active sites of the catalyst, and the second cycle confirming complete CO removal.

## 3. Results and Discussion

### *3.1 Physicochemical Characterization*

Fig. 1 shows the XRD patterns, measured at room temperature, of the Pd/C, $Pd_{(0.7)}Nb_2O_{5(0.3)}/C$, $Pd_{(0.5)}Nb_2O_{5(0.5)}/C$, and $Pd_{(0.3)}Nb_2O_{5(0.7)}/C$ catalysts. All samples display a diffraction peak near 25° (marked with #), attributed to the (002) plane of the hexagonal graphite structure in Vulcan XC-72 carbon (ICSD: 031170) [43]. Although Vulcan carbon is predominantly amorphous, it exhibits partial crystallinity [44]. In addition, reflections characteristic of face-centered cubic (fcc) Pd (ICSD: 064922) [45] and orthorhombic $Nb_2O_5$ (ICSD: 176081) [46] structures are observed. The latter phase is consistent with thermogravimetric analysis data (Fig. S4), which confirmed its formation upon calcination at approximately 600 °C. Unlike Pd/C, the $Pd_{(x)}Nb_2O_{5(y)}/C$ exhibits diffraction peaks from both Pd and $Nb_2O_5$, indicating the coexistence of these crystalline phases.

SEM images of $Pd_{(0.5)}Nb_2O_{5(0.5)}/C$ (Fig. S5) and $Nb_2O_5/C$ (Fig. S6) catalysts highlight the regions selected for elemental mapping and EDS spectra acquisition. The detection of Pd and Nb in both samples corroborates the XRD results, evidencing the simultaneous presence of both phases in the composite catalysts.

Furthermore, as shown in Fig. 1, the diffraction peak intensity associated with the $Nb_2O_5$ phase increases with its relative proportion in the catalyst, as expected due to the greater content of this phase in these composites. The Rietveld refinement was applied to obtain structural parameters, including phase quantification for each catalyst. Fig. 1 displays the observed ($I_{OBS}$) and calculated ($I_{CAL}$) diffractograms, along with their difference (Dif.), for the Pd/C, $Pd_{(0.7)}Nb_2O_{5(0.3)}/C$, $Pd_{(0.5)}Nb_2O_{5(0.5)}/C$, and $Pd_{(0.3)}Nb_2O_{5(0.7)}/C$ catalysts. The goodness-of-fit (S) values, all close to 1 (Table 1), confirm an excellent agreement between the experimental and calculated XRD patterns.

As shown in Table 1, the theoretical crystalline phase proportions agree with those obtained through Rietveld refinement. For instance, the Pd$_{(0.5)}$Nb$_2$O$_{5(0.5)}$/C catalyst contained 7.46% Pd and 7.38% Nb$_2$O$_5$, confirming that both crystalline phases are present in nearly equal amounts. This trend was consistent across the other binary catalysts.

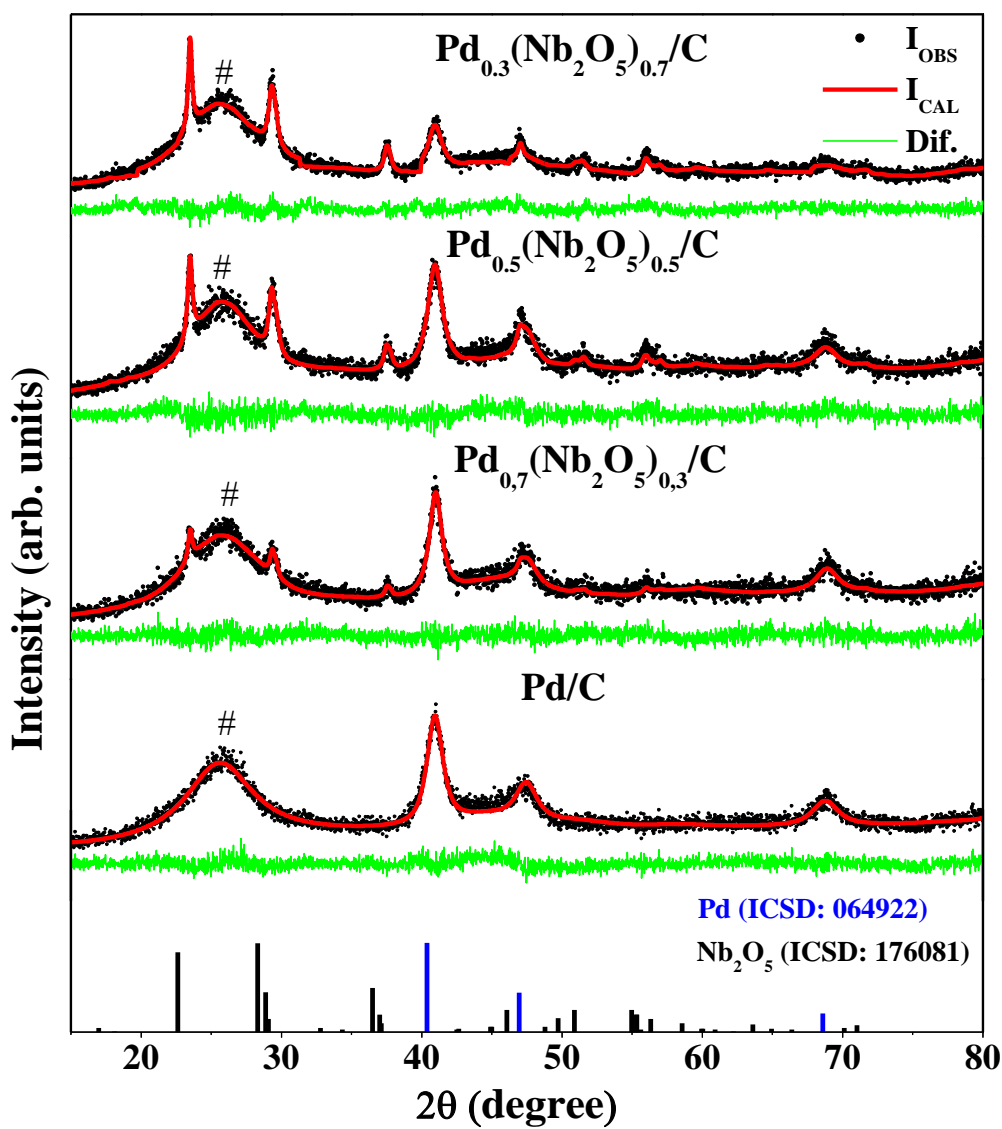

**Fig. 1.** Rietveld refinement (solid red line) of the XRD data (black circles) and difference between experimental and fitting data (green lines) for the Pd/C, Pd$_{(0.7)}$Nb$_2$O$_{5(0.3)}$/C, Pd$_{(0.5)}$Nb$_2$O$_{5(0.5)}$/C, and Pd$_{(0.3)}$Nb$_2$O$_{5(0.7)}$/C catalysts. The symbol # denotes the (002) peak of the Vulcan XC-72 carbon support. Blue and black vertical bars correspond to the ICSD indexation nº. 064922 (fcc Pd) and 1760081 (orthorhombic Nb$_2$O$_5$), respectively.

Pd crystallite sizes, calculated from the FWHM values through Rietveld refinement using the Scherrer equation (Eq. S1), were approximately 5 nm. Variation in $Nb_2O_5$ content did not significantly affect Pd crystallite size within the binary series; however, compared with Pd/C, a slight size increase was observed. Additionally, the Pd lattice parameters remained essentially unchanged. Similar observations were reported by Souza *et al.* [47,48] for PdTT– $Nb_2O_5$/C and Pd– $Nb_2O_{5(amorphous)}$/C catalysts prepared by a sol–gel method (calcined at 400 °C) and by chemical reduction with sodium borohydride, respectively. In both cases, $Nb_2O_5$ addition increased Pd crystallite size without peak broadening or shifts, indicating no incorporation of Nb into the Pd lattice.

**Table 1.** Quantitative phase composition (wt%), lattice parameters, crystallite size (d), and goodness-of-fit (S) values obtained from Rietveld refinement for the synthesized catalysts

|  | Phases | wt (%) | Lattice parameter (Å) | d (nm) | S |
|---|---|---|---|---|---|
| Pd/C | Pd | 3.95 | a = b = c = 3.897 (1) | 4.46 | 1.12 |
|  | C | 96.05 | a = b = 2.485 (4); c = 7.163 (8) | n.a. |  |
| $Pd_{(0.7)}Nb_2O_{5(0.3)}$/C | Pd | 3.15 | a = b = c = 3.897 (1) | 5.08 | 1.19 |
|  | $Nb_2O_5$ | 1.72 | a = 6.222 (7); b = 29.095 (30); c = 3.936 (3) | 18.17 |  |
|  | C | 95.13 | a = b = 2.456 (5); c = 7.174 (8) | n.a. |  |
| $Pd_{(0.5)}Nb_2O_{5(0.5)}$/C | Pd | 7.46 | a = b = c = 3.903 (1) | 5.00 | 1.23 |
|  | $Nb_2O_5$ | 7.38 | a = 6.221 (4); b = 29. 183 (19); c = 3.940 (2) | 17.65 |  |
|  | C | 85.16 | a = b = 2.464 (6); c = 7.168 (7) | n.a. |  |

| | Pd | 2.73 | a = b = c = 3.899 (2) | 4.92 | |
|---|---|---|---|---|---|
| Pd$_{(0.3)}$Nb$_2$O$_{5(0.7)}$/C | Nb$_2$O$_5$ | 7.46 | a = 6.229 (3); b = 29.134 (16); c = 3.934 (2) | 17.11 | 1.31 |
| | C | 89.81 | a = b = 2.454 (4); c = 7.211 (7) | – | |

For the Pd$_{(0.5)}$Nb$_2$O$_{5(0.5)}$/C catalyst, a slight variation in Pd lattice parameters was observed compared to Pd/C, possibly due to an increased interfacial area between Pd and Nb$_2$O$_5$. Similar alterations were reported by Rocha *et al*. [19], who synthesized PtNb$_2$O$_5$/C via chemical reduction with sodium borohydride for ethanol oxidation. They attributed such changes to strong metal–support interactions, as Nb$_2$O$_5$ can induce expansion or contraction in the metal's crystal lattice [19,49]. Likewise, Yang *et al*. [50] synthesized Pt–Nb$_2$O$_5$ by reflux reduction with ethylene glycol and, through FT-EXAFS analysis, detected Pt crystal lattice modifications after Nb$_2$O$_5$ incorporation. These alterations were attributed to the epitaxial growth of Pt on Nb$_2$O$_5$, indicating that mechanical interactions at the metal–support interface can also modify the crystal structure of noble metals.

According to Majumdar [51], an increase in crystallite size reduces the surface-related properties, including specific surface area. Thus, the Pd/C catalyst likely exhibits a larger Pd surface area than the binary catalysts, owing to its smaller average Pd crystallite size. In the case of Nb$_2$O$_5$, the Pd$_{(0.5)}$Nb$_2$O$_{5(0.5)}$/C and Pd$_{(0.3)}$Nb$_2$O$_{5(0.7)}$/C catalysts are expected to have a greater Nb$_2$O$_5$ surface area due to their smaller Nb$_2$O$_5$ crystallite sizes. As reported by Ullah *et al*. [52], semiconductors with smaller crystallites often exhibit superior photocatalytic performance as a result of quantum confinement effects and increased surface area. In this context, catalysts containing Nb$_2$O$_5$ with smaller crystallite sizes should perform better in photoassisted ethanol oxidation by facilitating electron–hole pair separation and enabling more efficient redox processes under UV irradiation.

Table 2 summarizes the texture coefficient values, calculated according to Eq. S2, for the various crystallographic planes. This parameter provides insight into the preferential orientation of surface-exposed planes in the catalyst structure [53,54]. Texture coefficient values above 1 indicate preferential orientation of the corresponding (hkl) planes, while values below 1 suggest low or no preferential exposure. The analysis was performed only for Pd, as it is the primary catalytic phase for ethanol oxidation.

The texture coefficient values in Table 2 indicate that Pd/C and $Pd_{(0.7)}Nb_2O_{5(0.3)}$/C exhibit a substantial preferential orientation toward the (111) and (220) planes. In $Pd_{(0.5)}Nb_2O_{5(0.5)}$/C, the (111) plane is likely the most exposed on the active surface, whereas $Pd_{(0.3)}Nb_2O_{5(0.7)}$/C displays dominant exposure of the (200) plane. Among all catalysts, Pd/C and $Pd_{(0.5)}Nb_2O_{5(0.5)}$/C show the highest preference for the (111) plane, suggesting greater availability of this surface facet for catalytic activity. Because the (111) plane is more exposed and promotes interactions between ethanol and surface OH groups, it likely governs the overall catalytic activity [55].

**Table 2.** Texture coefficient values for the crystallographic planes of Pd in the synthesized catalysts.

| *Texture Coefficient* | | | |
|---|---|---|---|
| **Crystallographic planes of Pd** | **(111)** | **(200)** | **(220)** |
| **Pd/C** | 1.11 | 0.73 | 1.17 |
| **$Pd_{(0.7)}Nb_2O_{5(0.3)}$/C** | 1.06 | 0.70 | 1.25 |
| **$Pd_{(0.5)}Nb_2O_{5(0.5)}$/C** | 1.13 | 0.92 | 0.95 |
| **$Pd_{(0.3)}Nb_2O_{5(0.7)}$/C** | 0.95 | 1.18 | 0.87 |

The chemical composition of the catalysts was determined by XPS. Fig. 2A–D shows the Pd 3d core-level spectra for Pd/C, $Pd_{(0.7)}Nb_2O_{5(0.3)}$/C, $Pd_{(0.5)}Nb_2O_{5(0.5)}$/C, and $Pd_{(0.3)}Nb_2O_{5(0.7)}$/C catalysts. Each spectrum was deconvoluted into two components corresponding to $Pd^0$ and $Pd^{2+}$, with doublets for the lower binding energy band (Pd $3d_{5/2}$) and higher binding energy band (Pd $3d_{3/2}$) [56]. High-resolution Nb 3d spectra for the $Pd_{(0.7)}Nb_2O_{5(0.3)}/C$, $Pd_{(0.5)}Nb_2O_{5(0.5)}/C$, and $Pd_{(0.3)}Nb_2O_{5(0.7)}/C$ catalysts (Fig. 2I–III, Table S1) display spin–orbit doublets for Nb $3d_{5/2}$ (~207.7 eV) and Nb $3d_{3/2}$ (~210.40 eV), separated by 2.7 eV. Thus, Nb is in the $Nb^{5+}$ oxidation state, thereby confirming the presence of $Nb_2O_5$ in the samples [57], as indicated by XRD (Fig. 1).

Data in Fig. 2 and Table S2 show a shift of the $Pd^0$ $3d_{5/2}$ peak toward lower binding energies for $Pd_{(0.7)}Nb_2O_{5(0.3)}/C$, $Pd_{(0.5)}Nb_2O_{5(0.5)}/C$, and $Pd_{(0.3)}Nb_2O_{5(0.7)}/C$ samples, compared to Pd/C. This behavior suggests a modification in the electronic environment of palladium by interaction with $Nb_2O_5$. Fu, Li, reported similar shifts and Zhang [58] for Pt nanoparticles supported on Nb-doped $TiO_2$ prepared by aerosol-assisted self-assembly. They observed that the addition of Nb promoted a shift of the $Pt^0$ $4f_{7/2}$ peaks to lower binding energies. The percentages of the Pd oxidation states for each catalyst are also presented in Table S2. Note that $Pd_{(0.5)}Nb_2O_{5(0.5)}/C$ possessed the highest $Pd^0$ content in its composition. Xu *et al*. [59] observed a comparable effect for the Pd/Nb–$TiO_2$–C catalyst in ethanol oxidation. The Nb–$TiO_2$ incorporation increased $Pd^0$ content by 15.7% relative to Pd/C.

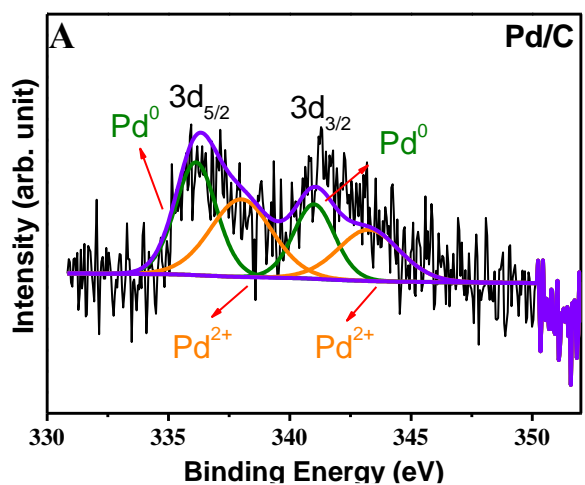

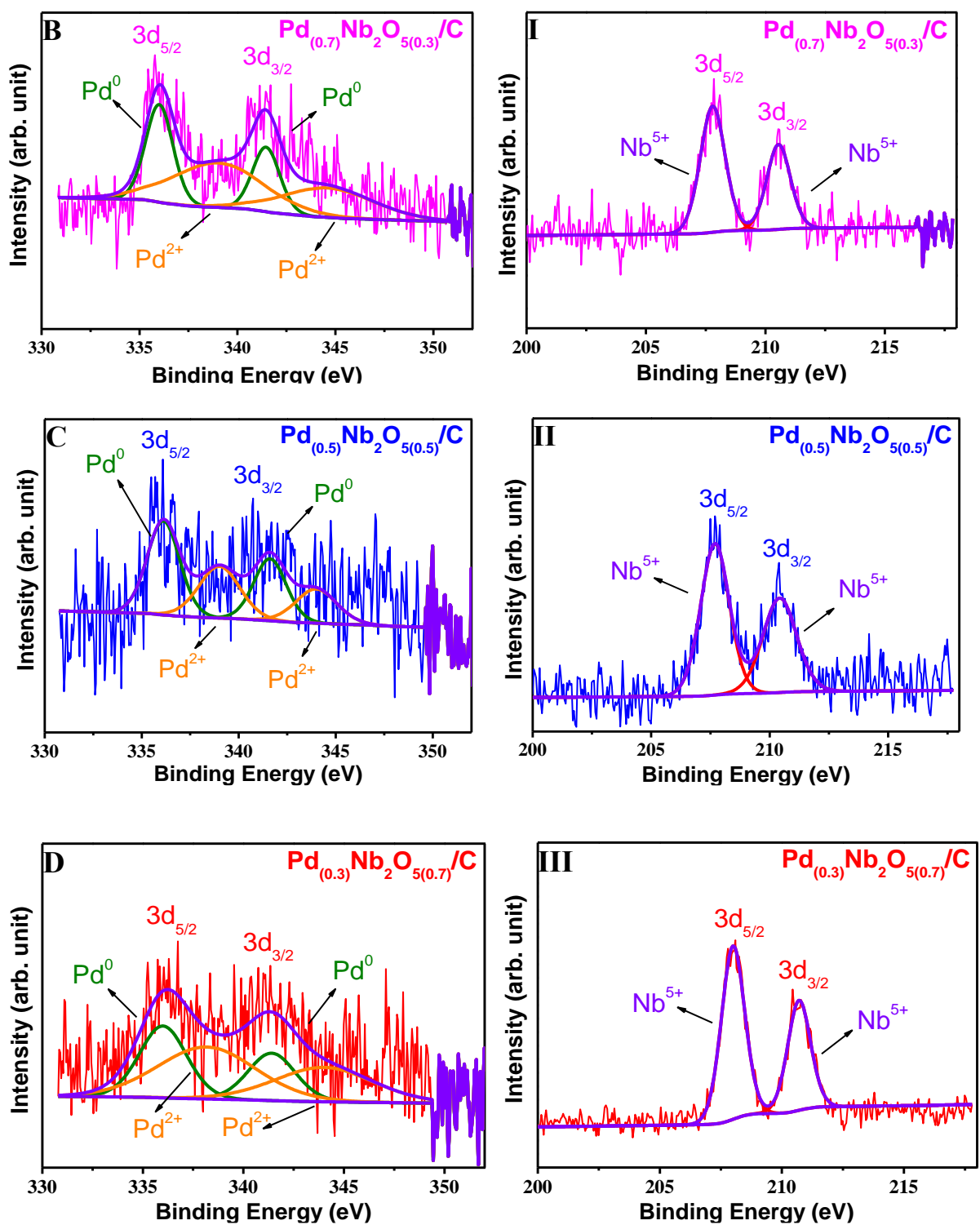

**Fig. 2.** High-resolution XPS spectra of the Pd 3d region (A-D) and Nb 3d region (I-III) for Pd/C, Pd$_{(0.7)}$Nb$_2$O$_{5(0.3)}$/C, Pd$_{(0.5)}$Nb$_2$O$_{5(0.5)}$/C, and Pd$_{(0.3)}$Nb$_2$O$_{5(0.7)}$/C catalysts.

This behavior can be understood from the intrinsic characteristics of $Nb_2O_5$, widely recognized in the literature as a reducible support that promotes strong metal–support interactions. Such interactions are attributed to the electron flow from $Nb_2O_5$ to the noble metal, favored by the difference in Fermi levels between the materials. $Nb_2O_5$ is characterized by a higher Fermi level than Pt, as well as by rapid surface redox reversibility ($Nb^{4+}/Nb^{5+}$) and the presence of oxygen vacancies that act as electron reservoirs [50]. These features promote electron transfer from $Nb_2O_5$ to Pt, increasing the electronic density in the metal, stabilizing its metallic state ($Pt^0$), and favoring the reduction of $Pt^{2+}$ to $Pt^0$ [50].

By analogy, electron transfer from $Nb_2O_5$ to Pd likely explains the decrease in Pd 3d binding energy and the higher $Pd^0$ fraction observed in the $Pd_{(0.5)}Nb_2O_{5(0.5)}/C$ catalyst. This interfacial electronic interaction contributes directly to the binding-energy shift detected in the XPS spectra and to the enhanced metallic character of Pd in the composite.

On the other hand, Pd/C, $Pd_{(0.7)}Nb_2O_{5(0.3)}/C$, and $Pd_{(0.3)}Nb_2O_{5(0.7)}/C$ exhibited a higher proportion of $Pd^{2+}$ (palladium oxide), which can inhibit the dissociative adsorption of ethanol and, consequently, compromise the catalytic activity of these catalysts [60]. According to Leybo *et al*. [61], systems with a high ratio of metallic atoms in contact with the support tend to exhibit greater electronic redistribution, especially when the support is a reducible oxide. This phenomenon favors charge transfer from the oxide to palladium, promoting the electronic enrichment of the metal and stabilizing the $Pd^0$ oxidation state. Thus, the results indicate that the higher proportion of $Pd^0$ observed in the $Pd_{(0.5)}Nb_2O_{5(0.5)}/C$ catalyst is directly associated with a more efficient metal–support interaction, resulting from a larger interfacial area available for electronic transfer.

To elucidate the morphological characteristics of the catalysts, Fig. 3 shows TEM micrographs of (A) $Nb_2O_5$, (B) Vulcan XC-72 carbon, and (C) $Pd_{(0.5)}Nb_2O_{5(0.5)}/C$. $Nb_2O_5$

appears as irregularly shaped particles with an average size of approximately 20 nm, consistent with the orthorhombic $Nb_2O_5$ reported by Shan *et al*. [62].

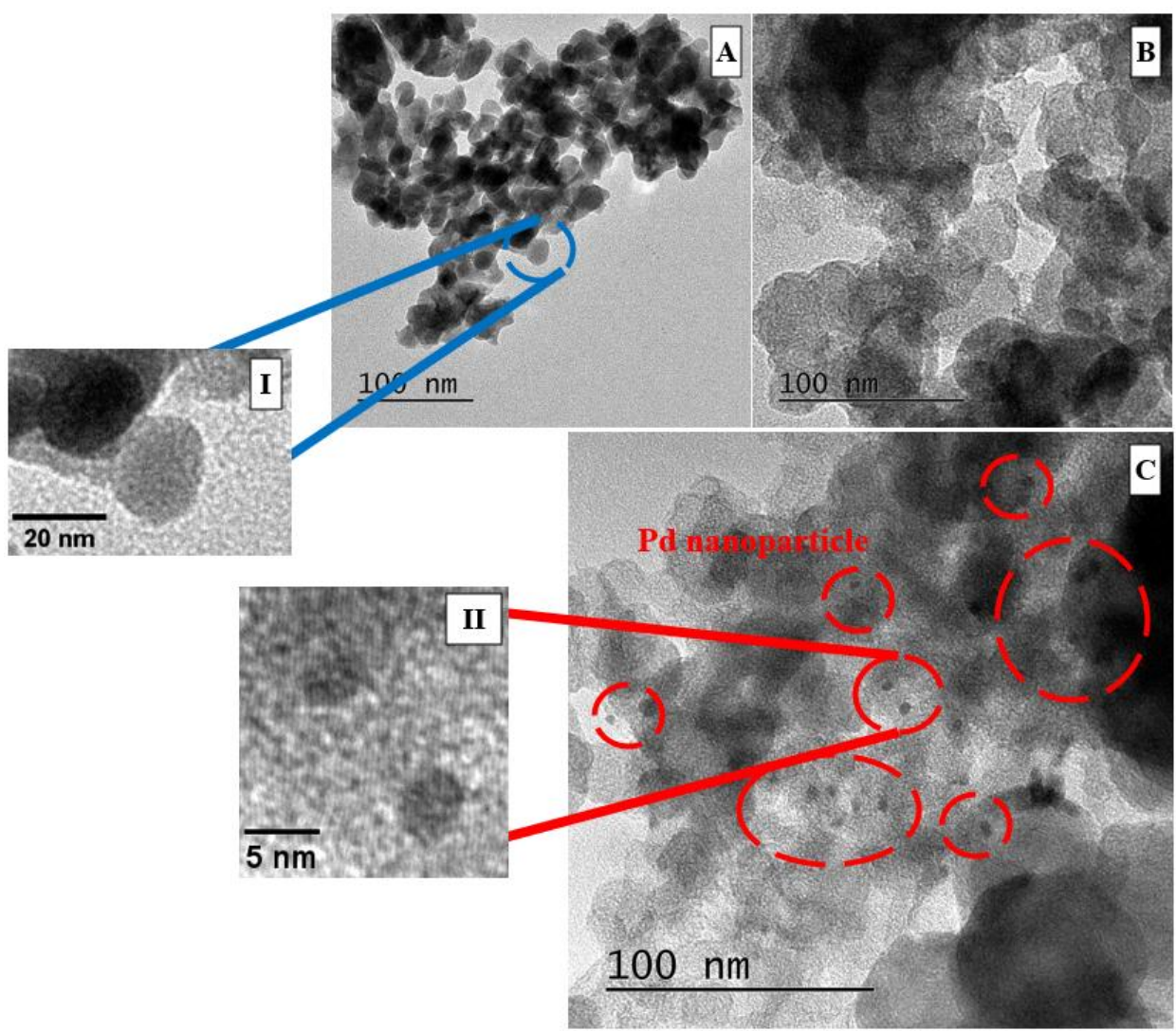

**Fig. 3.** TEM images of (A) $Nb_2O_5$, (B) Vulcan XC-72, and (C) $Pd_{(0.5)}Nb_2O_{5(0.5)}/C$. Insets: (I) representative $Nb_2O_5$ particle size; (II) representative Pd nanoparticle size.

In contrast, Vulcan XC-72 carbon (Fig. 3B) exhibits particles with more rounded and well-defined contours and larger dimensions compared to $Nb_2O_5$. The $Pd_{(0.5)}Nb_2O_{5(0.5)}/C$ catalyst (Fig. 3C) consists of Pd nanoparticles of about 5 nm in diameter uniformly dispersed on the support. Notably, no evidence of phase segregation or agglomeration between $Nb_2O_5$ and Vulcan XC-72 was observed, indicating a homogeneous distribution of these components in the catalysts. A similar morphological homogeneity was previously reported by Shao *et al*. [63] for Pd/Nb$_2$O$_5$/SiO$_2$ systems.

*3.2 Electrochemical characterization of the catalysts*

Fig. 4A shows the CO stripping curves for the Pd$_{(x)}$Nb$_2$O$_{5(y)}$/C electrocatalysts (x:y = 100:0, 70:30, 50:50, 30:70) used to investigate CO monolayer adsorption and oxidation on their surfaces [64]. Two main peaks are identified: peak "i", corresponding to CO oxidation Pd surface defect sites (twist or low-coordination sites), typically manifested as a broad signal [65]; and peak "ii", associated with the complete CO oxidation, which appears as a more intense current and is attributed to reactions occurring at more active sites, most likely extended terraces(flat), well-ordered domains of the Pd surface [60,66]. Koper *et al*. [65] investigated CO electrooxidation on Pt single-crystal surfaces in 0.1 mol L$^{-1}$ NaOH and identified up to four distinct peaks in the voltammogram. The feature between 0.72 and 0.80 V vs. RHE was associated with CO adsorbed on terrace sites, whereas the peak at 0.60 V was attributed to CO oxidation at step sites. A signal at 0.70 V was correlated with defects of specific orientation, while a broad peak emerging at approximately 0.35 V was attributed to twist sites. Comparable behavior has been observed on Pd surfaces. Chang *et al*. [67], combining Raman spectroscopy with DFT, demonstrated that Pd and Pt exhibit analogous CO adsorption characteristics, with similar peak positions and nearly equivalent shifts in the vibrational frequencies of adsorbed CO. Such assignments are in agreement with our results, confirming that the Pd–Nb$_2$O$_5$ catalysts exhibit CO oxidation behavior analogous to that reported for Pd and Pt surfaces.

As shown in Fig. 4A, the intensity of the complete CO oxidation peak (peak ii) correlates with the Pd content of the electrocatalysts, indicating that Pd serves as the primary active site for this reaction. This behavior is consistent with the intrinsically higher affinity of Pd for carbon relative to Nb$_2$O$_5$, which drives the preferential adsorption of CO on Pd surface sites.

In contrast, the oxophilic nature of $Nb_2O_5$ enables its participation through a bifunctional mechanism, in which complementary reactive species are supplied. Hence, Pd sites preferentially adsorb CO, while $Nb_2O_5$ provides oxygenated groups that facilitate its oxidation [68]. As shown in Fig. S7, the CO oxidation onset potential ($E_{COi}$) for all electrocatalysts was shifted to lower values in the binary electrocatalysts relative to Pd/C. This shift indicates that CO oxidation proceeds with lower energy demand, thereby mitigating the risk of Pd surface poisoning [69].

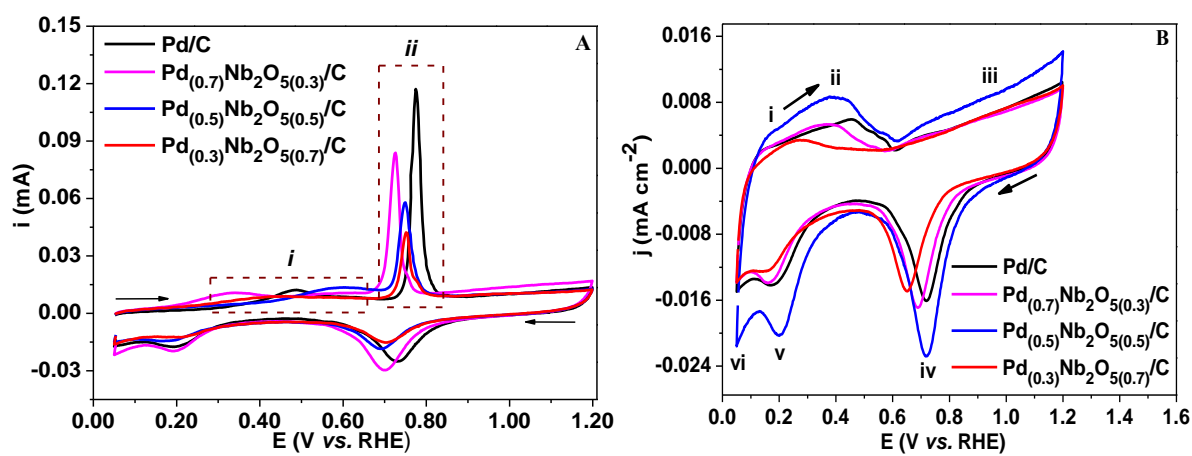

**Fig. 4.** A – CO stripping voltammograms and B – cyclic voltammograms (second cycle) of the catalysts, recorded in 1 mol $L^{-1}$ KOH at a scan rate of 20 mV $s^{-1}$. In B, the current density was normalized to the ECSA. (→) Anodic sweep; (←) Cathodic sweep.

Fig. 4B presents the cyclic voltammograms of the electrocatalysts recorded in 1.0 mol $L^{-1}$ KOH. The $Nb_2O_5$-containing catalysts displayed profiles similar to pure Pd/C, with well-defined voltammetric features. The Pd/C profile is consistent with previous reports, confirming the reliability of the synthesis procedure [70]. In the anodic (→) and cathodic (←) sweeps, six characteristic regions can be distinguished, each corresponding to a specific surface process. Regions (i and vi) are associated with hydrogen oxidation and $H_2$ formation/reduction, whereas regions (ii and v) correspond to hydrogen adsorption/desorption ($H_{ads}$) [71]. Above approximately 0.20 V *vs.* RHE, $OH^-$ adsorption begins, although it is not resolved as a distinct peak due to peak overlap [72]. In regions (iii and iv), palladium oxide formation and subsequent reduction occur [73].

The higher current densities observed for the binary PdNb$_2$O$_5$/C catalysts in regions (iii) and (iv), compared to Pd/C, highlight the contribution of Nb$_2$O$_5$ through a bifunctional mechanism. Owing to its pronounced oxophilicity, Nb$_2$O$_5$ promotes OH$^-$ adsorption on the catalyst surface, thereby enhancing the availability of oxygenated species and increasing the overall electrochemical activity [13,68].

Table 3 summarizes the ECSA values, calculated using Eq. S3, which reflects the number of accessible active sites on the catalyst surface. The Pd$_{(0.7)}$Nb$_2$O$_{5(0.3)}$/C and Pd$_{(0.3)}$Nb$_2$O$_{5(0.7)}$/C catalysts exhibited ECSA values comparable to those of Pd/C. In contrast, Pd$_{(0.5)}$Nb$_2$O$_{5(0.5)}$/C showed a lower ECSA (0.59 ± 0.02), likely due to partial encapsulation of Pd nanoparticles by Nb$_2$O$_5$ species. Such encapsulation, arising from strong metal–support interactions, reduces the fraction of exposed Pd sites by partially covering the metallic surface (see Table 1 and Fig. 2). A similar phenomenon was reported by Silva *et al.* [74] for Pt catalysts supported on carbon with mixed oxides (Ti$_{(1-X)}$Mo$_X$O$_2$), in which Mo-containing species encapsulated Pt particles and diminished their active area.

The reduced ECSA of Pd$_{(0.5)}$Nb$_2$O$_{5(0.5)}$/C may also be linked to preferential exposure of the Pd(111) facet (see Table 2), a crystallographic plane with high stability and strong affinity for oxide supports, which facilitates encapsulation [61]. Conversely, the Pd$_{(0.7)}$Nb$_2$O$_{5(0.3)}$/C and Pd$_{(0.3)}$Nb$_2$O$_{5(0.7)}$/C catalysts, despite containing less Pd, showed no significant loss of active area, suggesting weaker Pd–Nb$_2$O$_5$ interactions. In these samples, the irregular morphology of Nb$_2$O$_5$ observed in TEM (Fig. 3) likely promoted improved dispersion of Pd nanoparticles, thereby enhancing the ECSA.

**Table 3.** Electrochemically active surface area (ECSA), CO oxidation onset potential ($E_{COi}$), and CO oxidation peak potential ($E_{CO}$) of the electrocatalysts

| Electrocatalysts | $E_{COi}$ (V) | $E_{CO}$ (V) | ECSA (cm$^2$) |
|---|---|---|---|
| Pd/C | 0.49 | 0.77 | 0.72 ± 0.11 |
| Pd$_{(0.7)}$Nb$_2$O$_{5(0.3)}$/C | 0.39 | 0.72 | 0.80 ± 0.02 |
| Pd$_{(0.5)}$Nb$_2$O$_{5(0.5)}$/C | 0.43 | 0.75 | 0.59 ± 0.02 |
| Pd$_{(0.3)}$Nb$_2$O$_{5(0.7)}$/C | 0.33 | 075 | 0.67 ± 0.07 |

Comparison of the $E_{COi}$ and $E_{CO}$ values in Table 3 shows that the addition of Nb$_2$O$_5$ to the catalysts decreases the CO oxidation potential, thereby mitigating active-site poisoning. In addition to the bifunctional mechanism, an electronic effect also contributes to this behavior. The latter arises from electronic interactions between Pd and Nb$_2$O$_5$, which lower the CO adsorption energy on the Pd surface and consequently facilitate its oxidation [49]. Among the binary systems, however, the Pd$_{(0.5)}$Nb$_2$O$_{5(0.5)}$/C catalyst presented the highest CO oxidation onset potential. This performance may be related to its smaller ECSA and the greater exposure of the Pd(111) facet (Table 2). Li *et al*. [75] reported that Pd electrocatalysts with a high proportion of exposed (111) planes display enhanced intrinsic electrocatalytic activity but reduced CO tolerance. A similar trend is evident in Table 3, where catalysts with lower (111) surface exposure exhibit greater resistance to CO poisoning.

*3.3 Ethanol oxidation*

The cyclic voltammograms for ethanol oxidation are shown in Fig. 5A. Peak (i), observed during the anodic sweep, corresponds to the oxidation of ethanol and its chemically adsorbed intermediates, which are converted into oxidized products. Peak (ii), identified during the cathodic sweep, reflects the oxidation and removal of residual intermediates upon potential reversal [76]. Using peak (i) as reference, the Pd$_{(0.5)}$Nb$_2$O$_{5(0.5)}$/C electrocatalyst exhibited the highest current density (1.76 mA cm$^{-2}$), outperforming Pd/C (1.59 mA cm$^{-2}$). The Pd$_{(0.7)}$Nb$_2$O$_{5(0.3)}$/C and Pd$_{(0.3)}$Nb$_2$O$_{5(0.7)}$/C electrocatalysts reached current densities of 1.49 mA cm$^{-2}$ and 0.95 mA cm$^{-2}$, respectively.

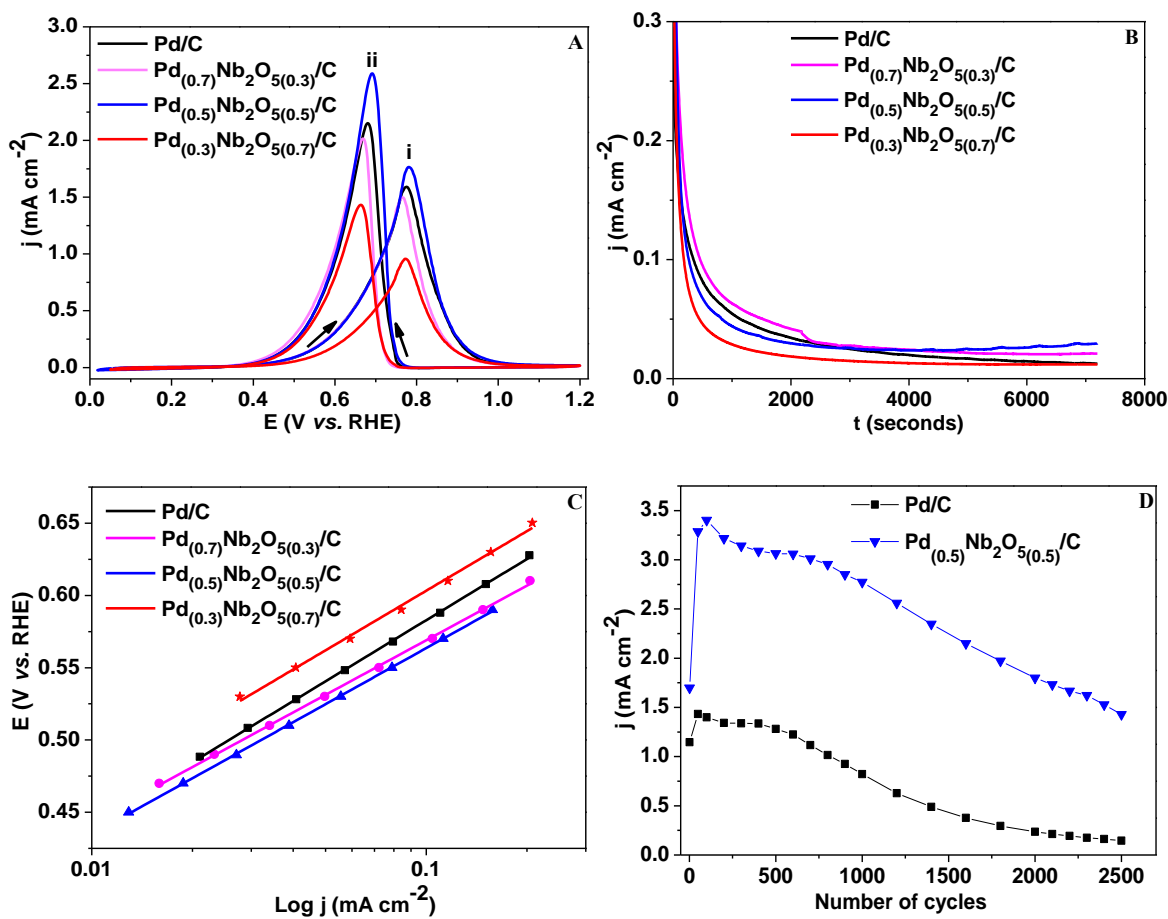

**Fig. 5.** Electrochemical studies for ethanol oxidation in the absence of light (current densities normalized to ECSA): C – Cyclic voltammograms (second cycle) recorded at 20 mV s$^{-1}$. (→) Anodic sweep and (←) Cathodic sweep. D – Chronoamperometric curves with the electrode polarized at 0.55 V. E – Tafel plots derived from the quasi-steady-state polarization curves (Fig. S8), obtained at 1 mV s$^{-1}$. F – Maximum current densities for ethanol oxidation on Pd/C and Pd$_{(0.5)}$Nb$_2$O$_{5(0.5)}$/C catalysts after 2500 consecutive cycles. All measurements were performed in 1 mol L$^{-1}$ KOH + 1 mol L$^{-1}$ ethanol aqueous solution.

Fig. 5B shows the chronoamperometric curves of the catalysts. All samples displayed an initial rapid decrease in current density, attributed to double-layer charging and the fast adsorption of carbonaceous intermediates, including CO-like species, on the Pd surface [77,78,79]. This early-stage decay is commonly observed during alkaline ethanol electrooxidation and reflects the rapid occupation of the most active sites by strongly bound adsorbates. Subsequently, the current decreased more gradually and reached a quasi-steady state after approximately 2500 s. This behavior indicates that surface poisoning does not progress to complete site blockage. Instead, a dynamic equilibrium is established between the continuous generation of strongly adsorbed intermediates and their oxidative removal by $OH^-$-derived surface species, which partially restores active sites and sustains a stable residual activity at a fixed potential [80]. At the end of the tests, Pd/C exhibited a specific activity of 0.012 mA cm$^{-2}$, while Pd$_{(0.5)}$Nb$_2$O$_{5(0.5)}$/C, Pd$_{(0.7)}$Nb$_2$O$_{5(0.3)}$/C, and Pd$_{(0.3)}$Nb$_2$O$_{5(0.7)}$/C achieved activities of 0.029 mA cm$^{-2}$, 0.021 mA cm$^{-2}$, and 0.011 mA cm$^{-2}$, respectively. Accordingly, Pd$_{(0.5)}$Nb$_2$O$_{5(0.5)}$/C and Pd$_{(0.7)}$Nb$_2$O$_{5(0.3)}$/C catalysts displayed activities 2.4- and 1.8-fold higher than Pd/C, indicating improved tolerance to poisoning under steady-state operation.

In addition to current density values, the oxidation onset potential provides further insight into electrocatalytic activity. As shown in the Tafel plots (Fig. 5C), the Pd$_{(0.5)}$Nb$_2$O$_{5(0.5)}$/C catalyst exhibited the lowest onset potential for ethanol oxidation (0.45 V), indicating a reduced energy barrier and enhanced catalytic efficiency. The corresponding kinetic parameters extracted from the Tafel plots are and coefficient of determination ($R^2$) summarized in Table S3.

The EIS analysis presented in Fig. S9 provides quantitative information on the charge-transfer resistance at the catalytic interface during the electrochemical oxidation of ethanol [81]. The decrease in the semicircle diameter in the Nyquist plots indicates a lower charge-transfer resistance (Rct), reflecting more favorable kinetics for ethanol oxidation [70]. The Pd$_{(0.5)}$Nb$_2$O$_{5(0.5)}$/C catalyst exhibited the smallest semicircle diameter, corresponding to a resistance of 68.1 Ω, whereas Pd/C, Pd$_{(0.7)}$Nb$_2$O$_{5(0.3)}$/C, and Pd$_{(0.3)}$Nb$_2$O$_{5(0.7)}$/C showed values of 92.4 Ω, 91.4 Ω, and 350 Ω, respectively. These results indicate that the 0.5:0.5 composition promotes ethanol oxidation and the removal of poisoning intermediates more effectively than the other catalysts under the conditions tested.

The superior performance of Pd$_{(0.5)}$Nb$_2$O$_{5(0.5)}$/C, both in terms of specific activity and tolerance to active-site poisoning, and the lowest resistance in charge transfer can be related to an optimized metal–oxide interfacial contact provided by the 0.5:0.5 ratio, which maximizes the synergistic interaction between Pd and Nb$_2$O$_5$ [61,82]. This proportion promotes stronger metal–support interactions, involving electronic transfer between Nb$_2$O$_5$ and Pd, and enhances the operation of the bifunctional mechanism [49]. This interpretation aligns with the higher metallic Pd fraction observed by XPS (Table S2) and the enhanced electrochemical activity shown in Fig. 4B. Furthermore, electronic transfer may induce a downward shift of the Pd *d*-band center, favoring desorption of strongly adsorbed species and thereby improving CO tolerance and catalytic activity [83].

By contrast, the Pd$_{(0.3)}$Nb$_2$O$_{5(0.7)}$/C catalyst, with lower Pd concentration and a higher Nb$_2$O$_5$ fraction, likely promoted excessive accumulation of hydroxyl groups on the surface. This led to high coverage of active sites, which hinders dissociative ethanol adsorption and accelerates surface poisoning [84].

Although the Pd$_{(0.7)}$Nb$_2$O$_{5(0.3)}$/C catalyst exhibited a high ECSA (Table 3), its lower Nb$_2$O$_5$ content likely reduced the interfacial area with Pd. This limitation diminishes the efficiency of both the bifunctional mechanism and the electronic effect, lowering the ability to oxidize intermediates and consequently reducing ethanol oxidation activity [85]. Another contributing factor negatively influencing ethanol oxidation may be the predominance of the C2 pathway, which produces little CO [86]. This is because catalysts with greater exposure of the Pd(111) facet are more active toward this pathway and display higher activity in ethanol oxidation [73,87]. Thus, the lower current density of $Pd_{(0.7)}Nb_2O_{5(0.3)}/C$ compared to $Pd_{(0.5)}Nb_2O_{5(0.5)}/C$ can be attributed primarily to reduced (111) surface exposure (Table 2) and diminished efficiency of the bifunctional mechanism and electronic effect.

To evaluate the surface stability of the catalysts (Fig. 5D), a durability test comprising 2500 cyclic voltammograms in the presence of ethanol was performed (Fig. S10). The evaluation included Pd/C and the best-performing catalyst, $Pd_{(0.5)}Nb_2O_{5(0.5)}/C$. Both materials initially showed an increase in catalytic activity, likely due to surface reorganization that optimized the distribution of active sites [88]. After this initial stage, Pd/C exhibited a sharp decline in performance, with current density decreasing by 89.8% after the $50^{th}$ cycle. In contrast, the binary catalyst maintained higher stability, with current density beginning to decrease only after the 100th cycle and showing a total loss of 57.9% by the $2500^{th}$ cycle.

This degradation is attributed to surface deactivation by strongly adsorbed intermediates, palladium oxide formation, and structural changes in the metallic nanoparticles [88,89]. The incorporation of $Nb_2O_5$ markedly enhanced stability during ethanol oxidation, primarily through the bifunctional mechanism. Where $OH^-$ species generated on the oxide assist in the oxidation of intermediates through an electronic effect. $Nb_2O_5$ acts then as an electron reservoir, donating charge to Pd during oxidation and thereby weakening the adsorption strength of intermediates. Together, these effects mitigate poisoning and improve the resistance of Pd toward oxidation processes [50].

### 3.4 Electrochemical Behavior of the $Nb_2O_5/C$ Catalyst

To investigate the photoelectrocatalytic properties of $Nb_2O_5/C$, photocurrent measurements were performed by applying 0.7 V (corresponding to the onset potential for oxygen oxidation) in 1 mol $L^{-1}$ KOH for 1200 s, with UV light pulses of 200 s (Fig. 6A). The absence of ethanol in the solution prevents interference from its oxidation, enabling an exclusive assessment of the effect of light on current generation by the oxide. A clear increase in current density was observed when the UV lamp was switched on and a decrease when it was turned off, confirming the sensitivity of $Nb_2O_5$ to ultraviolet irradiation. This behavior is consistent with the UV absorption band detected in the UV-Vis-DRS spectrum (Fig. S11). Similar photocurrent responses have been reported for $Nb_2O_5$, highlighting its potential as a photocatalyst [90,91].

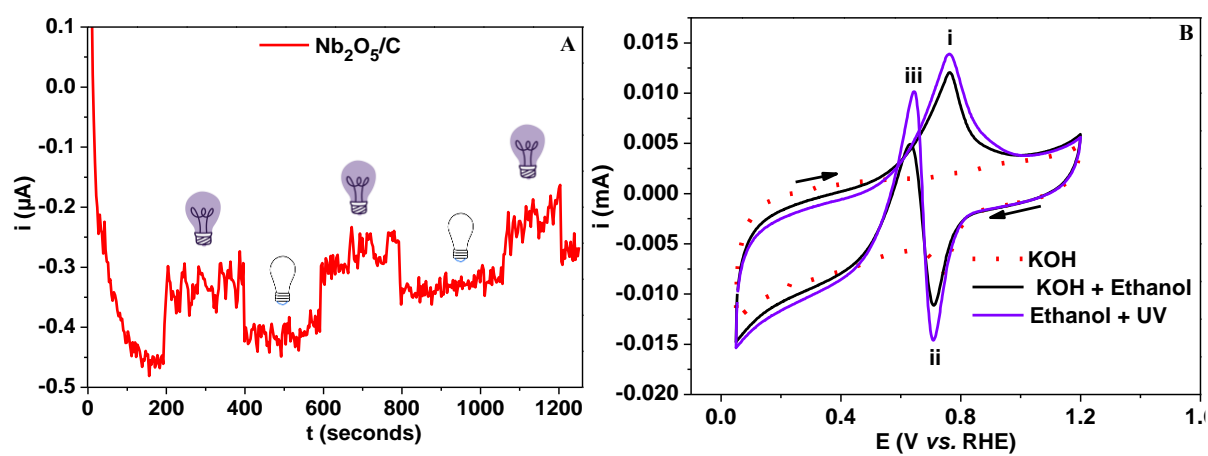

**Fig. 6.** A – Photocurrent response of $Nb_2O_5$/C recorded by applying 0.7 V in 1 mol $L^{-1}$ KOH for 1200 s with 200 s UV light pulses, indicated by purple lamp icons (on) and white lamp icons (off). B – Cyclic voltammograms of $Nb_2O_5$/C in 1 mol $L^{-1}$ KOH solution (red dotted curve), in 1 mol $L^{-1}$ KOH solution + ethanol (black curve), and in the same ethanol-containing solution under UV irradiation (purple curve). Scan rate: 20 mV $s^{-1}$. (→) Anodic sweep; (←) Cathodic sweep.

Fig. 6B shows the cyclic voltammetry results obtained for the $Nb_2O_5$/C catalyst, first in the absence and presence of ethanol (without UV excitation) and subsequently under light irradiation. In the cathodic sweep, a characteristic peak at ~0.73 V (peak ii), associated with oxygen reduction, appeared in all three voltammograms, consistent with literature reports [14,92,93]. The addition of ethanol to the electrolyte resulted in the emergence of two characteristic oxidation peaks (peaks i and iii), observed in both anodic and cathodic scans, indicating that $Nb_2O_5$/C displays modest electrocatalytic activity toward ethanol oxidation. Comparable activity has been documented for other metal oxides, including manganese oxide ($Mn_3O_4$) and cobalt oxide ($Co_3O_4$) [94], manganese–molybdenum oxide ($MnMoO_4$) [95], and titanium dioxide ($TiO_2$) [96].

As shown in the CV curves, UV irradiation enhanced the current density of the ethanol oxidation peaks by 8%, confirming the photoelectrocatalytic activity of $Nb_2O_5$. This enhancement is attributed to the generation of electron–hole pairs that participate in charge transfer to the external circuit and/or to the formation of hydroxyl radicals (•OH), which promote ethanol oxidation [97].

*3.5 Electrocatalytic Activity in Photoassisted Ethanol Oxidation*

Given the photoelectrocatalytic properties of $Nb_2O_5$, $Pd_{(x)}Nb_2O_{5(y)}$/C catalysts were investigated for photoassisted ethanol electrooxidation under UV illumination. Their performance was evaluated by cyclic voltammetry with and without light excitation (Fig. 7A). In both conditions, two characteristic ethanol oxidation peaks (i and ii) were observed. In the absence of light, $Pd_{(0.5)}Nb_2O_{5(0.5)}$/C exhibited peak (i) current densities of 1.44 mA cm$^{-2}$, which increased to 2.10 mA cm$^{-2}$ under UV irradiation, corresponding to a 1.5-fold enhancement. For $Pd_{(0.7)}Nb_2O_{5(0.3)}$/C, the peak (i) current density increased from 1.07 mA cm$^{-2}$ to 1.27 mA cm$^{-2}$ (1.2-fold), whereas $Pd_{(0.3)}Nb_2O_{5(0.7)}$/C displayed the lowest activity, with peak (i) values of 0.68 mA cm$^{-2}$ (dark) and 0.97 mA cm$^{-2}$ (UV illumination), corresponding to a 1.4-fold increase.

The current enhancement observed for $Pd_{(0.5)}Nb_2O_{5(0.5)}$/C and $Pd_{(0.3)}Nb_2O_{5(0.7)}$/C upon UV illumination can be attributed to the higher $Nb_2O_5$ content, which promotes photocatalytic activity [98]. Nevertheless, Pd plays a crucial role, as the increase in current density for the Pd-containing photoelectrocatalysts was substantially higher than for $Nb_2O_5$/C alone, which showed only a 1.1-fold increment. Notably, even the least effective Pd-based photoelectrocatalyst outperformed bare $Nb_2O_5/C$, suggesting that Pd enhances the photoelectrocatalytic process. This synergistic effect is consistent with findings by Murcia *et al.* [99], who demonstrated improved photocatalytic oxidative dehydrogenation of ethanol with $Pt/TiO_2$ compared to $TiO_2$ alone. The presence of Pt facilitated hole–electron separation by serving as electron-trapping sites, thereby reducing recombination and enhancing photocatalysis. A similar mechanism was proposed in another study [100].

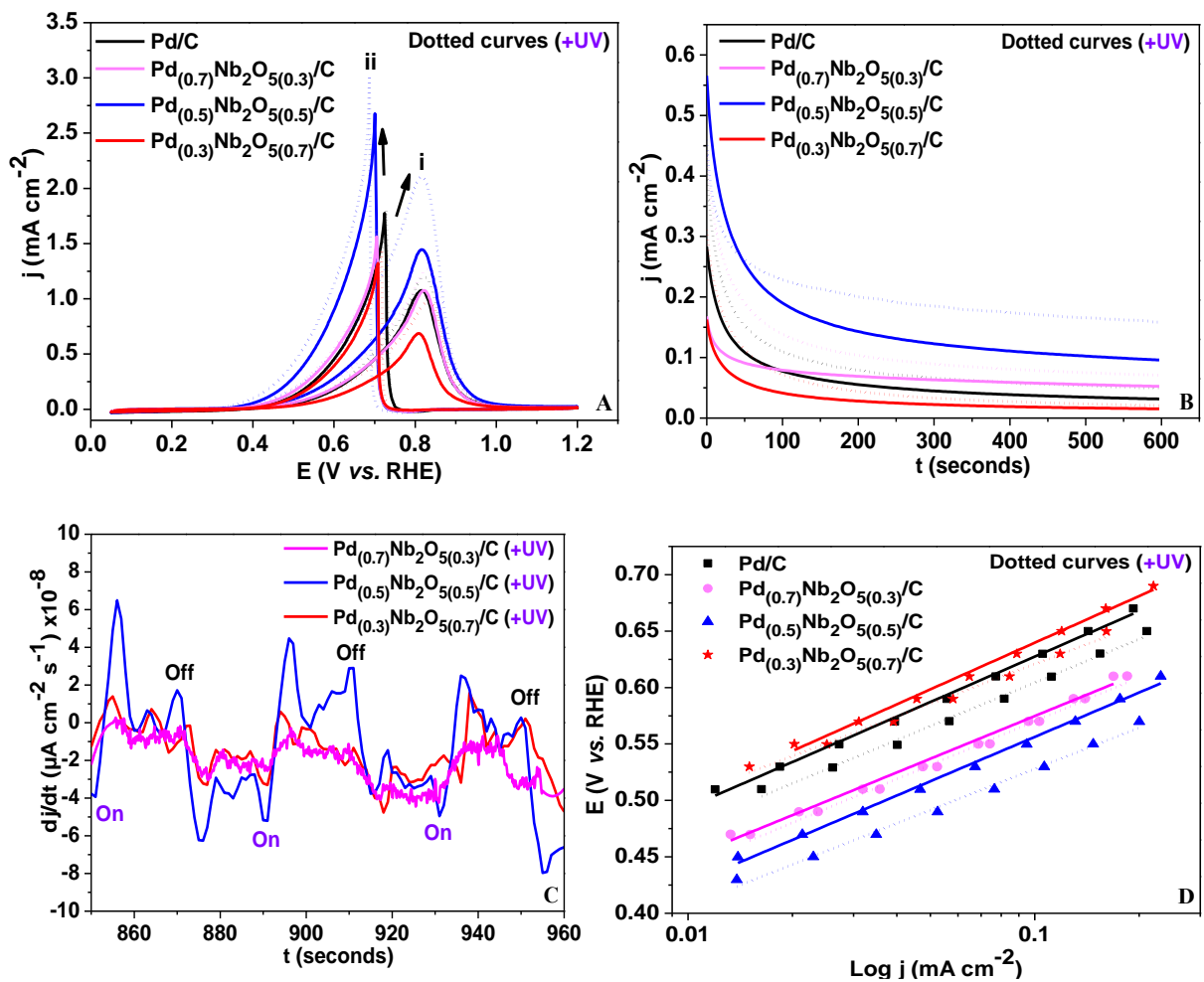

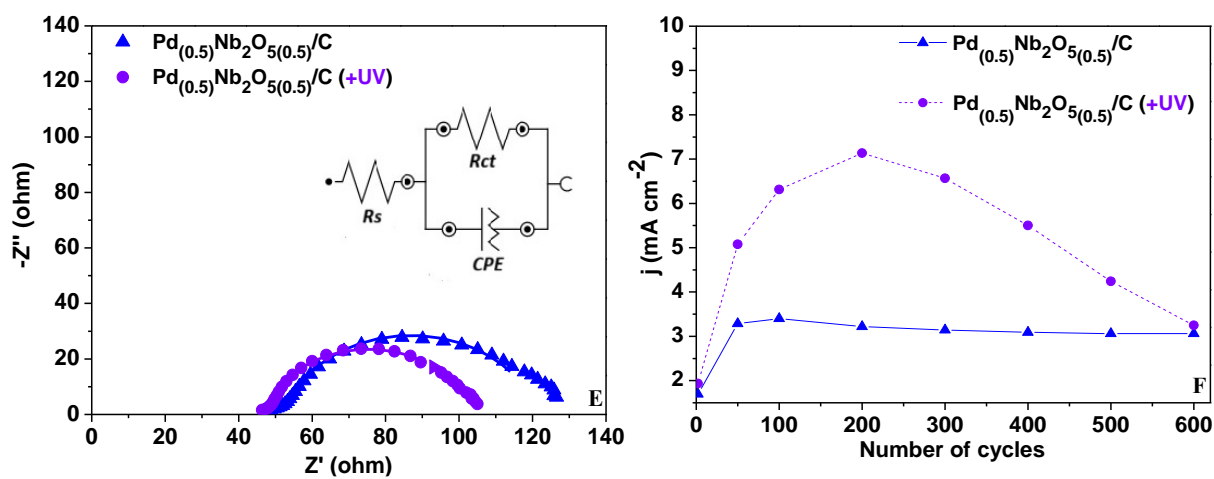

**Fig. 7.** Electrochemical studies of ethanol oxidation under UV irradiation (dotted curves) and in the dark (solid curves) in aqueous solution containing 1 mol L$^{-1}$ KOH + 1 mol L$^{-1}$ ethanol (current densities normalized by ECSA): A – Cyclic voltammograms (second cycle) recorded at 20 mV s$^{-1}$. (→) Anodic sweep; (←) Cathodic sweep. B – Chronoamperometric responses with electrodes polarized at 0.55 V. C – Time-derivative photocurrent responses (dj/dt) obtained UV on/off chronoamperometric experiments (Fig. S12) at 0.55 V vs. RHE for Pd$_{(0.7)}$Nb$_2$O$_{5(0.3)}$/C, Pd$_{(0.5)}$Nb$_2$O$_{5(0.5)}$/C, and Pd$_{(0.3)}$Nb$_2$O$_{5(0.7)}$/C. The labels "On" and "Off" denote the periods when UV irradiation was applied or removed, respectively. D – Tafel plots derived from quasi-steady-state polarization curves. E – Nyquist plots of Pd$_{(0.5)}$Nb$_2$O$_{5(0.5)}$/C recorded in the dark (blue curve) and under UV illumination (purple curve). Measurements were performed using ten points per decade in the frequency range of 1000 to 0.1 Hz, at 0.77 V and an amplitude of 5 mV. The inset shows the equivalent circuit used for fitting (Rs, Rct, CPE). F – Maximum current densities of Pd$_{(0.5)}$Nb$_2$O$_{5(0.5)}$/C after 600 consecutive cycles.

The Pd/C electrocatalyst exhibited a peak (i) current density of 1.07 mA cm$^{-2}$, which increased to 1.19 mA cm$^{-2}$ under photoassisted electrooxidation. Similar enhancements under UV irradiation have been reported for Pt/C catalysts in the photoassisted electrocatalytic oxidation of methanol [101,102] and ethanol [103]. This improvement is often attributed to the plasmonic effect, wherein the interaction of light with free electrons in metals induces collective electron oscillations, thereby enhancing charge transfer processes [104]. For Pd, the plasmonic resonance typically lies in the ultraviolet region, supporting its contribution to improved photoelectrocatalytic activity under illumination [105]. When comparing Pd/C with the Pd$_{(0.5)}$Nb$_2$O$_{5(0.5)}$/C photoelectrocatalyst, light irradiation produced an approximately twofold increase in catalytic activity, further highlighting the role of photocatalysis in enhancing the electrochemical performance of these materials.

Chronoamperometric measurements (Fig. 7B) were performed with and without UV light to evaluate catalyst poisoning during ethanol oxidation over 600 s. Under UV illumination, Pd$_{(0.5)}$Nb$_2$O$_{5(0.5)}$/C achieved a final current of 0.158 mA cm$^{-2}$, corresponding to a 1.6-fold increase relative to the dark condition (0.096 mA cm$^{-2}$) and a 5.1-fold enhancement compared with Pd/C (0.031 mA cm$^{-2}$). This improvement is attributed to the higher generation of •OH under UV irradiation, which facilitates the removal of adsorbed intermediates from the catalyst surface, thereby enhancing tolerance to poisoning and sustaining higher photoelectrocatalytic activity [106,107]. Notably, all catalysts demonstrated improved resistance to poisoning in the presence of ultraviolet light.

The time-derivative under successive UV on/off cycles is shown in Fig. 7C for the binary electrocatalysts in a solution containing 1 mol L$^{-1}$ KOH and 1 mol L$^{-1}$ ethanol. When the UV light is switched on, positive dj/dt peaks reflect a prompt increase in catalytic current, consistent with a fast photoinduced contribution that facilitates the oxidation of ethanol and its intermediates. Among the evaluated catalysts, Pd$_{(0.5)}$Nb$_2$O$_{5(0.5)}$/C and Pd$_{(0.3)}$Nb$_2$O$_{5(0.7)}$/C exhibited the most intense responses, indicating higher efficiency in electron–hole separation under the tested conditions [108]. After the light is turned off, dj/dt becomes negative, reflecting an abrupt decay in current. This behavior supports a predominantly photoinduced electronic effect rather than a slow photothermal contribution [109].

For the best-performing catalyst, $Pd_{(0.5)}Nb_2O_{5(0.5)}$/C, we extracted a characteristic time constant from the exponential fit of the photocurrent rise (Eq. S4), obtaining τ = 1.65 s with excellent fitting quality ($R^2$ = 0.9936). This result indicates that the system reaches a steady photo-assisted regime rapidly upon irradiation. The obtained τ value lies within the range reported for oxide-metal photoelectrocatalytic systems, which varies from 4.94 s down to 0.09 s, supporting the kinetic plausibility of the observed UV-enhanced response [110].

The electrolyte temperature was monitored during the UV-assisted experiments using a digital thermometer. In short-duration measurements, the solution temperature remained stable at 24 ± 1 °C. For longer tests, including polarization curves, chronoamperometry, and stability cycling, a thermostatic bath was used to control the electrolyte temperature, which was maintained at 27 ± 1 °C. These measures minimize bulk photothermal contributions to the measured current response. In this context, Li, Li, and Liu [102] compared cyclic voltammograms for methanol oxidation recorded at 25 °C under UV–visible irradiation with those recorded at 27 °C in the dark. Their results show that such a modest increase in temperature does not induce abrupt changes or significantly modify the overall electrochemical response.

The Tafel plots for ethanol oxidation with and without UV irradiation (Fig. 7D) were derived from the quasi-steady-state polarization curves (Fig. S13), recorded at 1 mV s$^{-1}$. The results reveal the influence of light on the onset potential ($E_i$) of ethanol oxidation. As summarized in Table S4, $Pd_{(0.5)}Nb_2O_{5(0.5)}$/C exhibited the lowest $E_i$ under UV illumination (0.43 V), corresponding to a 0.02 V decrease relative to the dark value (0.45 V). In contrast, Pd/C displayed $E_i$ = 0.51 V with no detectable shift under illumination and was 0.08 V higher than that of $Pd_{(0.5)}Nb_2O_{5(0.5)}$/C. $Pd_{(0.7)}Nb_2O_{5(0.3)}$/C and $Pd_{(0.3)}Nb_2O_{5(0.7)}$/C exhibited $E_i$ values of 0.47 V and 0.55 V, respectively, with only $Pd_{(0.3)}Nb_2O_{5(0.7)}$/C showing a 0.02 V decrease under illumination, indicating a reduction in the energy required to initiate the reaction, although this effect was insufficient to surpass the other catalysts. In contrast, $Pd_{(0.7)}Nb_2O_{5(0.3)}/C$ did not exhibit any change in $E_i$, indicating an essentially negligible photoassisted response compared to the other binary catalysts.

This light-induced shift contrasts with literature reports in which only the temperature was varied. In those studies, the current density increased with temperature, but the onset potential remained essentially unchanged even when the temperature was raised from 15 to 55 °C, demonstrating that heating alone does not affect $E_i$ [111]. Therefore, the change observed in our system cannot be ascribed solely to a photothermal effect but must instead be dominated by a light-activated process, consistent with photocatalysis being the primary mechanism underlying the enhanced electrocatalytic activity.

Overall, catalysts with higher $Nb_2O_5$ content in their composition showed improved onset potentials when exposed to UV irradiation. This behavior may be attributed to more efficient generation of photogenerated holes ($h^+$), which interact with adsorbed hydroxyl ions to produce hydroxyl radicals. These radicals facilitate electron removal from ethanol, thereby promoting its oxidation [100,112]. Similar behavior was reported by Cai *et al*. [113], who observed that the reduction in onset potential for ethanol oxidation on a $Pt/ZrO_2$/n-Si catalyst under visible light was due to electron donation from ethanol, which filled photogenerated holes in the semiconductor, suppressed charge recombination, and enabled oxidation at lower potentials.

To further understand the influence of UV irradiation on interfacial kinetics, EIS was also recorded for $Pd_{(0.5)}Nb_2O_{5(0.5)}/C$ in the dark and under UV light (Fig. 7E). Under illumination, the Nyquist plots revealed a clear reduction in semicircle diameter, and the fitted Rct decreases from 68.1 Ω to 55.2 Ω. This reduction indicates that UV irradiation enhances the interfacial kinetics of ethanol oxidation, making the catalytic interface more efficient under irradiation and corroborates the photoassisted activity increase observed in the voltammetric measurements [100].

To evaluate catalyst stability under UV excitation (Fig. 7F), 600 consecutive cyclic voltammograms were recorded in the presence of ethanol (Fig. S10) using the best-performing material, $Pd_{(0.5)}Nb_2O_{5(0.5)}$/C. In the absence of light, a slight decrease in current density was observed beginning at approximately the 100th cycle. Under UV irradiation, however, the current density initially increased, reaching a maximum near the 200th cycle. This behavior highlights the beneficial effect of UV light on catalyst stability and performance during ethanol oxidation. Previous studies have similarly reported that the presence of light increases the stability and prolongs the catalyst lifetime, which is an essential factor for practical applications [108,114].

Nevertheless, a sharp decline in current density was observed after the 200th cycle under illumination. This effect is likely associated with degradation of the carbon support, exacerbated by the generation of reactive oxygen species, particularly •OH, generated through the interaction of UV light with $Nb_2O_5$ and Pd. This interpretation aligns with the mechanism proposed by Cai *et al.* [115], who showed that the simultaneous presence of Pt, $O_2$, and water promotes the formation of oxidizing species (•OH and •OOH) which corrode the carbon support and lead to up to a 70% loss of electrochemically active Pt area after 500 potential cycles between 0 and 1.2 V (NHE).

To investigate the mass transport mechanism of ethanol oxidation on the $Pd_{(0.5)}Nb_2O_{5(0.5)}$/C catalyst, cyclic voltammetry measurements were carried out at different scan rates, both with and without UV irradiation (Fig. 8A–C). In both conditions, the peak (i) current density increased with increasing scan rate. Plots of the maximum peak (i) current density as a function of the square root of the scan rate (Fig. 8B–D) revealed strong linearity, indicating that the process is diffusion-controlled. This behavior suggests that ethanol transport to the electrode surface is the rate-limiting step, rather than surface activation or chemical transformation [116,117].

The ratio of the diffusion coefficients under illumination ($D_{light}$) and in the dark ($D_{dark}$), calculated using Eq. S5 was 2.35. This ratio indicates that UV light accelerates the movement of charge carriers, facilitating faster electron transfer at the electrode surface. Consequently, the kinetics of the electrochemical process are enhanced, which directly translates into an improved efficiency for ethanol oxidation [107].

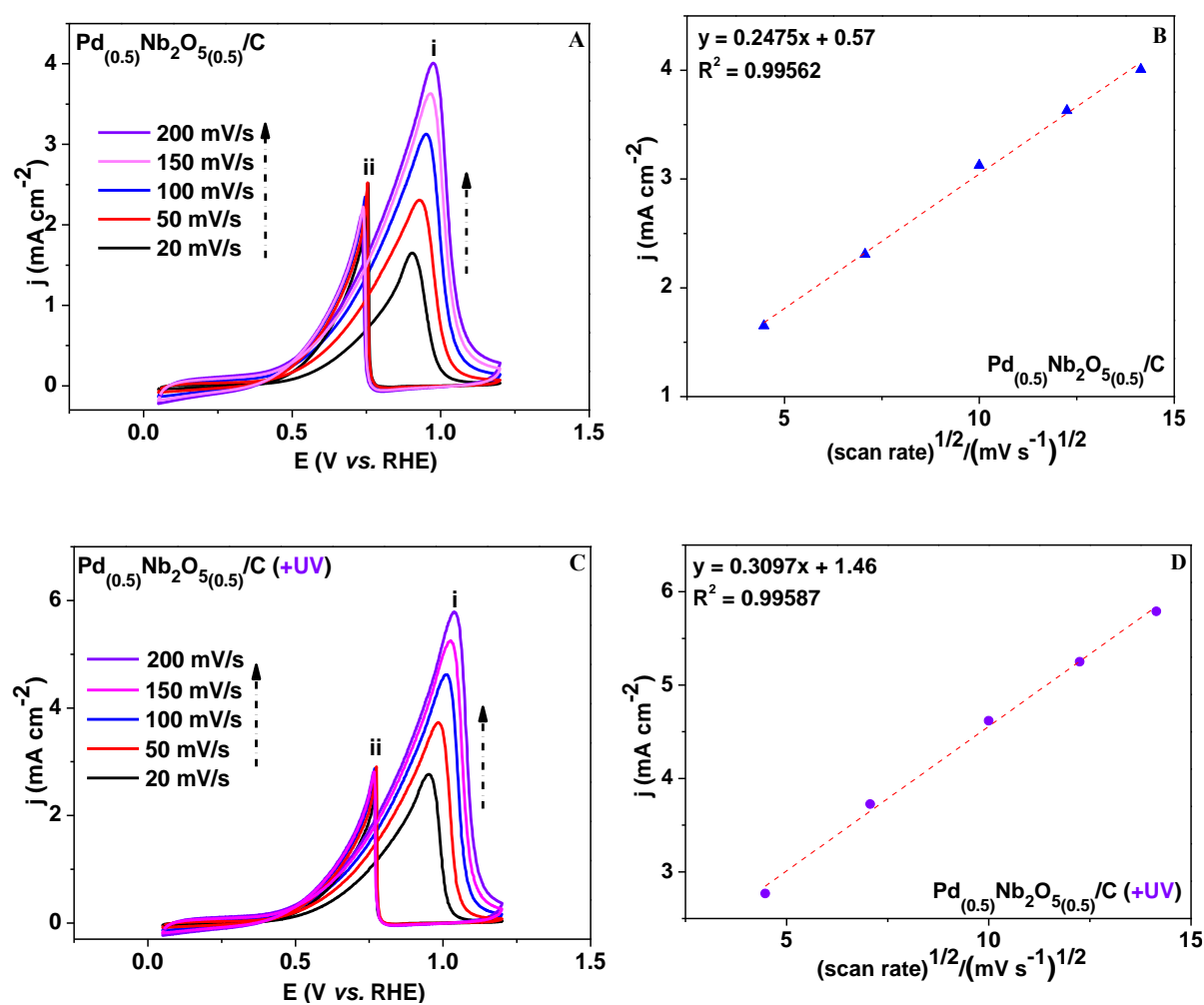

**Fig. 8.** Cyclic voltammograms (second cycle) of $Pd_{(0.5)}Nb_2O_{5(0.5)}/C$ recorded at different scan rates in 1 mol L$^{−1}$ KOH + 1 mol L$^{−1}$ ethanol solution: A – without UV excitation and C – with UV excitation. Linear relationship between the peak current density for ethanol oxidation and the square root of the scan rates: B – without UV excitation and D – with UV excitation.

Table 4 provides a comparative overview of the activity of the $Pd_{(0.5)}Nb_2O_{5(0.5)}/C$ catalyst developed in this study and previously reported Pd-based and/or metal–oxide catalysts used for the photoassisted electrocatalytic oxidation of ethanol. The table lists the current ratios obtained under illumination ($I_{light}$) and in the dark ($I_{dark}$), along with catalyst composition, electrolyte medium, ethanol concentration, and the type of light source employed. This comparison enables an assessment of the effect of photoactivation across different catalytic systems.

**Table 4.** Comparison of the $Pd_{(0.5)}Nb_2O_{5(0.5)}/C$ catalyst developed in this study with Pd-based and/or metal oxide-based catalysts reported for the photoassisted electrocatalytic oxidation of ethanol. The table includes the $I_{light}/I_{dark}$ ratio, catalyst composition, electrolyte medium, ethanol concentration, and type of light source employed

| Catalyst | Electrolyte (mol L$^{-1}$) | Ethanol concentration (mol L$^{-1}$) | $I_{light}/I_{dark}$ | Type of light | Reference |
|---|---|---|---|---|---|
| $Pd_{(0.5)}Nb_2O_{5(0.5)}/C$ | KOH (1.0) | 1.0 | 1.46 | UV | This study |
| PtNiRu/TiO$_2$ | H$_3$PO$_4$ (0.5) | 1.0 | 2.48 | Xenon | [118] |
| Au/ZnO | NaOH (1.0) | 1.5 | 2.20 | Xenon | [119] |
| Pt/Fe$_2$O$_3$/C | KOH (1.0) | 1.0 | 1.13 | Xenon | [103] |
| Au/rGo/TiO$_2$ | NaOH (1.0) | 1.5 | 1.61 | Xenon | [120] |
| ZnO@Pd@Pt | NaOH (1.0) | 0.25 + 0.25 (Methanol) | 1.21 | UV | [121] |
| IrO$_2$/Pd/TaON | NaOH (0.2) | 5% volume | 1.33 | Xenon | [122] |
| Pd/BP-CB | KOH (1.0) | 1.0 | 1.34 | Xenon | [100] |
| Pt/ZrO$_2$/n-Si | KOH (1.0) | 1.0 | 2.21 | Xenon | [113] |

The $Pd_{(0.5)}Nb_2O_{5(0.5)}/C$ catalyst achieved an $I_{light}/I_{dark}$ ratio of 1.46, demonstrating a marked enhancement in activity under UV illumination and performance comparable to or superior to that of most catalysts tested in the same electrolyte medium (1 mol L$^{-1}$ KOH). Reports of higher current ratios typically involved different light sources, alternative electrolytes, or more noble metals such as Pt, Ru, or Au, which substantially increase material cost. Although $Nb_2O_5$ has been scarcely explored in photoassisted fuel cells, the present findings underscore its potential as an efficient and economical catalyst.

*3.6 Proposed Mechanism for the Photoassisted Electrocatalytic Oxidation of Ethanol*

Direct elucidation of the reaction intermediates and confirmation of potential alterations in the oxidation pathway—including a possible contribution of $Nb_2O_5$ to C–C bond cleavage—would require operando techniques such as in situ FTIR or online mass spectrometry, which lie beyond the scope of the present study. Accordingly, the proposed mechanism integrates our electrochemical results with the established knowledge in the literature on Pd-catalyzed alcohol oxidation systems, providing a coherent framework to rationalize the trends observed.

The literature establishes two widely accepted reaction pathways for ethanol oxidation, which may contribute to the process observed in this photoassisted electrocatalytic system: the electrocatalytic pathway (Eqs. 1–4) and the photoelectrocatalytic pathway (Eqs. 5–9). Pd acts as the main active site for ethanol oxidation due to its strong affinity for carbon-containing species. Meanwhile, $Nb_2O_5$, owing to its pronounced oxophilicity, adsorbs hydroxyl ions ($OH_{ads}$) and facilitates the approach of hydroxyl ions ($OH^-$) to the catalyst surface. This dual action enhances ethanol oxidation by lowering the potential required for the reaction and mitigating Pd surface poisoning by intermediates such as $CH_3CO$ and CO, thereby leading to higher current densities [117].

Under UV excitation, photons (hv) interact with Pd* [104], inducing the plasmonic effect that enhances charge transfer. Simultaneously, $Nb_2O_5$ promotes electron–hole ($e^-/h^+$) separation. The photogenerated electrons are either transferred to Pd or trapped in $Nb_2O_5$ structural defects, which decreases recombination losses. The photogenerated holes react with OH$_{ads}$ to generate hydroxyl radicals (•OH), species with greater oxidative power that accelerate ethanol oxidation and more effectively remove poisoning intermediates compared to hydroxyl ions [100,112,114].

Electrocatalytic mechanism:

$$Nb_2O_5 + OH^- \rightarrow Nb_2O_5\text{-}OH_{ads} + e^- \tag{1}$$

$$Pd + CH_3CH_2OH \rightarrow Pd(CH_3CH_2OH)_{ads} \tag{2}$$

$$Pd(CH_3CH_2OH)_{ads} + 3OH^- \rightarrow Pd(CH_3CO)_{ads} + 3H_2O + 3e^- \tag{3}$$

$$Pd(CH_3CO)_{ads} + Nb_2O_5\text{-}OH_{ads} + OH^- \rightarrow Nb_2O_5 + Pd + CH_3COO^- + H_2O \tag{4}$$

Photoelectrocatalytic mechanism:

$$Nb_2O_5 + h\nu \rightarrow e^- + h^+ \tag{5}$$

$$h^+ + Nb_2O_5\text{-}OH_{ads} \rightarrow \bullet OH \tag{6}$$

$$Pd\text{-}(CH_3CH_2OH)_{ads} + \bullet OH + OH^- \rightarrow CO_2 + H_2O + e^- \tag{7}$$

$$\text{Intermediates}(CO_{ads}) + \bullet OH \rightarrow CO_2 + H^+ + e^- \tag{8}$$

$$Pd + h\nu \rightarrow Pd^* + e^- \tag{9}$$

Fig. 9 presents a schematic representation of the possible interactions between the catalysts, the solution species, and UV irradiation, constructed based on the electrochemical trends and the physicochemical characterizations observed. Panel A depicts the working electrode immersed in the electrolyte solution, with its surface coated by the catalyst composed of Pd, Nb$_2$O$_5$, and carbon.

Panel B of Fig. 9 illustrates the interaction of the Pd/C catalyst with ethanol and/or its intermediates under UV irradiation. In this representation, Pd particles appear in greater quantity and with smaller crystallite size compared to those in the binary catalysts, as confirmed by the XRD results (Table 1). These features likely enhance the exposure of active sites, consistent with the high ECSA values obtained (Table 3). However, the illustration also highlights a higher concentration of ethanol relative to hydroxyl ions in the vicinity of Pd, reflecting its low oxophilicity. This limitation hinders sufficient hydroxyl ion adsorption, thereby reducing the capacity to remove poisoning intermediates [13]. Consequently, Pd exhibits low tolerance to poisoning, as evidenced by the chronoamperometric profiles (Fig. 5B and Fig. 7B) and CO-stripping data (Table 3), in addition to requiring higher overpotentials to initiate ethanol oxidation, as indicated by the Tafel plots (Fig. 5C and Fig. 7D).

Panel C of Fig. 9 depicts the $Pd_{(0.7)}Nb_2O_{5(0.3)}$/C catalyst. The material contains a higher Pd content and a lower proportion of $Nb_2O_5$ relative to the other binary catalysts. Despite this composition, its ECSA was comparable to that of Pd/C (Table 3). This behavior is likely associated with the irregular surface morphology of $Nb_2O_5$ (Fig. 3), which enhances the exposure of Pd active sites. The schematic also indicates a higher concentration of ions and hydroxyl radicals near the catalyst compared with Pd/C. This effect arises from the greater oxophilicity [68] and photocatalytic activity [26] of $Nb_2O_5$, which promote ethanol oxidation. As a result, under UV irradiation, $Pd_{(0.7)}Nb_2O_{5(0.3)}$/C exhibited improved tolerance to poisoning (Fig. 7B) and higher specific activity than Pd/C (Fig. 7A).

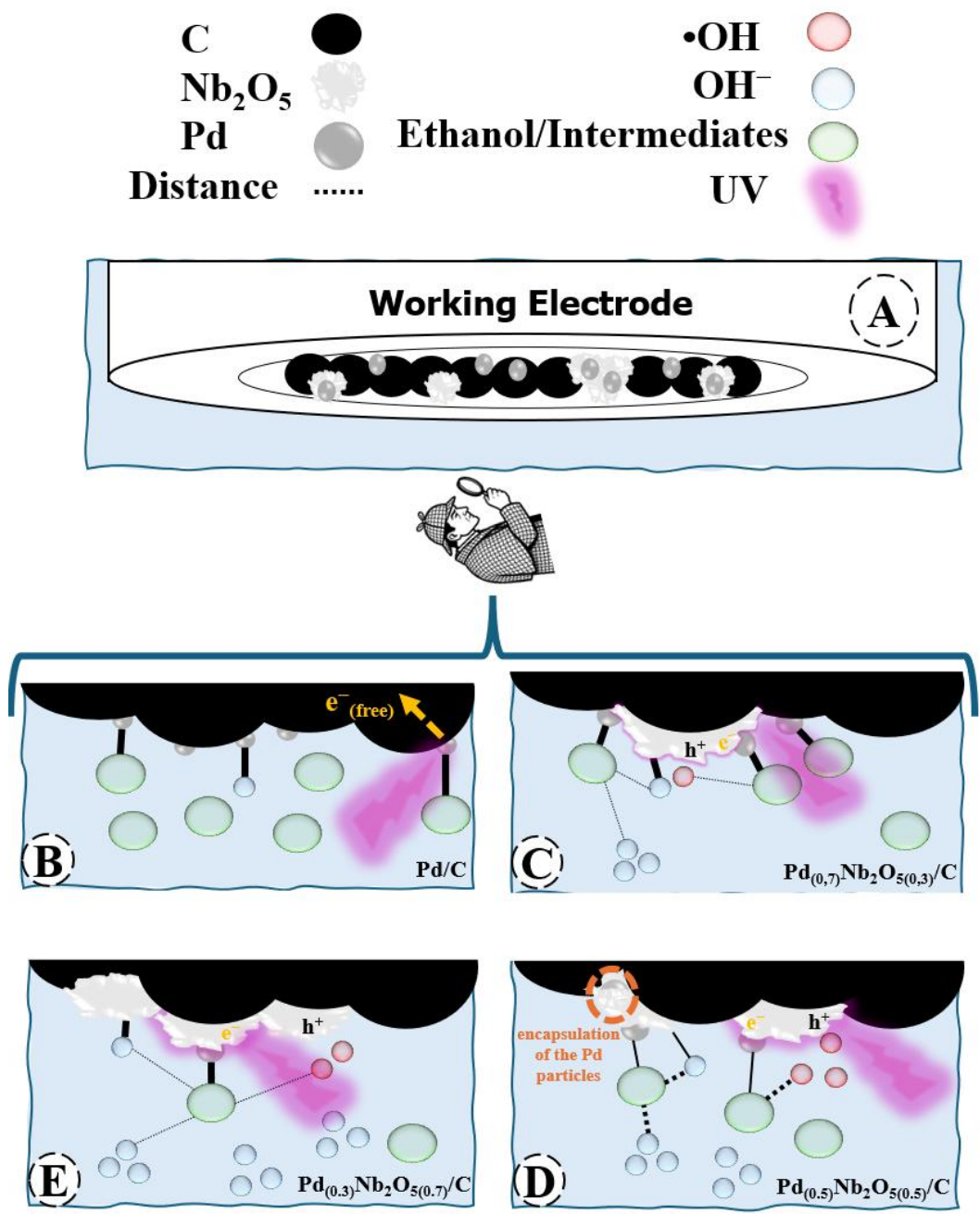

**Fig. 9.** A – Schematic representation of the working electrode surface coated with carbon, $Nb_2O_5$, and Pd nanoparticles, immersed in the electrolyte. Comparative behavior of the catalysts under illumination: for Pd/C (B), plasmonic excitation generates free electrons ($e^-$), but the low oxophilicity of Pd limits $OH^-$ adsorption, reducing tolerance to poisoning; in $Pd_{(0.7)}Nb_2O_{5(0.3)}/C$ (C), the smaller $Nb_2O_5$ content restricts bifunctional and electronic effects; in $Pd_{(0.3)}Nb_2O_{5(0.7)}/C$ (E), excess $Nb_2O_5$ favors hydroxyl accumulation; and in $Pd_{(0.5)}Nb_2O_{5(0.5)}/C$ (D), the partial encapsulation of Pd decreases active site exposure, but the balanced ratio optimizes electron transfer and OH• radical formation, maximizing the removal of intermediates and catalytic performance.

Panel D of Fig. 9 presents the $Pd_{(0.5)}Nb_2O_{5(0.5)}/C$ catalyst. XPS analysis (Fig. 2) revealed stronger $Pd–Nb_2O_5$ interactions in this material than in the other binary systems. However, this catalyst displayed the lowest ECSA (Table 3), possibly due to the equimolar $Pd:Nb_2O_5$ ratio that favors a larger interfacial contact area. This inference is supported by the slight variation observed in the Pd lattice parameter (Table 1), suggesting intensified electronic interactions between the metal and the support. Such interactions may have led to partial encapsulation of Pd particles, thereby reducing the electrochemically active area. Nonetheless, the enhanced metal–support interaction facilitated the formation of metallic $Pd^o$ and likely decreased the adsorption strength of ethanol and its intermediates, reducing the distance between these species and hydroxyl ions or radicals. The schematic also illustrates a higher concentration of hydroxyl species than in Panel B, consistent with the higher $Nb_2O_5$ fraction in the composition, which promotes the oxidation of intermediates. The close association between Pd and $Nb_2O_5$ likely strengthened the bifunctional mechanism, intensified electronic effects, and promoted hole separation [82]. As a result, the $Pd_{(0.5)}Nb_2O_{5(0.5)}/C$ catalyst exhibited higher tolerance to poisoning (Fig. 7B), required lower energy to initiate ethanol oxidation (Fig. 7D), achieved higher current densities than the other catalysts (Fig. 7A), demonstrated greater efficiency in hole–electron separation (Fig. 7C), and, under UV irradiation, showed lower charge-transfer resistance (Fig. 7E).

Panel E of Fig. 9 illustrates the $Pd_{(0.3)}Nb_2O_{5(0.7)}/C$ catalyst. According to the XRD data (Table 1) and XPS results (Fig. 2), this catalyst exhibits structural characteristics and electronic interactions similar to those of the $Pd_{(0.7)}Nb_2O_{5(0.3)}/C$ catalyst. In both cases, the interfacial interaction between Pd and $Nb_2O_5$ appears relatively weak, suggesting that the bifunctional mechanism predominates during ethanol oxidation. Besides, the higher amount of $Nb_2O_5$ in $Pd_{(0.3)}Nb_2O_{5(0.7)}/C$, compared to the other catalysts, likely increased the surface concentration of hydroxyl ions. However, this excess of $Nb_2O_5$ accelerated catalyst poisoning and hindered ethanol diffusion to Pd active sites [84], resulting in impaired performance and low current density (Fig. 5 and Fig. 7). Conversely, under UV irradiation, the elevated $Nb_2O_5$ content may have favored the generation of hydroxyl radicals, which facilitated the removal of adsorbed intermediates from the catalyst surface. This effect resulted in enhanced photoelectrocatalytic activity, reflected in the second-highest $I_{light}/I_{dark}$ ratio among the catalysts studied and in the presence of slightly defined photocurrent transients (Fig. 7A and Fig. 7C).

## 4. Conclusions

This study demonstrates the successful development of binary photoelectrocatalysts composed of Pd nanoparticles combined with $Nb_2O_5$ supported on carbon. We demonstrate that integrating Pd (5 nm diameter) with orthorhombic $Nb_2O_5$ nanoparticles on carbon support yields efficient binary photoelectrocatalysts for ethanol oxidation. By systematically evaluating the role of $Nb_2O_5$ incorporation and UV irradiation, we established that the Pd–$Nb_2O_5$ interface enhances electrocatalytic activity through complementary bifunctional and electronic effects, while simultaneously mitigating surface poisoning. Under illumination, $Nb_2O_5$ promoted charge separation and the generation of reactive oxygen species, further lowering the onset potential, increasing tolerance to intermediates, and improving catalyst stability. Among the synthesized materials, $Pd_{(0.5)}Nb_2O_{5(0.5)}/C$ exhibited the most favorable balance between metal–support interaction and photocatalytic contribution, delivering superior current density, a reduced onset potential, and a fivefold improvement in poisoning resistance compared with Pd/C. Importantly, $Nb_2O_5$, a relatively underexplored and cost-effective oxide, demonstrated significant promise as a photoactive co-catalyst in alcohol electrooxidation. In summary, Pd–$Nb_2O_5$ interfaces effectively couple electrocatalysis and photocatalysis, enabling the rational design of photo-assisted anodes for alkaline fuel cells and related oxidation reactions. The insights gained here provide a foundation for developing cost-effective, light-responsive catalysts beyond Pd systems.

**Declaration of competing interest**

The authors declare that they have no known competing financial interests or personal relationships that could have appeared to influence the work reported in this paper.

**Data availability**

Data will be made available on request.

**Acknowledgments**

This study was financed by CAPES (grant 88887.807550/2023-00), CNPq (grants 305287/2022-2, 383416/2023-0, and 307866/2022-0), and the FAPITEC/SE.

**References**

[1] Ahmad, T.; Zhang, D. A critical review of comparative global historical energy consumption and future demand: The story told so far, Energy Reports. 6 (2020) 1973–1991, https://doi.org/10.1016/j.egyr.2020.07.020.
[2] Balezentis, T. Shrinking ageing population and other drivers of energy consumption and $CO_2$ emission in the residential sector: A case from Eastern Europe, Energy Policy. 140 (2020) 111433, https://doi.org/10.1016/j.enpol.2020.111433.
[3] Mendonça, A. K. de S.; *et al*. Hierarchical modeling of the 50 largest economies to verify the impact of GDP, population and renewable energy generation in $CO_2$ emissions, Sustainable Production and Consumption. 22 (2020) 58–67, https://doi.org/10.1016/j.spc.2020.02.001.
[4] Nwaka, I. D.; *et al*. Agricultural production and $CO_2$ emissions from two sources in the ECOWAS region: New insights from quantile regression and decomposition analysis, Science of The Total Environment. 748 (2020) 141329, https://doi.org/10.1016/j.scitotenv.2020.141329.
[5] Paramati, S. R.; Shahzad, U.; Doğan, B. The role of environmental technology for energy demand and energy efficiency: Evidence from OECD countries, Renewable and Sustainable Energy Reviews. 153 (2022) 111735, https://doi.org/10.1016/j.rser.2021.111735.
[6] Kasaeian, A.; *et al*. Integration of solid oxide fuel cells with solar energy systems: A review, Applied Thermal Engineering. 224 (2023) 120117, https://doi.org/10.1016/j.applthermaleng.2023.120117.

[7] Singla, M. K.; Nijhawan, P.; Oberoi, A. S. Hydrogen fuel and fuel cell technology for cleaner future: A review, Environmental Science and Pollution Research. 28 (2021), https://doi.org/10.1007/s11356-020-12231-8.

[8] Thomas, J. M.; *et al*. Decarbonising energy: The developing international activity in hydrogen technologies and fuel cells, Journal of Energy Chemistry. 51 (2020) 405–415, https://doi.org/10.1016/j.jechem.2020.03.087.

[9] Shaari, N.; *et al*. Progress and challenges: Review for direct liquid fuel cell, International Journal of Energy Research. 45 (2021) 6644–6688, https://doi.org/10.1002/er.6353.

[10] Bai, J.; *et al*. Nanocatalysts for electrocatalytic oxidation of ethanol, ChemSusChem. 12 (2019) 2117–2132, https://doi.org/10.1002/cssc.201803063.

[11] Badwal, S. P. S.; *et al*. Direct ethanol fuel cells for transport and stationary applications – A comprehensive review, Applied Energy. 145 (2015) 80–103, https://doi.org/10.1016/j.apenergy.2015.02.002.

[12] Akhairi, M. A. F.; Kamarudin, S. K. Catalysts in direct ethanol fuel cell (DEFC): An overview, International Journal of Hydrogen Energy. *41* (2016) 4214–4228, https://doi.org/10.1016/j.ijhydene.2015.12.145.

[13] Ma, L.; Chu, D.; Chen, R. Comparison of ethanol electro-oxidation on Pt/C and Pd/C catalysts in alkaline media, International Journal of Hydrogen Energy. *37* (2012) 11185–11194, https://doi.org/10.1016/j.ijhydene.2012.04.132.

[14] Hassan, K. M.; *et al*. Electrocatalytic oxidation of ethanol at Pd, Pt, Pd/Pt and Pt/Pd nanoparticles supported on poly(1,8-diaminonaphthalene) film in alkaline medium, RSC Advances. (2018), *8* (28), 15417–15426, https://doi.org/10.1039/C7RA13694C.

[15] Fu, X.; *et al*. Noble metal-based electrocatalysts for alcohol oxidation reactions in alkaline media, Advanced Functional Materials, *32* (2022) 2106401, https://doi.org/10.1002/adfm.202106401.

[16] Chae, E. P.; *et al*. Bimetallic nickel–palladium nanoparticles with low Ni content and their enhanced ethanol oxidation performance: Using a pulsed laser as modification machinery, Fuel. 321 (2022) 124108, https://doi.org/10.1016/j.fuel.2022.124108.

[17] Alvarenga, G. M.; Villullas, H. M. Transition metal oxides in the electrocatalytic oxidation of methanol and ethanol on noble metal nanoparticles, Current Opinion in Electrochemistry. *4* (2017) 39–44, https://doi.org/10.1016/j.coelec.2017.09.004.

[18] Liang, Q.; *et al*. Ultra-efficient electrooxidation of ethylene glycol enabled by Pd-loaded Fe-doped $Nb_2O_5$ with abundant oxygen vacancies, Chemical Engineering Journal. *475* (2023) 146050, https://doi.org/10.1016/j.cej.2023.146050.

[19] Rocha, T. A.; *et al*. Electrocatalytic activity of platinum–niobium nanoparticles for ethanol oxidation, Journal of The Electrochemical Society. 159 (2012) F650–F658, DOI: 10.1149/2.040210jes.

[20] Kang, S.; *et al*. Sustainable production of fuels and chemicals from biomass over niobium-based catalysts: A review, Catalysis Today. 374 (2021) 61–76, https://doi.org/10.1016/j.cattod.2020.10.029.

[21] Prado, A. G. S.; *et al*. $Nb_2O_5$ as efficient and recyclable photocatalyst for indigo carmine degradation, Applied Catalysis B: Environmental. 82 (2008) 219–224, https://doi.org/10.1016/j.apcatb.2008.01.024.

[22] Falk, G. S.; *et al*. Microwave-assisted synthesis of $Nb_2O_5$ for photocatalytic application of nanopowders and thin films, Journal of Materials Research. 32 (2017) 3271–3278, https://doi.org/10.1557/jmr.2017.93.

[23] Kotesha, K. M.; *et al*. $Nb_2O_5$–$SnS_2$–CdS heteronanostructures as efficient visible-light-harvesting materials for production of $H_2$ under solar light irradiation, Journal of Alloys and Compounds. 835 (2020) 155399, https://doi.org/10.1016/j.jallcom.2020.155399.

[24] Silva, M. A.; *et al*. Photocatalytic $Nb_2O_5$-doped $TiO_2$ nanoparticles for glazed ceramic tiles. Ceramics International. 42 (2016) 5113–5122, https://doi.org/10.1016/j.ceramint.2015.12.029.

[25] Jiao, X.; *et al*. Photocatalytic conversion of waste plastics into $C_2$ fuels under simulated natural environment conditions, Angewandte Chemie International Edition. 59 (2020) 15497–15501, https://doi.org/10.1002/anie.201915766.

[26] Su, K.; *et al*. $Nb_2O_5$-based photocatalysts, Advanced Science. 8 (2021) 2003156, https://doi.org/10.1002/advs.202003156.

[27] Ding, H.; *et al*. Niobium-based oxide anodes toward fast and safe energy storage: A review, Materials Today Nano. 11 (2020) 100082, https://doi.org/10.1016/j.mtnano.2020.100082.

[28] Pang, R.; *et al*. Polymorphs of $Nb_2O_5$ compound and their electrical energy storage applications, Materials. 16 (2023) 6956, https://doi.org/10.3390/ma16216956.

[29] Polo, A. S.; *et al*. Pt–Ru–$TiO_2$ photoelectrocatalysts for methanol oxidation. Journal of Power Sources. 196 (2011) 872–876, https://doi.org/10.1016/j.jpowsour.2010.06.076.

[30] John, S.; *et al*. Synergy of photocatalysis and fuel cells: A chronological review on efficient designs, potential materials and emerging applications, Frontiers in Chemistry. 10 (2022) https://doi.org/10.3389/fchem.2022.1038221.

[31] Hu, J.; Zhai, C.; Zhu, M. Photo-responsive metal/semiconductor hybrid nanostructure: A promising electrocatalyst for solar light enhanced fuel cell reaction, Chinese Chemical Letters. 32 (2021) 1348–1358, https://doi.org/10.1016/j.cclet.2020.09.049.

[32] Valim, R. B.; *et al*. Using carbon black modified with $Nb_2O_5$ and $RuO_2$ for enhancing selectivity toward $H_2O_2$ electrogeneration, Journal of Environmental Chemical Engineering. 9 (2021) 106787, https://doi.org/10.1016/j.jece.2021.106787.

[33] Raba, A. M.; Barba-Ortega, J.; Joya, M. R. The effect of the preparation method of $Nb_2O_5$ oxide influences the performance of the photocatalytic activity, Applied Physics A. 119 (2015) 923–928, https://doi.org/10.1007/s00339-015-9041-3.

[34] Kowal, A.; *et al*. Synthesis, characterization and electrocatalytic activity for ethanol oxidation of carbon-supported Pt, Pt–Rh, Pt–$SnO_2$ and Pt–Rh–$SnO_2$ nanoclusters, Electrochemistry Communications. 11 (2009) 724–727, https://doi.org/10.1016/j.elecom.2009.01.022.

[35] Jiang, L.; *et al*. Size-controllable synthesis of monodispersed $SnO_2$ nanoparticles and application in electrocatalysts, The Journal of Physical Chemistry B. 109 (2005) 8774–8778, https://doi.org/10.1021/jp050334g.

[36] Young, R. A. *The Rietveld Method*; International Union of Crystallography; Oxford University Press: New York, (1995).

[37] Larson, A. C.; Von Dreele, R. B. General Structure Analysis System (GSAS); Los Alamos National Laboratory, Report LAUR 86-748, (2004).

[38] Toby, B. H. EXPGUI, a graphical user interface for GSAS, Journal of Applied Crystallography 34 (2001) 210–213, https://doi.org/10.1107/S0021889801002242.

[39] Han, Q.; *et al*. Accelerating carrier separation to boost the photocatalytic $CO_2$ reduction performance of ternary heterojunction Ag–$Ti_3C_2T_x$/ZnO catalysts, RSC Advances. 14 (2024) 13719–13733, https://doi.org/10.1039/D4RA01985G.

[40] Kamble, G. S.; Ling, Y.-C. Solvothermal synthesis of facet-dependent $BiVO_4$ photocatalyst with enhanced visible-light-driven photocatalytic degradation of organic pollutant: Assessment of toxicity by zebrafish embryo, Scientific Reports. 10 (2020), https://doi.org/10.1038/s41598-020-69706-4.

[41] Zhang, D.; *et al*. Role of oxygen active species in the photocatalytic degradation of phenol using polymer-sensitized $TiO_2$ under visible light irradiation, Journal of Hazardous Materials. 163 (2009) 843–847, https://doi.org/10.1016/j.jhazmat.2008.07.036.

[42] Bhanawat, A.; Zhu, K.; Pilon, L. How do bubbles affect light absorption in photoelectrodes for solar water splitting?, Sustainable Energy & Fuels. 6 (2022) 910–924, https://doi.org/10.1039/D1SE01730F.

[43] Hassel, O.; Mark, H. Über die Kristallstruktur des Graphits. Zeitschrift für Physik. 25 (1924) 317–337, https://doi.org/10.1007/BF01327534.

[44] Pérez-Rodríguez, S.; *et al*. Pd catalysts supported onto nanostructured carbon materials for $CO_2$ valorization by electrochemical reduction, Applied Catalysis B: Environmental. 163 (2015) 83–95, https://doi.org/10.1016/j.apcatb.2014.07.031.

[45] Davey, W. P. Precision measurements of the lattice constants of twelve common metals, Physical Review. 25 (1925) 753–761, https://doi.org/10.1103/PhysRev.25.753.

[46] Raji-Adefila, B.; *et al*. Mechanochemically enabled metastable niobium tungsten oxides, Journal of the American Chemical Society. 146 (2024) 10498–10507, https://doi.org/10.1021/jacs.3c14275.

[47] Souza, F. M.; *et al*. Niobium: A promising Pd co-electrocatalyst for ethanol electrooxidation reactions, Journal of Solid State Electrochemistry. 22 (2017) 1495–1506, https://doi.org/10.1007/s10008-017-3802-1.

[48] Souza, F. M.; *et al*. Niobium increasing the electrocatalytic activity of palladium for alkaline direct ethanol fuel cell, Journal of Electroanalytical Chemistry. 858 (2020) 113824, https://doi.org/10.1016/j.jelechem.2020.113824.

[49] Rocha, T. A.; *et al*. Nb as an influential element for increasing the CO tolerance of PEMFC catalysts, Journal of Applied Electrochemistry. 43 (2013) 817–827, https://doi.org/10.1007/s10800-013-0572-z.

[50] Yang, L.; *et al*. Stabilizing Pt electrocatalysts via introducing reducible oxide support as reservoir of electrons and oxygen species, ACS Catalysis. 12 (2022) 13523–13532. https://doi.org/10.1021/acscatal.2c04158.

[51] Majumdar, S. The effects of crystallite size, surface area and morphology on the sensing properties of nanocrystalline $SnO_2$-based system, Ceramics International. 41 (2015) 14350–14358, https://doi.org/10.1016/j.ceramint.2015.07.068.

[52] Ullah, S.; *et al*. Supported nanostructured photocatalysts: The role of support–photocatalyst interactions, Photochemical & Photobiological Sciences. 22 (2022) 219–240, https://doi.org/10.1007/s43630-022-00299-9.

[53] Wadge, M. D.; *et al*. Developing highly nanoporous titanate structures via wet chemical conversion of DC magnetron sputtered titanium thin films, Journal of Colloid and Interface Science. 566 (2020) 271–283, https://doi.org/10.1016/j.jcis.2020.01.073.

[54] Yang, X.; *et al*. Facet-dependent catalytic activity of shape-controllable palladium nanocrystals with clean surface in acetophenone selective hydrogenation, Journal of Catalysis. 428 (2023) 115159, https://doi.org/10.1016/j.jcat.2023.115159.

[55] Guo, Y.; *et al*. Potential-dependent mechanistic study of ethanol electro-oxidation on palladium, ACS Applied Materials & Interfaces. 13 (2021) 16602–16610, https://doi.org/10.1021/acsami.1c04513.

[56] Xing, X.-L.; *et al*. High performance Ag-rich Pd–Ag bimetallic electrocatalyst for ethylene glycol oxidation in alkaline media, Journal of The Electrochemical Society. 165 (2018) J3259–J3265, DOI: 10.1149/2.0311815jes.

[57] Mohite, N.; *et al*. $Nb_2O_5$ porous nanotubes: Potential approach as photoanode material for dye-sensitized solar cells, Bulletin of Materials Science. 47 (2024), https://doi.org/10.1007/s12034-024-03168-6.

[58] Fu, X.; Li, R.; Zhang, Y. High electrocatalytic activity of Pt on porous Nb-doped $TiO_2$ nanoparticles prepared by aerosol-assisted self-assembly, RSC Advances. 12 (2022) 22070–22081, https://doi.org/10.1039/D2RA03821H.

[59] Xu, W.; *et al*. Niobium-doped titanium dioxide on a functionalized carbon supported palladium catalyst for enhanced ethanol electro-oxidation, RSC Advances. 7 (2017) 34618–34623, https://doi.org/10.1039/C7RA05208A.

[60] Santos, D. S.; *et al*. Selectivity and catalytic performance of Pd@Pt/C nanoparticles for methanol electrooxidation. Electrochimica Acta. 467 (2023) 143018, https://doi.org/10.1016/j.electacta.2023.143018.

[61] Leybo, D.; *et al*. Metal–support interactions in metal oxide-supported atomic, cluster, and nanoparticle catalysis, Chemical Society Reviews (2024), https://doi.org/10.1039/D4CS00527A.

[62] Shan, Y.; *et al*. Niobium pentoxide: A promising surface-enhanced Raman scattering active semiconductor substrate, npj Computational Materials. 3 (2017), https://doi.org/10.1038/s41524-017-0008-0.

[63] Shao, Y.; *et al*. $Pd/Nb_2O_5/SiO_2$ catalyst for the direct hydrodeoxygenation of biomass-related compounds to liquid alkanes under mild conditions, ChemSusChem. 8 (2015) 1761–1767, https://doi.org/10.1002/cssc.201500053.

[64] Ciapina, E. G.; Santos, S. F.; Gonzalez, E. R. Electrochemical CO stripping on nanosized Pt surfaces in acid media: A review on the issue of peak multiplicity, Journal of Electroanalytical Chemistry. 815 (2018) 47–60, https://doi.org/10.1016/j.jelechem.2018.02.047.

[65] Koper, M. T. M. Introductory lecture: Electrocatalysis—Theory and experiment at the interface, Faraday Discussions. 140 (2009) 11–24, https://doi.org/10.1039/B812859F.

[66] Obradović, M. D.; *et al*. Palladium–copper bimetallic surfaces as electrocatalysts for the ethanol oxidation in an alkaline medium, Journal of Electroanalytical Chemistry. 944 (2023) 117673, https://doi.org/10.1016/j.jelechem.2023.117673.

[67] Chang, X.; *et al*. Understanding the complementarities of surface-enhanced infrared and Raman spectroscopies in CO adsorption and electrochemical reduction, Nature Communications. 13 (2022), https://doi.org/10.1038/s41467-022-30262-2.

[68] Kayode, G. O.; Montemore, M. M. Factors controlling oxophilicity and carbophilicity of transition metals and main group metals, Journal of Materials Chemistry A. 9 (2021), 22325–22333, https://doi.org/10.1039/D1TA06453C.

[69] Zhang, B.; *et al*. Construction of graphene-wrapped Pd/TiO$_2$ hollow spheres with enhanced anti-CO poisoning capability toward photoassisted methanol oxidation reaction, ACS Sustainable Chemistry & Engineering. 9 (2021), 1352–1360, https://doi.org/10.1021/acssuschemeng.0c08129.

[70] Almeida, C. V. S.; *et al*. Rhodium-decorated palladium nanocubes supported on Ni(OH)$_2$ nanosheets/C for enhanced ethanol oxidation, Journal of Electroanalytical Chemistry. (2024) 118437, https://doi.org/10.1016/j.jelechem.2024.118437.

[71] Puglia, M. K.; Bowen, P. K. Cyclic voltammetry study of noble metals and their alloys for use in implantable electrodes, ACS Omega. 7 (2022) 34200–34212, https://doi.org/10.1021/acsomega.2c03563.

[72] Bambagioni, V.; *et al*. Energy efficiency enhancement of ethanol electrooxidation on Pd–CeO$_2$/C in passive and active polymer electrolyte-membrane fuel cells, ChemSusChem. 5 (2012) 1266–1273, https://doi.org/10.1002/cssc.201100738.

[73] Wang, L.; et al. Deactivation of palladium electrocatalysts for alcohols oxidation in basic electrolytes, Electrochimica Acta. 177 (2015) 100–106. https://doi.org/10.1016/j.electacta.2015.02.026.

[74] Silva, C.; et al. Effect of the reductive treatment on the state and electrocatalytic behavior of Pt in catalysts supported on Ti$_{0.8}$Mo$_{0.2}$O$_2$–C composite, Reaction Kinetics, Mechanisms and Catalysis. 135 (2021) 29–47, https://doi.org/10.1007/s11144-021-02131-4.

[75] Li, G.; et al. Dependence of palladium catalysts on the crystal plane in the methanol oxidation reaction, The Journal of Physical Chemistry C. 127 (2023) 1822–1827, https://doi.org/10.1021/acs.jpcc.2c07445.

[76] Liu, C.; et al. One-step synthesis of AuPd alloy nanoparticles on graphene as a stable catalyst for ethanol electro-oxidation, International Journal of Hydrogen Energy. 41 (2016) 13476–13484, https://doi.org/10.1016/j.ijhydene.2016.05.194.

[77] Miao, Z.; *et al*. Morphology-control and template-free fabrication of bimetallic Cu–Ni alloy rods for ethanol electro-oxidation in alkaline media, Journal of Alloys and Compounds. 855 (2020) 157438, https://doi.org/10.1016/j.jallcom.2020.157438.

[78] Hajnajafi, M. *et al*. Nanoscale Engineering of Building Blocks to Synthesize a Three-Dimensional Architecture of Pd Aerogel as a Robust Self-Supporting Catalyst toward Ethanol Electrooxidation, Energy & Fuels. 35 (2021) 3396–3406, https://doi.org/10.1021/acs.energyfuels.0c04213.

[79] Farsadrooh, M.; Noroozifar, M.; Modarresi-alam, A. R. An easy and eco-friendly method to fabricate three-dimensional Pd-M (Cu, Ni) nanonetwork structure decorated on the graphene nanosheet with boosted ethanol electrooxidation activity in alkaline medium. International, Journal of Hydrogen Energy. 44 (2019) 28821–28832, https://doi.org/10.1016/j.ijhydene.2019.09.048.

[80] BARBOSA, A. F. B. *et al*. Effect of the Random Defects Generated on the Surface of Pt(111) on the Electro-oxidation of Ethanol: An Electrochemical Study, ChemPhysChem. 20 (2019) 3045–3055, https://doi.org/10.1002/cphc.201900544.

[81] GUO, J. *et al*. New understandings of ethanol oxidation reaction mechanism on Pd/C and Pd$_2$Ru/C catalysts in alkaline direct ethanol fuel cells, Applied Catalysis B: Environmental. 224 (2018) 602–611, https://doi.org/10.1016/j.apcatb.2017.10.037.

[82] Singh, R.; *et al*. Synthesis of CeO$_x$-decorated Pd/C catalysts by controlled surface reactions for hydrogen oxidation in anion exchange membrane fuel cells, Advanced Functional Materials. 30 (2020), https://doi.org/10.1002/adfm.202002087.

[83] Liu, R.; *et al*. Reversed charge transfer to modulate the d-band center of Pd for efficient direct H$_2$O$_2$ synthesis, ACS Catalysis. 14 (2024) 3955–3965, https://doi.org/10.1021/acscatal.3c05910.

[84] Cermenek, B.; *et al*. Alkaline ethanol oxidation reaction on carbon supported ternary PdNiBi nanocatalyst using modified instant reduction synthesis method, Electrocatalysis. 11 (2020) 203–214, https://doi.org/10.1007/s12678-019-00577-8.

[85] Olmos, C. M.; *et al*. Size, nanostructure, and composition dependence of bimetallic Au–Pd supported on ceria–zirconia mixed oxide catalysts for selective oxidation of benzyl alcohol, Journal of Catalysis. 375 (2019) 44–55, https://doi.org/10.1016/j.jcat.2019.05.002.

[86] Yerga, M. D.; Henriksson, G.; Cornell, A. Insights on the ethanol oxidation reaction at electrodeposited PdNi catalysts under conditions of increased mass transport, International Journal of Hydrogen Energy. 46 (2021) 1615–1626, https://doi.org/10.1016/j.ijhydene.2020.10.103.

[87] Neto, V. E. S.; *et al*. Realising the activity benefits of Pt preferential (111) surfaces for ethanol oxidation in a nanowire electrocatalyst, Electrochimica Acta. 348 (2020) 136206, https://doi.org/10.1016/j.electacta.2020.136206.

[88] Obradović, M. D.; *et al*. Electrochemical oxidation of ethanol on palladium–nickel nanocatalyst in alkaline media, Applied Catalysis B: Environmental. 189 (2016) 110–118, https://doi.org/10.1016/j.apcatb.2016.02.039.

[89] El Attar, A.; *et al*. Effect of electrochemical activation on the performance and stability of hybrid (PPy/$Cu_2O$ nanodendrites) for efficient ethanol oxidation in alkaline medium, Journal of Electroanalytical Chemistry. 885 (2021) 115042, https://doi.org/10.1016/j.jelechem.2021.115042.

[90] De Araújo, M. A.; *et al*. Contrasting transient photocurrent characteristics for thin films of vacuum-doped "grey" $TiO_2$ and "grey" $Nb_2O_5$, Applied Catalysis B: Environmental. 237 (2018) 339–352, https://doi.org/10.1016/j.apcatb.2018.05.065.

[91] Zhang, H.; *et al*. Photocatalytic selective oxidation of 5-hydroxymethylfurfural to 2,5-diformylfuran over $Nb_2O_5$ under visible light, ACS Sustainable Chemistry & Engineering. 5 (2017) 3517–3523, https://doi.org/10.1016/j.apcatb.2018.05.065.

[92] Zhang, X.; *et al*. Catalytically active single-atom niobium in graphitic layers, Nature Communications. 4 (2013), https://doi.org/10.1038/ncomms2929.

[93] Hazarika, K. K.; *et al*. $Fe_xCo_{3-x}O_4$ nanohybrids anchored on a carbon matrix for high-performance oxygen electrocatalysis in alkaline media, ChemElectroChem. 9 (2022), https://doi.org/10.1002/celc.202200867.

[94] Askari, N.; Askari, M. B.; Di Bartolomeo, A. Electrochemical alcohol oxidation and biological properties of $Mn_3O_4$–$Co_3O_4$–rGO, Journal of The Electrochemical Society. 169 (2022) 106511, DOI: 10.1149/1945-7111/ac96b2.

[95] Salarizadeh, P.; *et al*. Electrocatalytic performance of $MnMoO_4$–rGO nano-electrocatalyst for methanol and ethanol oxidation, Molecules. 28 (2023) 4613, https://doi.org/10.3390/molecules28124613.

[96] Pereira, V. S.; *et al*. Effects of $TiO_2$ in Pd–$TiO_2$/C for glycerol oxidation in a direct alkaline fuel cell, Journal of Fuel Chemistry and Technology. 50 (2022) 474–482, https://doi.org/10.1016/S1872-5813(21)60171-8.

[97] Li, S.; Morrissey, K. H.; Bartlett, B. M. Strategies in photochemical alcohol oxidation on noble-metal free nanomaterials as heterogeneous catalysts, Chemical Communications. 60 (2024) 10295–10305, https://doi.org/10.1039/D4CC01204F.

[98] Bera, K. K.; *et al*. Synergistic photo-enhanced electrocatalysis of Pt–ZnO–$Bi_2O_3$ heterojunction for methanol oxidation under visible light illumination, Energy Advances. 1 (2022) 908–925, https://doi.org/10.1039/D2YA00166G.

[99] Murcia, J. J.; *et al*. Photocatalytic ethanol oxidative dehydrogenation over Pt/$TiO_2$: Effect of the addition of blue phosphors, International Journal of Photoenergy. 2012 (2011) 687262, https://doi.org/10.1155/2012/687262.

[100] Liu, H.; *et al*. Few-layered black phosphorus/cucurbit[6]uril as a Pd catalyst support for photo-assisted electrocatalytic ethanol oxidation reaction, Colloids and Surfaces A: Physicochemical and Engineering Aspects. 644 (2022) 128817, https://doi.org/10.1016/j.colsurfa.2022.128817.

[101] Odetola, C.; Trevani, L. N.; Easton, E. B. Photo-enhanced methanol electrooxidation: Further insights into Pt and $TiO_2$ nanoparticle contributions, Applied Catalysis B: Environmental. 210 (2017) 263–275, https://doi.org/10.1016/j.apcatb.2017.03.027.

[102] Li, P.; Li, Z. P.; Liu, B. H. Methanol electrooxidation promoted by UV–vis light irradiation, Journal of Power Sources. 199 (2012) 146–149, https://doi.org/10.1016/j.jpowsour.2011.10.015.

[103] Kang, S.; Shen, P. K. Facile synthesis of porous hematite supported Pt catalyst and its photo-enhanced electrocatalytic ethanol oxidation performance, Electrochimica Acta. 168 (2015) 104–110, https://doi.org/10.1016/j.electacta.2015.03.203.

[104] Yang, Y.; *et al*. A Pd-based plasmonic photocatalyst for nitrogen fixation through an antenna–reactor mechanism, Chemical Science. 14 (2023) 10953–10961, https://doi.org/10.1039/D3SC02862C.

[105] Raut, W. R.; *et al.* Rapid biosynthesis of platinum and palladium metal nanoparticles using root extract of *Asparagus racemosus* Linn, Advanced Materials Letters. 4 (2013) 650–654, https://doi.org/10.5185/amlett.2012.11470.

[106] Zhai, C.; *et al*. Construction of Pt/graphitic $C_3N_4$/$MoS_2$ heterostructures on photo-enhanced electrocatalytic oxidation of small organic molecules, Applied Catalysis B: Environmental. 243 (2018) 283–293, https://doi.org/10.1016/j.apcatb.2018.10.047.

[107] Papaderakis, A.; *et al*. The effect of carbon content on methanol oxidation and photo-oxidation at Pt–$TiO_2$–C electrodes, Catalysts. 10 (2020) 248, https://doi.org/10.3390/catal10020248.

[108] Gao, H.; *et al*. Snowflake-like $Cu_2S$ as visible-light-carrier for boosting Pd electrocatalytic ethylene glycol oxidation under visible light irradiation, Electrochimica Acta. 330 (2020) 135214, https://doi.org/10.1016/j.electacta.2019.135214.

[109] Al-amin, M. D. *et al*. Quantification and description of photothermal heating effects in plasmon-assisted electrochemistry, Communications Chemistry. 7 (2024), https://doi.org/10.1038/s42004-024-01157-8.

[110] Nallabala, R. N. K. *et al*. Enhanced self-driven ultraviolet photodetection performance of high-k $Ta_2O_5$/GaN heterostructure, Materials Science in Semiconductor Processing. 170 (2024) 107954, https://doi.org/10.1016/j.mssp.2023.107954.

[111] Contreras, E. *et al*. Anodes for Direct Alcohol Fuel Cells Assisted by Plasmon-Accelerated Electrochemical Oxidation Using Gold Nanoparticle-Decorated Buckypapers, ACS applied energy materials. 3 (2020) 8755–8764, https://doi-org.ez29.periodicos.capes.gov.br/10.1021/acsaem.0c01293.

[112] Shi, J.; Wang, B.; Hu, S. From photo-assisted methanol catalytic oxidation to direct methanol fuel cells: Applications of semiconductor-based electrode, Surfaces and Interfaces. 46 (2024) 103970, https://doi.org/10.1016/j.surfin.2024.103970.

[113] Cai, Q.; *et al*. A silicon photoanode for efficient ethanol oxidation under alkaline conditions, RSC Advances. 7 (2017) 21809–21814, https://doi.org/10.1039/C7RA02848B.

[114] He, Z.-L.; *et al*. Significantly enhanced photoelectrocatalytic alcohol oxidation performance of CdS nanowire-supported Pt via the "bridge" role of nitrogen-doped graphene quantum dots, ACS Sustainable Chemistry & Engineering. 8 (2020) 12331–12341, https://doi.org/10.1021/acssuschemeng.0c05097.

[115] Cai, M.; *et al*. Investigation of thermal and electrochemical degradation of fuel cell catalysts, Journal of Power Sources. 160 (2006) 977–986. https://doi.org/10.1016/j.jpowsour.2006.03.033.

[116] Aoki, K. J.; *et al*. Peak potential shift of fast cyclic voltammograms owing to capacitance of redox reactions, Journal of Electroanalytical Chemistry. 856 (2020) 113609, https://doi.org/10.1016/j.jelechem.2019.113609.

[117] Yaqoob, L.; Noor, T.; Iqbal, N. A comprehensive and critical review of the recent progress in electrocatalysts for the ethanol oxidation reaction, RSC Advances. 11 (2021) 16768–16804, https://doi.org/10.1039/D1RA01841H.

[118] Chu, D.; *et al*. Anode catalysts for direct ethanol fuel cells utilizing directly solar light illumination, ChemSusChem. 2 (2009) 171–176, https://doi.org/10.1002/cssc.200800158.

[119] Leelavathi, A.; Madras, G.; Ravishankar, N. New insights into electronic and geometric effects in the enhanced photoelectrooxidation of ethanol using ZnO nanorod/ultrathin Au nanowire hybrids, Journal of the American Chemical Society. 136 (2014) 14445–14455, https://doi.org/10.1021/ja5059444.

[120] Leelavathi, A.; Madras, G.; Ravishankar, N. Ultrathin Au nanowires supported on rGO/$TiO_2$ as an efficient photoelectrocatalyst, Journal of Materials Chemistry A. 3 (2015) 17459–17468, https://doi.org/10.1039/C5TA03988F.

[121] Yang, H.; *et al*. Enhanced photocatalytic activity from mixture-fuel cells by ZnO template-assisted Pd–Pt hollow nanorods, ChemistrySelect. 2 (2017) 9842–9846. https://doi.org/10.1002/slct.201701406.

[122] Queiroz, A. C.; Ticianelli, E. A. Photoelectrochemical oxidation of ethanol under visible light irradiation on TaON-based catalysts, Journal of The Electrochemical Society. 165 (2018) F123–F131, DOI: 10.1149/2.0131803jes.

# 1. Materials and methods

*Reagents*

To ensure high purity and experimental reproducibility, we used the following reagents (purity or concentration and supplier in parentheses): palladium(II) chloride ($PdCl_2$, 99.0%), citric acid ($C_6H_8O_7$, 99.0%), and potassium hydroxide (KOH, 90.0%) were purchased from Sigma-Aldrich (St. Louis, MO, USA). Hydrochloric acid (HCl, 37.0%, Sigma-Aldrich) was acquired from Sigma-Aldrich (Austria). Niobium(V) chloride ($NbCl_5$, 99.0%), ethylene glycol ($C_2H_6O_2$, 99.0%), and 2-propanol (($CH_3$)$_2$CHOH, 99.5%) were bought from Sigma-Aldrich (Steinheim, Germany). Nafion® ($C_7HF_{13}O_5S.C_2F_4$, 5wt% in a mixture of lower aliphatic alcohols and water) was acquired from Sigma-Aldrich (Switzerland). Ethanol ($CH_3CH_2OH$, 99.9%) was purchased from Sigma-Aldrich (Duque de Caxias, Brazil). Sodium hydroxide (NaOH, 98.0%) was bought from Sigma-Aldrich (Czech Republic). Nitrogen gas ($N_2$, 99.99%, White Martins) and carbon monoxide (CO, 99.99%, White Martins) were acquired from White Martins (Brazil). Vulcan carbon (XC-72) was purchased from Cabot (Boston, MA, USA). And ultrapure water (18 MΩ cm at 25 °C) was obtained using a Gehaka purification system (São Paulo, Brazil).

*Electrochemical setup*

Electrochemical experiments were performed using an Autolab PGSTAT 302N potentiostat/galvanostat (Metrohm Autolab, Utrecht, The Netherlands). Data were acquired and processed using NOVA (version 2.1.7, Metrohm Autolab, Utrecht, The Netherlands). Experiments employed a single-compartment Pyrex® glass cell configured in a three-electrode arrangement with gas inlet/outlet ports. A glassy carbon disk (geometric area 0.071 cm²) served as the working electrode, a platinum plate (10 × 10 mm) as the counter electrode, and a reversible hydrogen electrode (RHE) assembled in the same electrolyte as the reference. All potentials are reported versus RHE.

Photoassisted measurements were conducted inside a Faraday cage. As illustrated in Fig. S10, the electrochemical cell was positioned above a low-pressure mercury lamp (14 W, 253.7 nm, G-light, Feira de Santana, Brazil) placed beneath the cell for the UV-assisted oxidation tests. The temperature of the electrolyte solution was monitored at the end of the experiments under UV irradiation using a digital thermometer. For long-duration measurements, including polarization curves, chronoamperometry, and stability tests, a thermostatic bath was employed to cool the electrolyte solution.

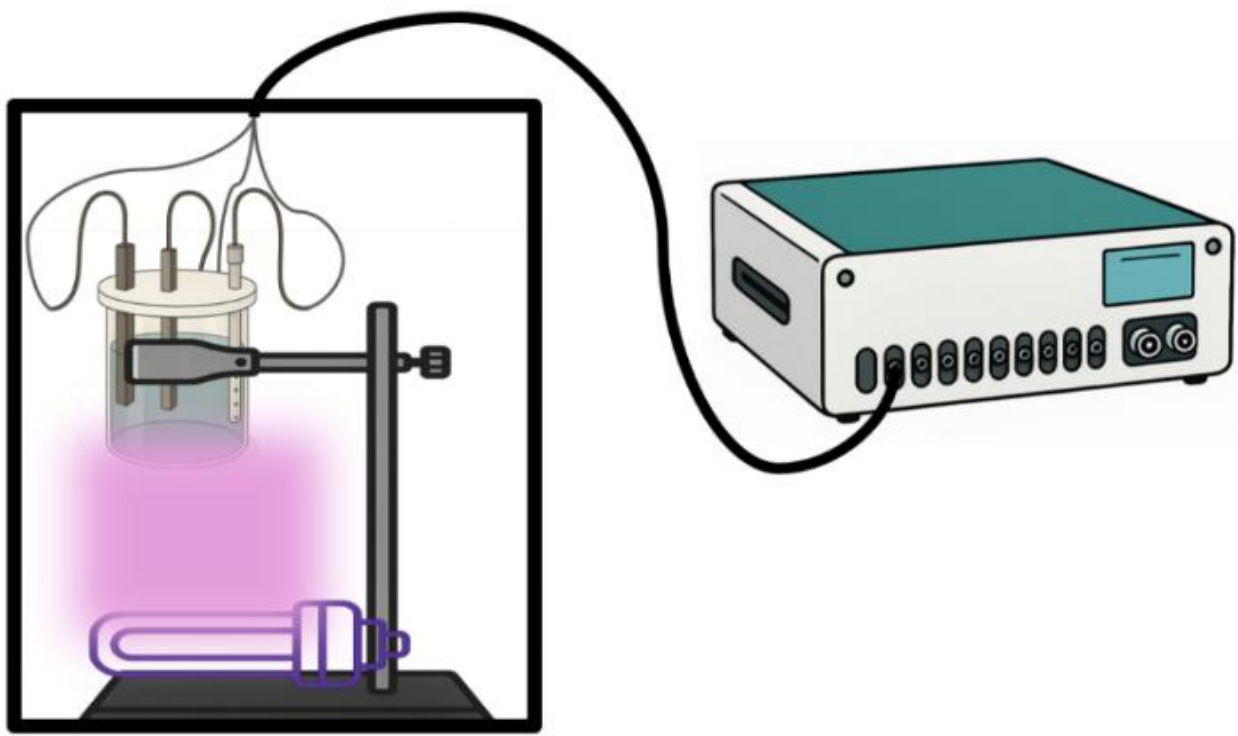

**Fig. S10.** Experimental setup of the photo-assisted process, in which the suspended electrochemical cell is irradiated with UV light from below.

To prepare the catalyst-coated glassy carbon electrode, we first made a homogeneous ink by mixing catalyst powder (3 mg), 2-propanol (1.0 mL), and Nafion® solution (30 μL). The suspension was sonicated for 10 min using a probe-type ultrasonic homogenizer (Desruptor DE500, Eco-Sonics). An aliquot (5 μL) of the ink was drop-cast onto the glassy carbon electrode and allowed to dry in air at room temperature.

## Synthesis of Niobium Pentoxide (Nb$_2$O$_5$)

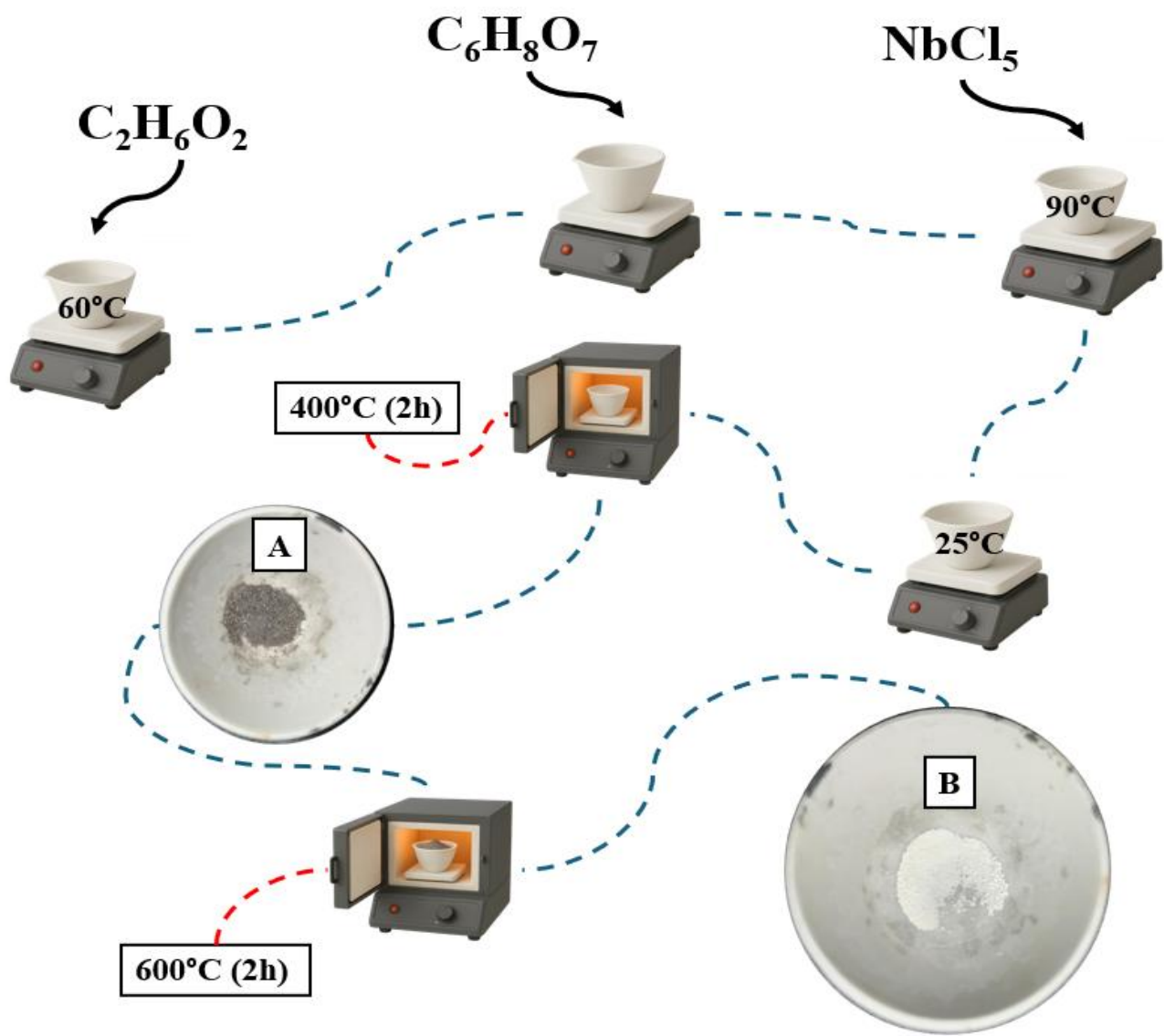

**Fig. S11.** Schematic representation of the Nb$_2$O$_5$ synthesis, from precursor dissolution to the calcined powder. Visual appearance of Nb$_2$O$_5$ particles at (A) the pre-calcination stage (gray) and (B) after calcination (white).

*Synthesis of Pd$_{(x)}$Nb$_2$O$_{5(y)}$/C Catalysts*

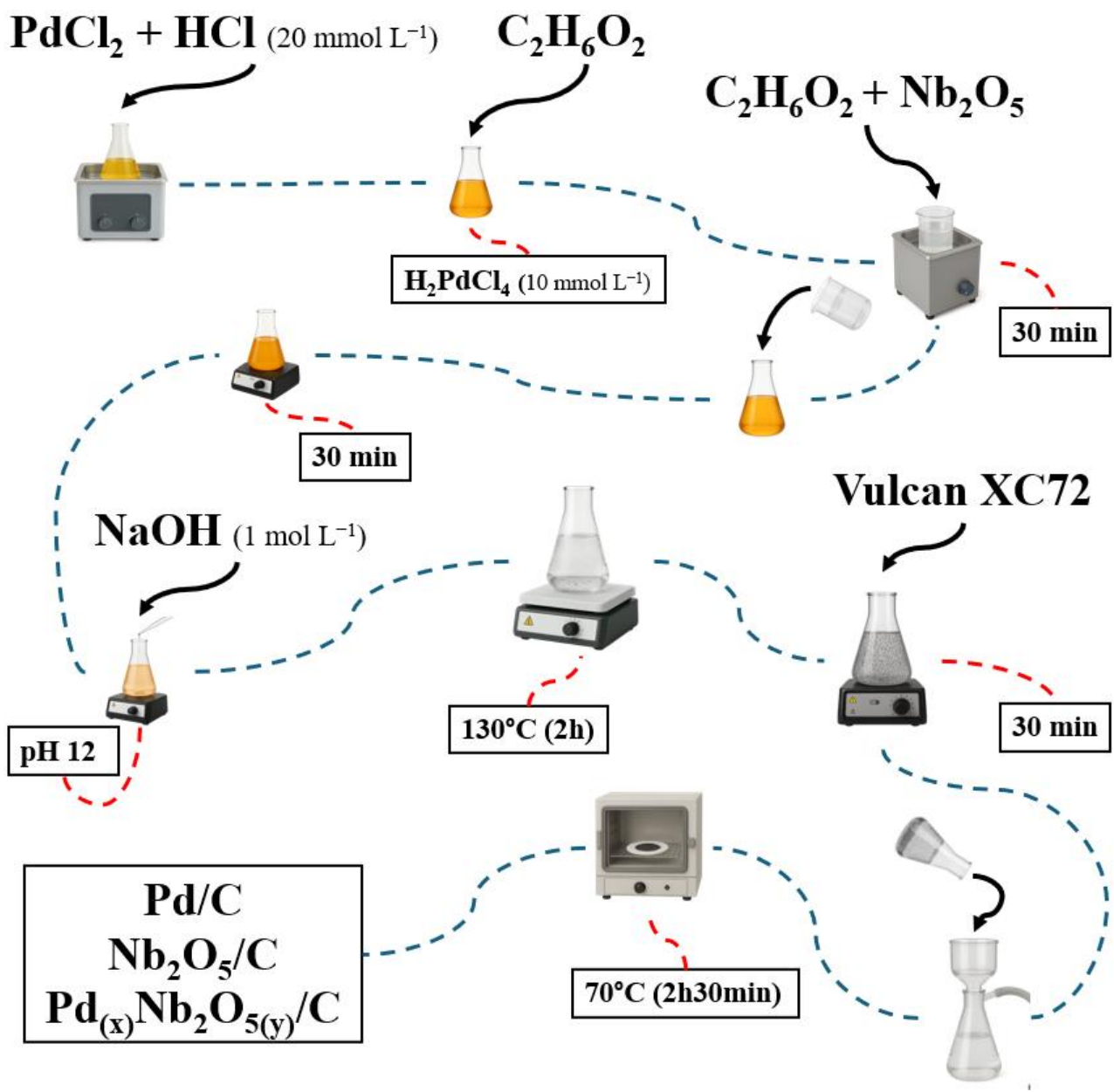

**Fig. S12.** Flowchart illustrating the experimental steps for preparing the Pd/C, Nb$_2$O$_5$/C, Pd$_{(0,7)}$Nb$_2$O$_{5(0,3)}$/C, Pd$_{(0,5)}$Nb$_2$O$_{5(0,5)}$/C, and Pd$_{(0,3)}$Nb$_2$O$_{5(0,7)}$/C catalysts, from precursor dispersion to their incorporation onto the carbon support.

*Crystallite size determination*

The crystallite size (d) was estimated using the Scherrer equation (Eq. S1), where λ is the X-ray wavelength, β is the full width at half maximum (FWHM) of the selected diffraction peak, expressed in radians after correction for instrumental broadening (assuming a Gaussian profile), θ is the Bragg angle in radians, and K is the Scherrer shape factor [1].

$$d = \frac{K\lambda}{\beta \cos\theta} \qquad \text{(Eq. S1)}$$

*Electrochemical parameters*

The texture coefficient (TC), a dimensionless metric, was calculated using the Harris method (Eq. S2). In this equation, I(*h k l*) is the measured XRD intensity of each peak, $I_0$(*h k l*) is the XRD intensity of a reference material (provided by the International Centre for Diffraction Data, ICDD), and *n* is the number of peaks used in the calculation.

$$TC_{(h\,k\,l)} = \frac{I_{(h\,k\,l)}}{I_{o(h\,k\,l)}} * \left[\frac{1}{n} * \sum_{n=1}^{n} \frac{I_{(h\,k\,l)}}{I_{o(h\,k\,l)}}\right]^{-1} \qquad \text{(Eq. S2)}$$

Electrochemical currents were normalized to the electrochemically active surface area (ECSA), estimated from the integrated charge of the CO stripping peak ($Q_{int}$), using Eq. S3. Here, $Q_{CO}$ = 420 μC cm$^{-2}$ is the charge required to oxidize a compact CO monolayer [2].

$$\text{ECSA (cm}^2) = \frac{Q_{int}(\mu C)}{Q_{CO}\,(\mu C\,cm^{-2})} \qquad \text{(Eq. S3)}$$

The time constant τ was obtained by fitting the photocurrent rise to the exponential model described by Eq. S4, where I(t) is the photocurrent at time t, $I_0$ represents the steady-state photocurrent, and τ is the characteristic time constant of the photoinduced response [3].

$$I(t) = I_0 * \left[1 - e^{-t/\tau}\right] \qquad \text{(Eq. S4)}$$

Eq. S5 is an expression used to compare the diffusion coefficients under illuminated and dark conditions, based on electrochemical measurements (Fig. 8) [4,5].

Where:

D – Diffusion coefficient;

$i_p$ – Peak current density i;

$v^{1/2}$ – Square root of the scan rate.

$$\frac{D_{light}}{D_{dark}} = \left[\frac{\left(\frac{i_p}{v^{1/2}}\right)_{light}}{\left(\frac{i_p}{v^{1/2}}\right)_{dark}}\right]^2 \quad \text{(Eq. S5)}$$

## 2. Results and discussion

Fig. S13 shows thermogravimetric analysis (TGA) of pre-calcined $Nb_2O_5$, segmented into five regions. Regions (i), (ii), and (iii) correspond to: (i) desorption of physisorbed water; (ii) a slow, continuous mass loss attributed to partial decomposition of residual organics; and (iii) combustion of more refractory organic residues, centered near 480 °C [6,7]. In region (iv), $Nb_2O_5$ crystallizes from an amorphous phase to the pseudo-hexagonal (TT) polymorph, and in region (v), the orthorhombic (T) phase forms near 600 °C [8]. Approaching ≈750 °C, a further phase transition consistent with thermal stabilization is indicated, with no additional mass loss detected [9].

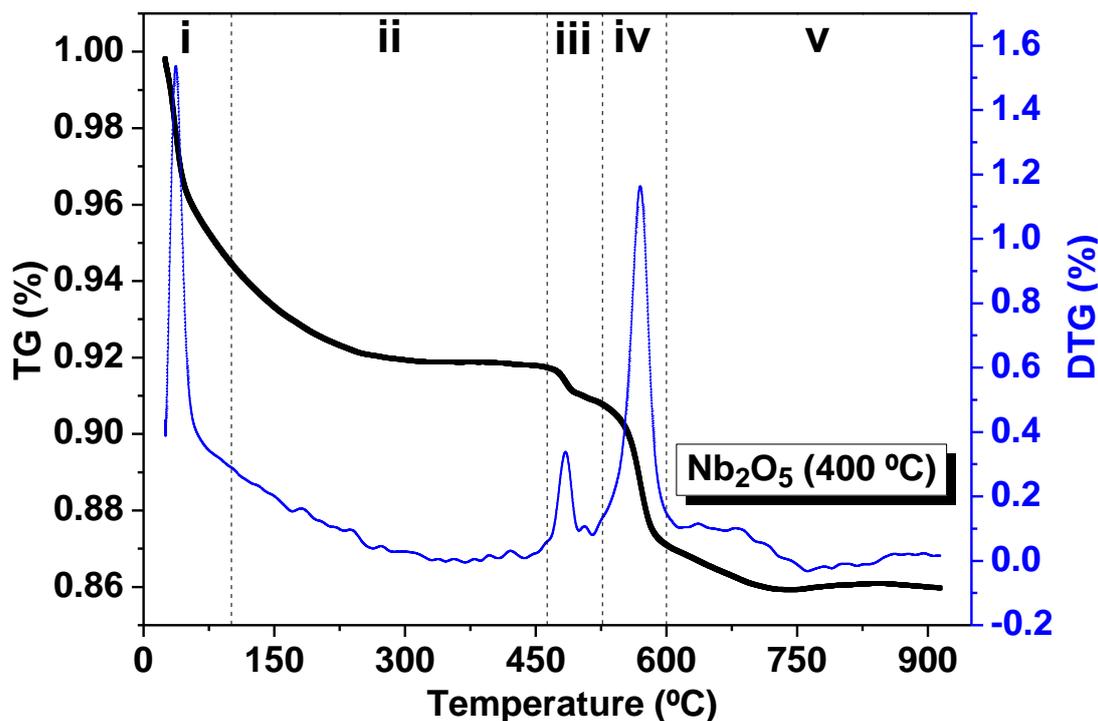

**Fig. S13.** Thermogravimetric analysis of $Nb_2O_5$ pre-calcined at 400 °C. The graph is segmented into five regions, labeled as (i), (ii), (iii), (iv), and (v).

Fig. S14A shows an SEM image highlighting the region used for elemental mapping and EDS spectrum acquisition for $Pd_{(0.5)}Nb_2O_{5(0.5)}/C$. The EDS spectrum Fig. S14B reveals the presence of niobium (Nb), palladium (Pd), oxygen (O), and carbon (C) in the catalyst, consistent with the expected composition. The identification of these elements indicates successful catalyst synthesis by the adopted methodology. These EDS results are consistent with the XRD patterns (Figure 4), which are compatible with the presence of the corresponding phases.

Elemental maps were acquired to examine the spatial distributions of Pd, Nb, O, and C. Fig. S14C, S5D, and S5F show the distribution of Nb, O, and C, respectively. Regions with a strong Nb signal coincide with areas of a strong O signal, indicating $Nb_2O_5$ domains. Furthermore, some areas that appear white in the C map correspond to regions rich in Nb and O, suggesting $Nb_2O_5$ domains not supported on carbon.

The Pd map (Fig. S14E) overlaps with both C-rich and Nb/O-rich regions, indicating Pd nanoparticles dispersed on carbon and on $Nb_2O_5$. However, localized regions of intense Pd signal suggest partial nanoparticle agglomeration, which can reduce the electrochemically active surface area and adversely affect ethanol electrooxidation activity [10].

Fig. S15 shows a scanning electron microscopy (SEM) micrograph of the $Nb_2O_5/C$ catalyst with the area selected for elemental mapping and EDS spectrum acquisition. The EDS results identify Nb, O, and C, consistent with $Nb_2O_5$ dispersed on the carbon support.

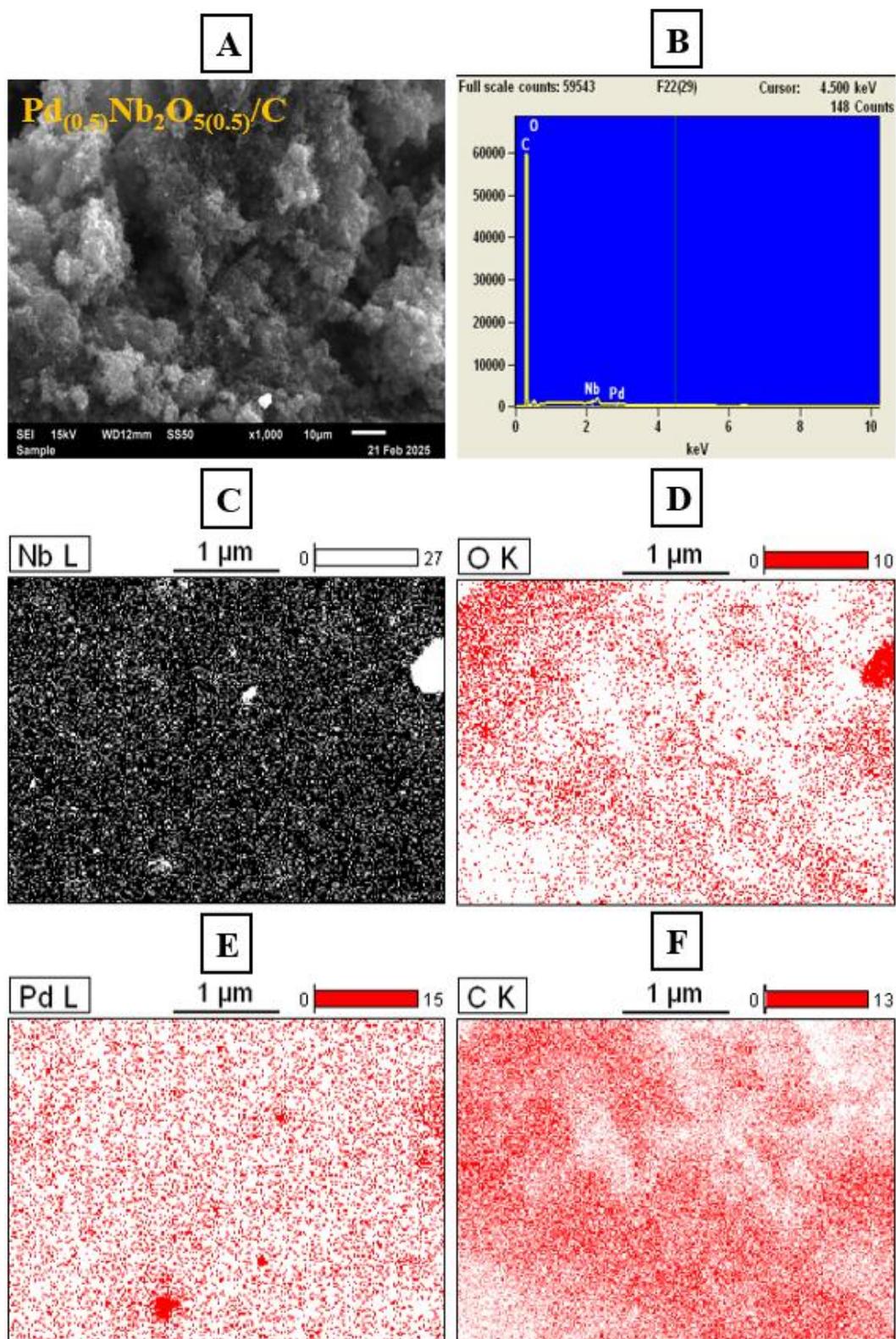

**Fig. S14.** A – SEM image (1000×) of the $Pd_{(0.5)}Nb_2O_{5(0.5)}/C$ catalyst, B – EDS spectrum. (C–F) Elemental maps for niobium, oxygen, palladium, and carbon.

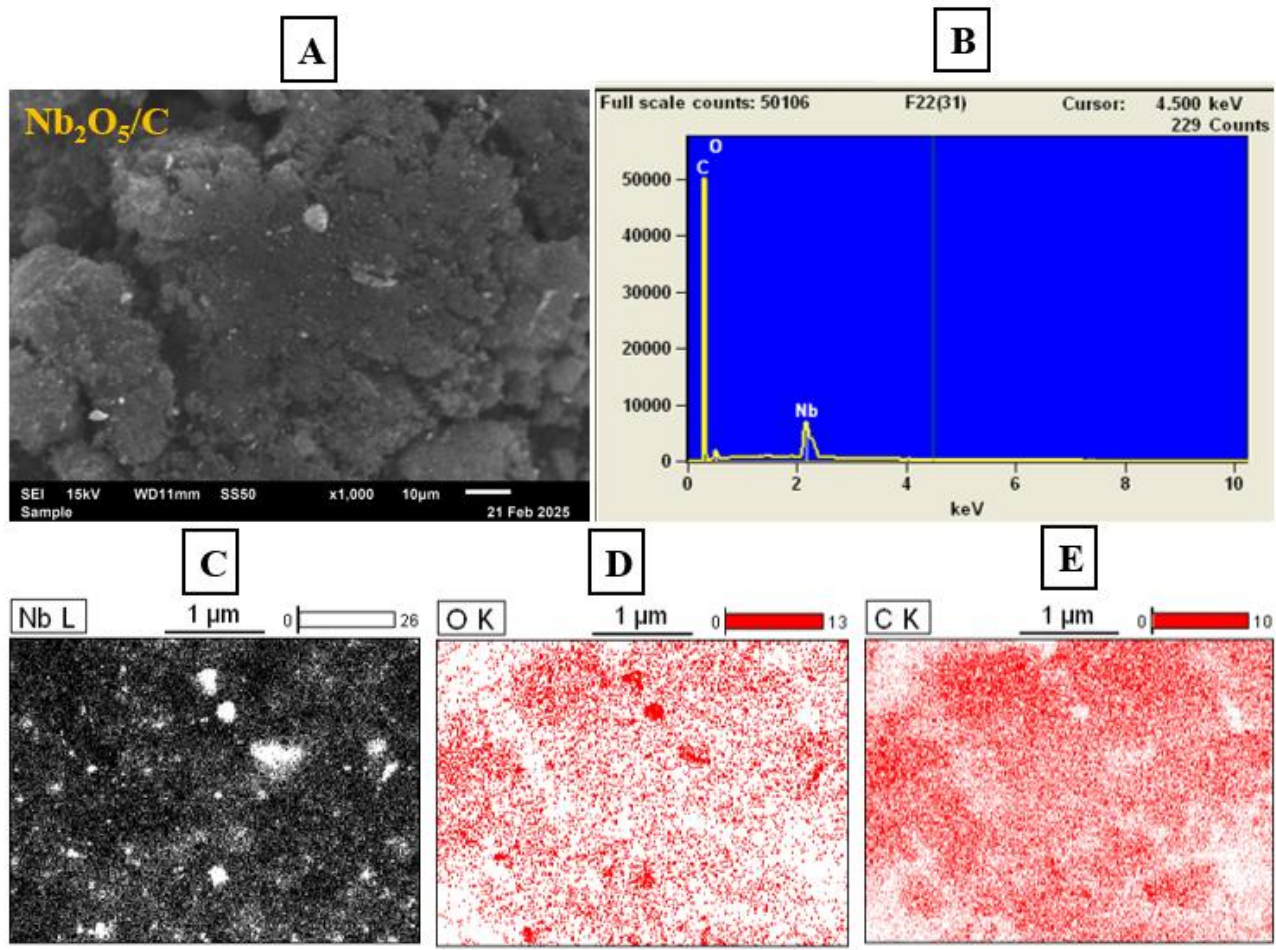

**Fig. S15.** A – Scanning electron microscopy image (1000×), B – EDS spectrum, and C–E elemental maps of the $Nb_2O_5/C$ catalyst.

**Table S5.** High-resolution XPS results for the Nb 3d region: binding energy position (eV), peak area, and relative fraction of Nb chemical states obtained from deconvolution of the Nb 3d doublet.

| Catalyst | Element | Nb $3d_{3/2}$ (eV) | Nb $3d_{5/2}$ (eV) |
|---|---|---|---|
| Pd/C | $Nb^{5+}$ | --- | --- |
| $Pd_{(0.5)}Nb_2O_{5(0.5)}/C$ | $Nb^{5+}$ | 210.40 | 207.67 |
| $Pd_{(0.7)}Nb_2O_{5(0.3)}/C$ | $Nb^{5+}$ | 210.54 | 207.79 |
| $Pd_{(0.3)}Nb_2O_{5(0.7)}/C$ | $Nb^{5+}$ | 210.70 | 207.99 |

**Table S6.** High-resolution XPS results for the Pd 3d region: binding energy position (eV), peak area, and relative fraction of Pd chemical states obtained from deconvolution of the Pd 3d doublet.

| Pd/C | Pd 3d | | | |
|---|---|---|---|---|
| | Position | | Area | % Pd |
| | Pd $3d_{5/2}$ | Pd $3d_{3/2}$ | | |
| $Pd^0$ | 336.13 | 341.00 | 481.40 | 49.17 |
| $Pd^{2+}$ | 338.00 | 343.26 | 498.00 | 50.83 |
| **$Pd_{(0.5)}Nb_2O_{5(0.5)}/C$** | Pd 3d | | | |
| | Position | | Area | % Pd |
| | Pd $3d_{5/2}$ | Pd $3d_{3/2}$ | | |
| $Pd^0$ | 336.11 | 341.61 | 372.50 | 58.99 |
| $Pd^{2+}$ | 339.00 | 344.00 | 259.1 | 41.01 |
| **$Pd_{(0.7)}Nb_2O_{5(0.3)}/C$** | Pd 3d | | | |
| | Position | | Area | % Pd |
| | Pd $3d_{5/2}$ | Pd $3d_{3/2}$ | | |
| $Pd^0$ | 336.01 | 341.46 | 547.9 | 39.66 |
| $Pd^{2+}$ | 339.00 | 344.61 | 834.6 | 60.34 |
| **$Pd_{(0.3)}Nb_2O_{5(0.7)}/C$** | Pd 3d | | | |
| | Position | | Area | % Pd |
| | Pd $3d_{5/2}$ | Pd $3d_{3/2}$ | | |
| $Pd^0$ | 336.00 | 341.41 | 553.40 | 40.02 |
| $Pd^{2+}$ | 338.17 | 344.00 | 830.00 | 59.98 |

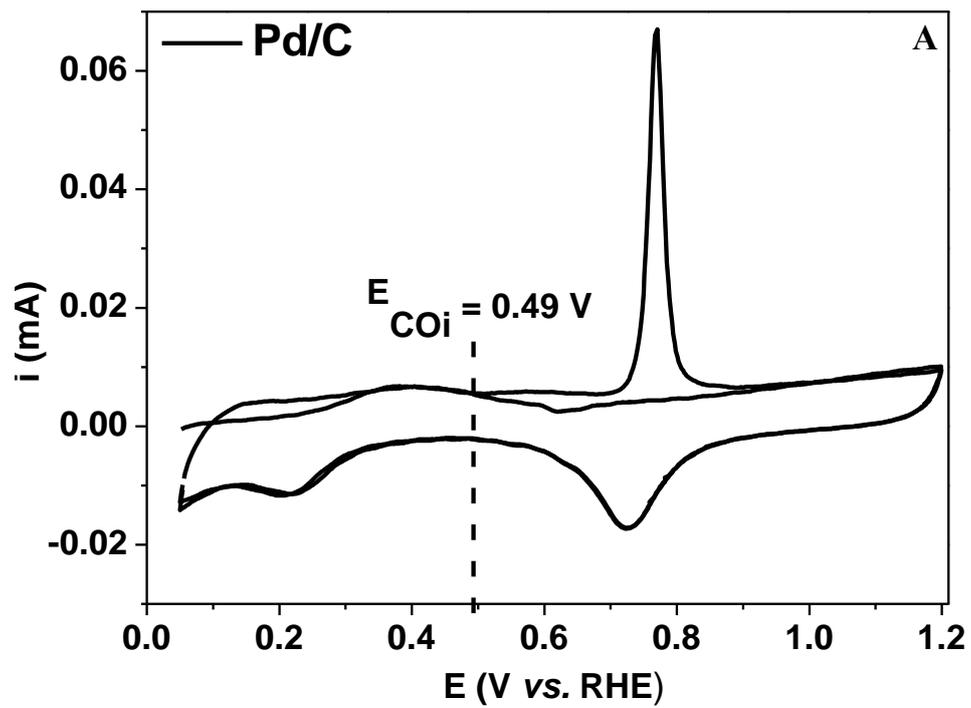
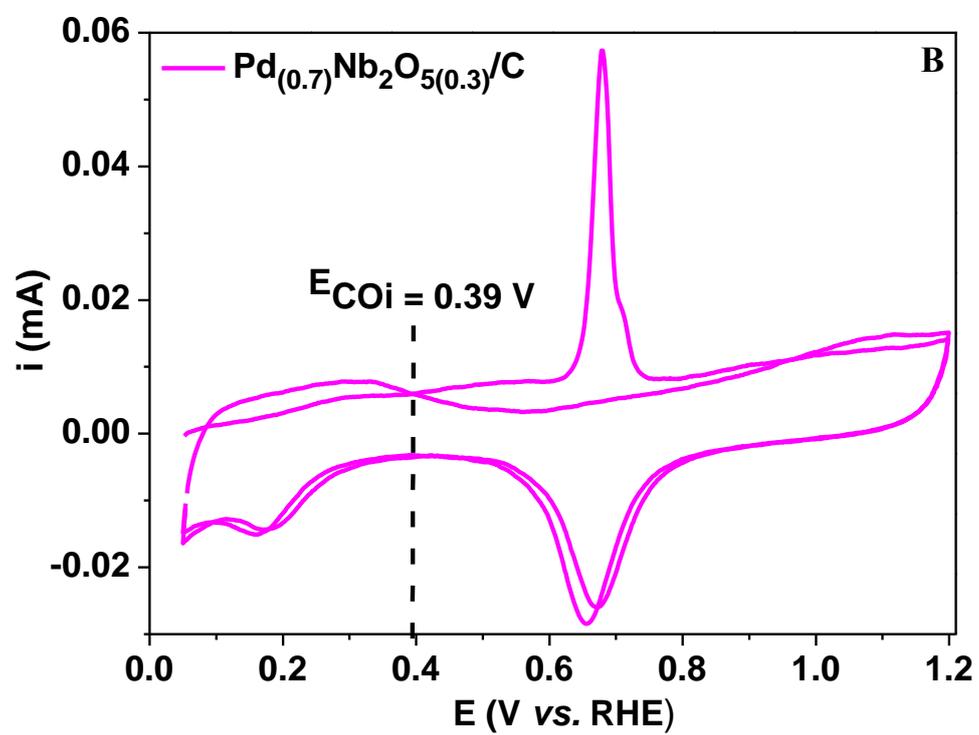

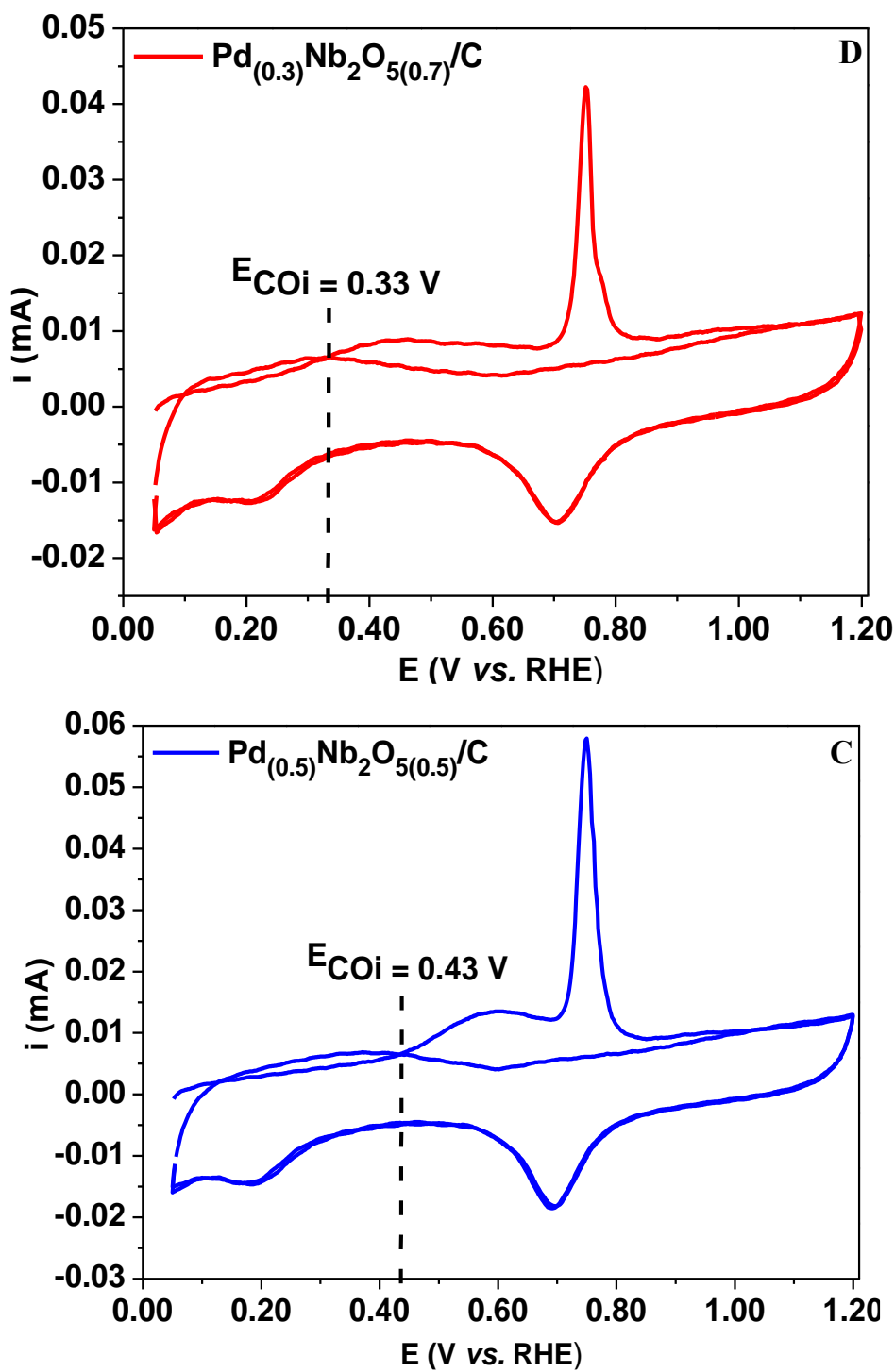

**Fig. S16.** CO stripping cyclic voltammograms (first and second cycles) recorded for the synthesized electrocatalysts (A – Pd/C, B – $Pd_{(0.7)}Nb_2O_{5(0.3)}$/C, C – $Pd_{(0.5)}Nb_2O_{5(0.5)}$/C, and D – $Pd_{(0.3)}Nb_2O_{5(0.7)}$/C) in aqueous 1 mol $L^{-1}$ KOH at 20 mV $s^{-1}$. Dashed lines indicate the onset potential for CO oxidation ($E_{COi}$).

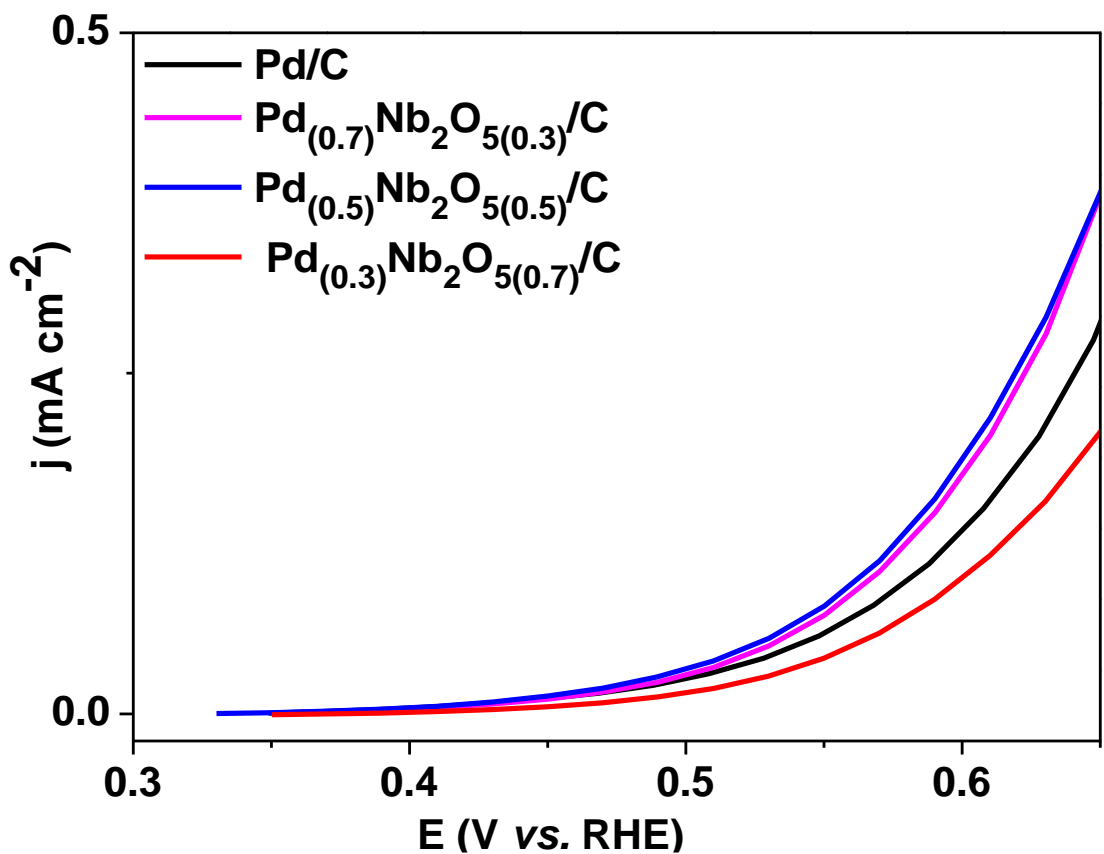

**Fig. S17.** Quasi-steady-state polarization curves for the catalysts recorded without UV irradiation.

**Table S7.** Electrochemical parameters derived from the Tafel plots: Onset potentials for the ethanol oxidation reaction ($E_i$) and coefficient of determination ($R^2$).

| Catalyst | $E_i$ (V vs. RHE) | $R^2$ |
|---|---|---|
| Pd/C | 0.49 | 0.9996 |
| Pd$_{(0.7)}$Nb$_2$O$_{5(0.3)}$/C | 0.47 | 0.9992 |
| Pd$_{(0.5)}$Nb$_2$O$_{5(0.5)}$/C | 0.45 | 0.9997 |
| Pd$_{(0.3)}$Nb$_2$O$_{5(0.7)}$/C | 0.53 | 0.9959 |

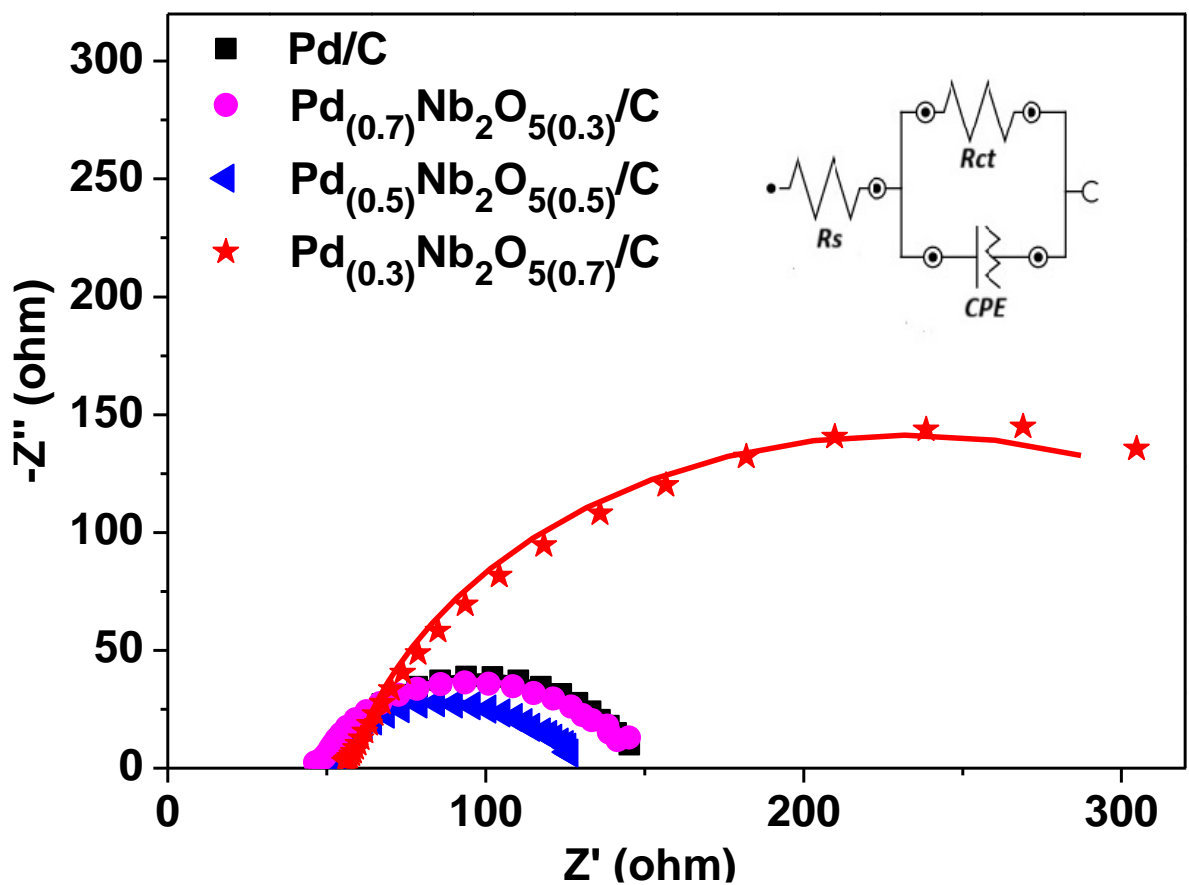

**Fig. S18.** Nyquist plots obtained for Pd/C catalyst (black square), $Pd_{(0.7)}Nb_2O_{5(0.3)}$/C (pink circle), $Pd_{(0.5)}Nb_2O_{5(0.5)}$/C (blue triangle), and $Pd_{(0.3)}Nb_2O_{5(0.7)}$/C (red star) during ethanol electrooxidation in 1 mol L$^{-1}$ KOH + 1 mol L$^{-1}$ ethanol. Measurements were performed at 0.77 V vs RHE using ten points per decade over 1000–0.1 Hz with a 5 mV AC amplitude. Solid lines represent the fitted curves. The inset shows the equivalent circuit used for fitting, consisting of Rs (solution resistance), Rct (charge-transfer resistance), and CPE (constant phase element).

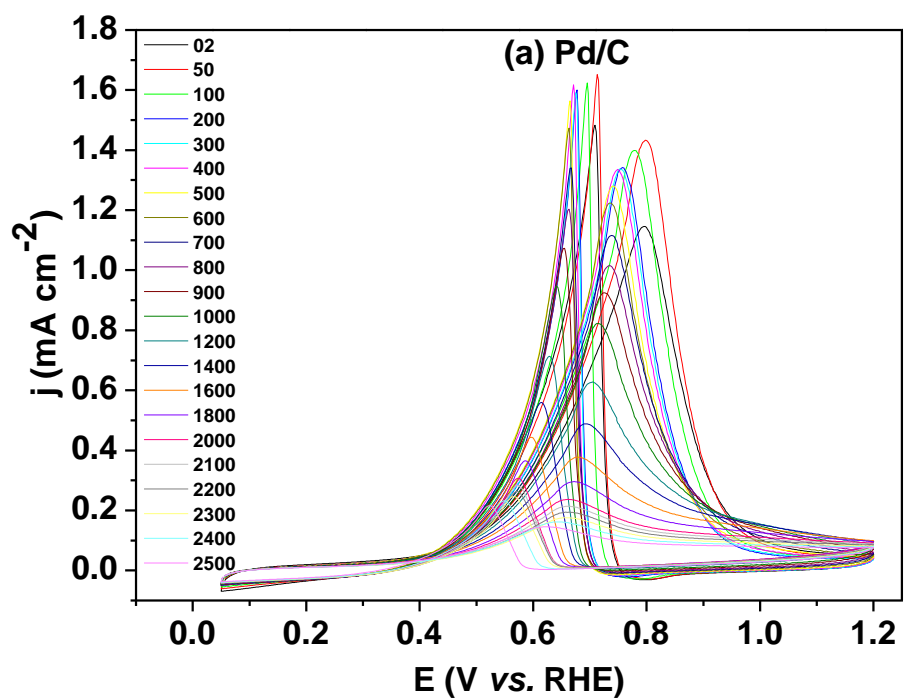

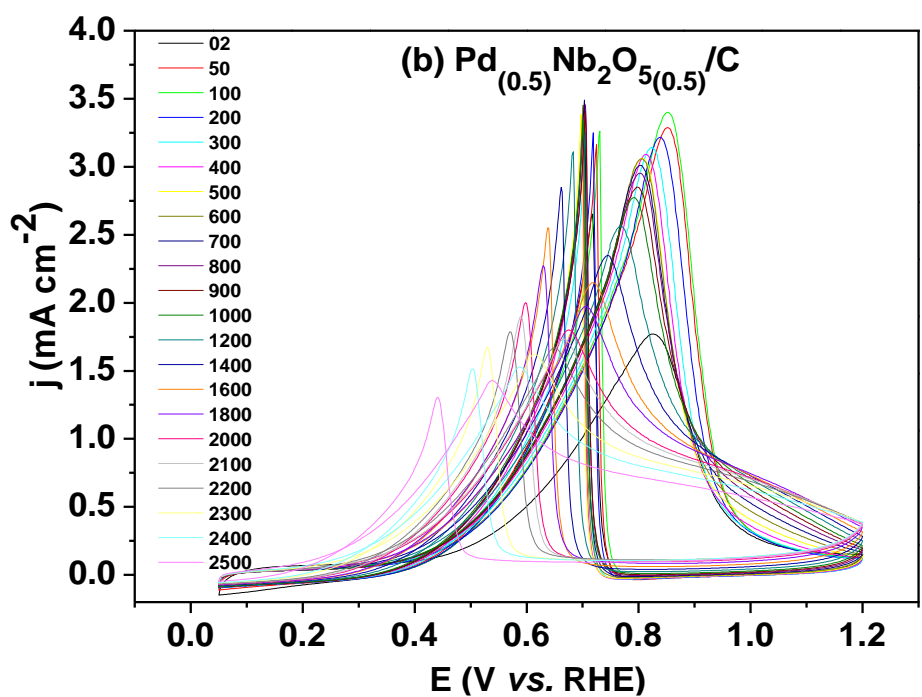

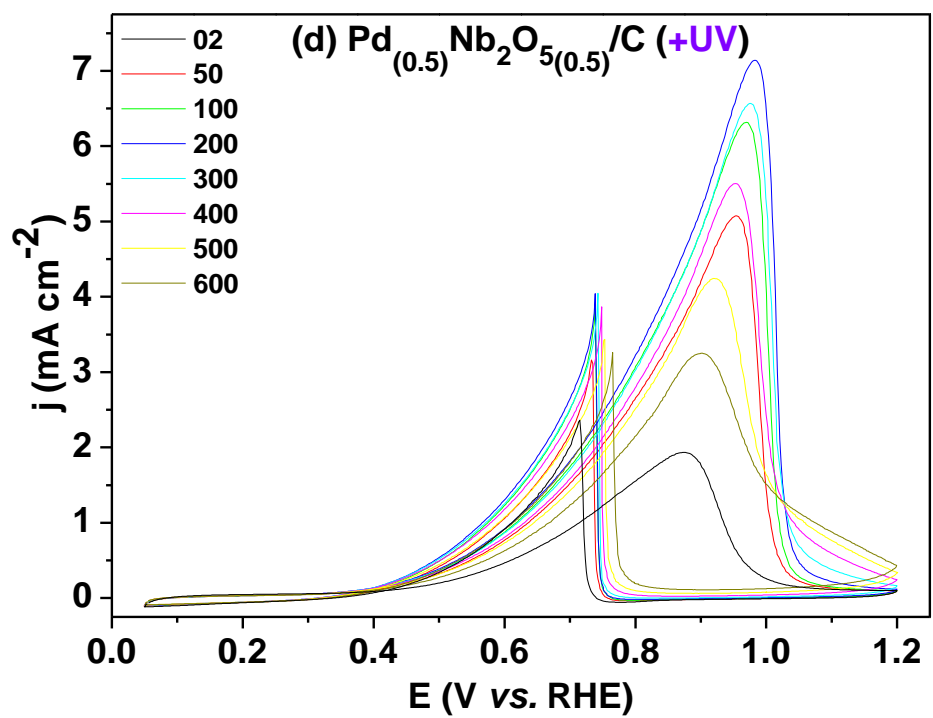

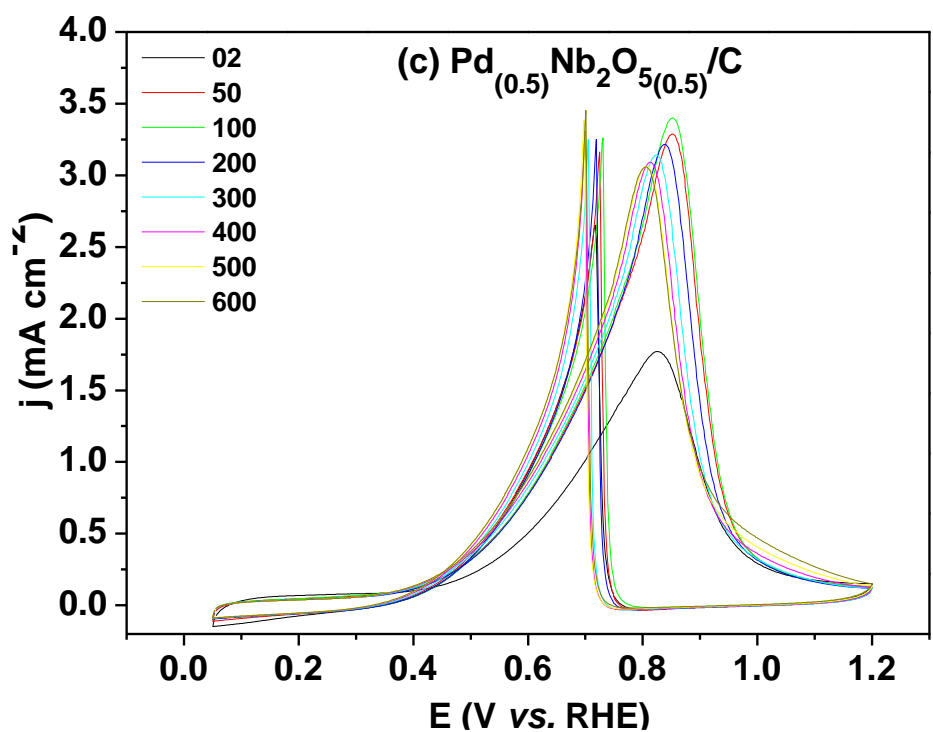

**Fig. S19.** Accelerated stability tests for (a) Pd/C, (b) Pd$_{(0.5)}$Nb$_2$O$_{5(0.5)}$/C, (c) Pd$_{(0.5)}$Nb$_2$O$_{5(0.5)}$/C under dark conditions, and (d) Pd$_{(0.5)}$Nb$_2$O$_{5(0.5)}$/C under UV illumination. Voltammograms were recorded in aqueous 1 mol L$^{-1}$ KOH + 1 mol L$^{-1}$ ethanol at a scan rate of 100 mV s$^{-1}$.

The optical properties of $Nb_2O_5$ were investigated by diffuse reflectance spectroscopy in the ultraviolet–visible region (UV–vis–DRS). As shown in Fig. S20, $Nb_2O_5$ exhibits a pronounced ultraviolet absorption band ranging between 200 and 400 nm. This behavior is consistent with literature reports [11-15] and supports the potential of this material for application in photoassisted oxidation of ethanol. Tauc analysis (inset) yields a band-gap energy of 3.10 eV, in agreement with reported values [16,17]. Furthermore, note that $Nb_2O_5$ has a band gap comparable to that of titanium dioxide–based photocatalysts [18,19], further underscoring the suitability of $Nb_2O_5$ for photocatalytic processes.

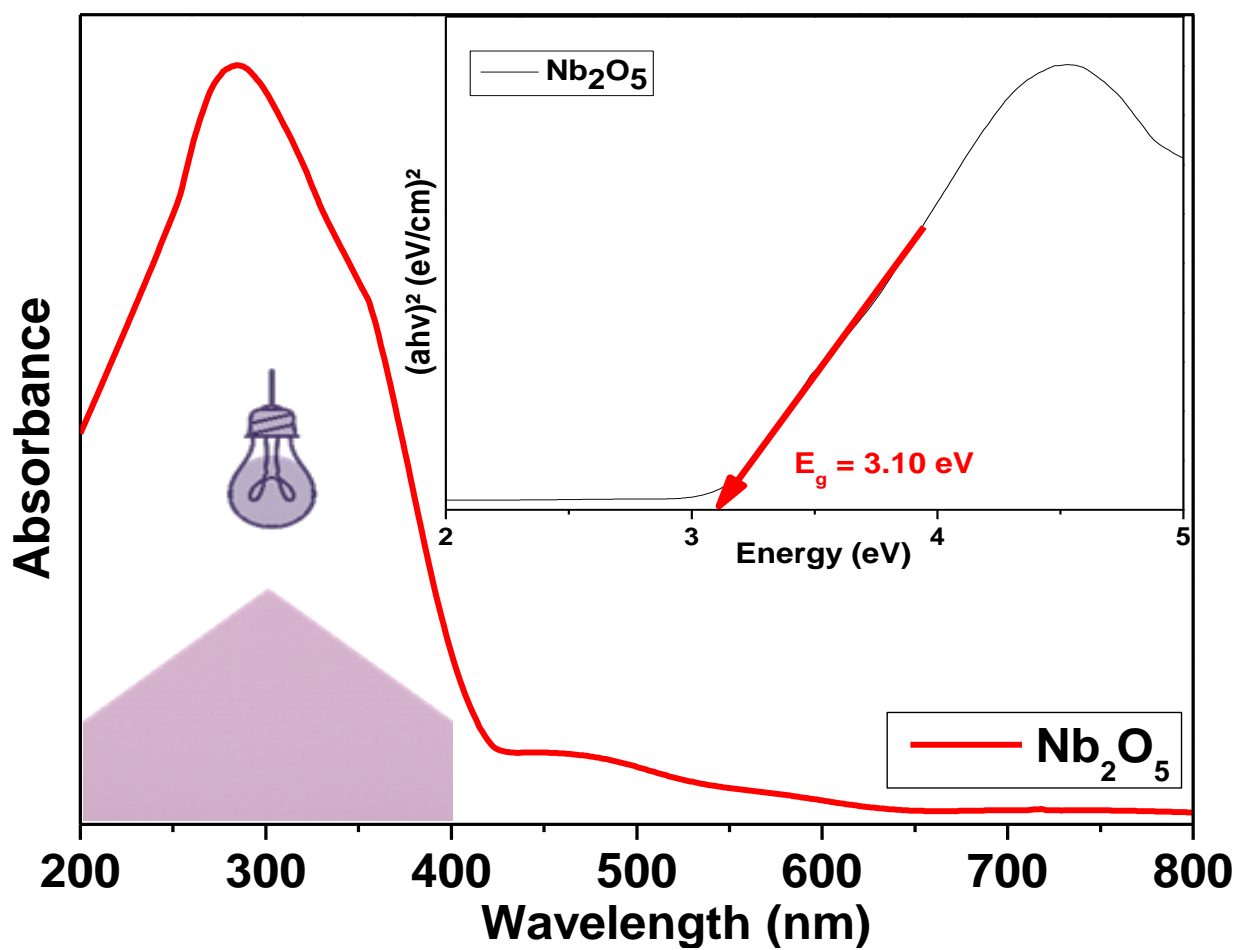

**Fig. S20.** UV–vis–DRS spectrum of $Nb_2O_5$ calcined at 600 °C. Inset: Tauc plot used to estimate the optical band gap.

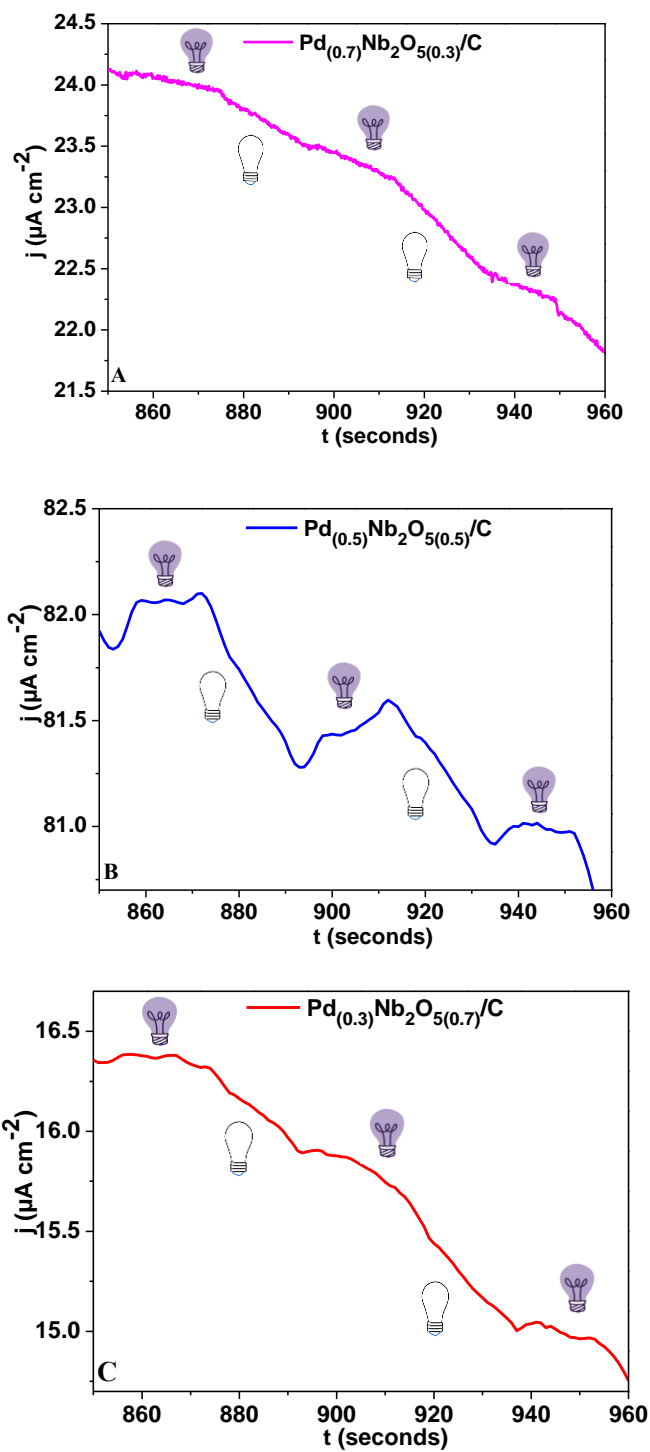

**Fig. S21.** Photocurrent responses obtained by chronoamperometry at 0.55 V vs. RHE under successive UV on/off cycles for (A) Pd$_{(0.7)}$Nb$_2$O$_{5(0.3)}$/C, (B) Pd$_{(0.5)}$Nb$_2$O$_{5(0.5)}$/C, and (C) Pd$_{(0.3)}$Nb$_2$O$_{5(0.7)}$/C in a solution containing 1 mol L$^{-1}$ KOH + 1 mol L$^{-1}$ ethanol. The lamp icons indicate the periods during which UV irradiation was applied.

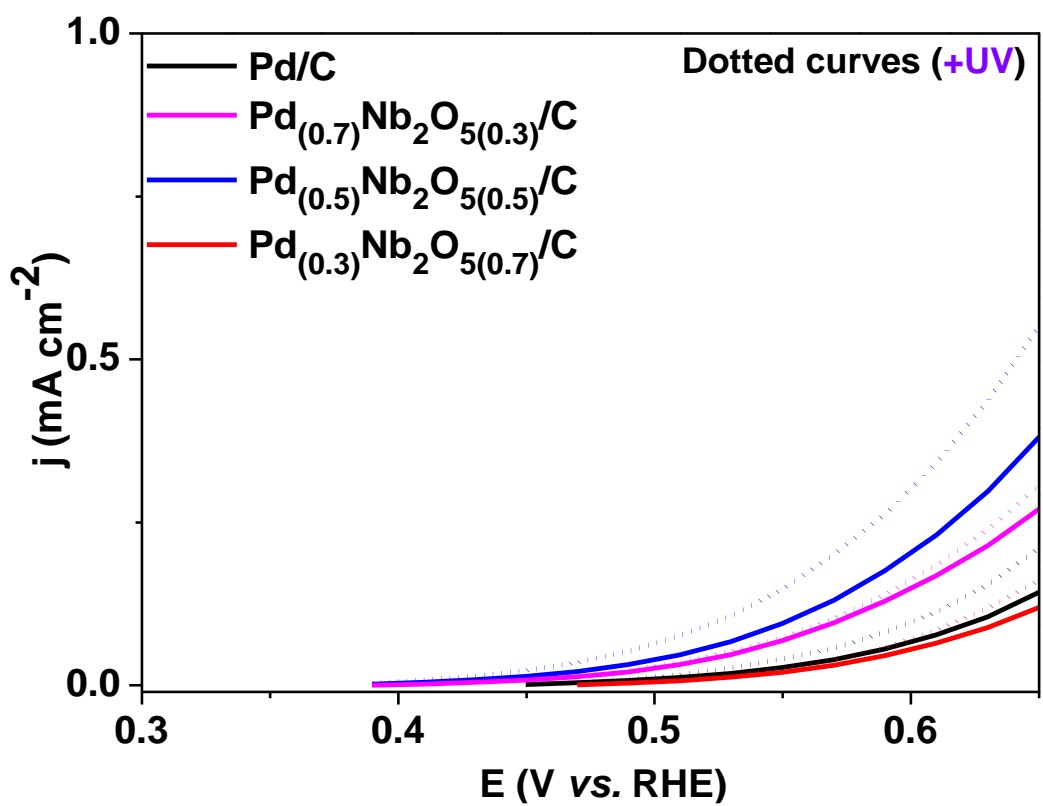

**Fig. S22.** Quasi-steady-state polarization curves for the catalysts, recorded under dark and UV illumination.

**Table S8.** Electrochemical parameters derived from Tafel plots: onset potentials for the ethanol oxidation reaction ($E_i$) and coefficient of determination ($R^2$).

| Catalyst | $E_i$ (V vs. RHE) | $R^2$ |
|---|---|---|
| Pd/C | 0.51 | 0.9955 |
| Pd/C (UV) | 0.51 | 0.9931 |
| Pd$_{(0.7)}$Nb$_2$O$_{5(0.3)}$/C | 0.47 | 0.9911 |
| Pd$_{(0.7)}$Nb$_2$O$_{5(0.3)}$/C (UV) | 0.47 | 0.9930 |
| Pd$_{(0.5)}$Nb$_2$O$_{5(0.5)}$/C | 0.45 | 0.9944 |
| Pd$_{(0.5)}$Nb$_2$O$_{5(0.5)}$/C (UV) | 0.43 | 0.9923 |
| Pd$_{(0.3)}$Nb$_2$O$_{5(0.7)}$/C | 0.55 | 0.9946 |
| Pd$_{(0.3)}$Nb$_2$O$_{5(0.7)}$/C (UV) | 0.53 | 0.9914 |

# References

[123] Hassanzadeh-tabrizi, S. A. Precise calculation of crystallite size of nanomaterials: A review, Journal of Alloys and Compounds. 968 (2023) 171914, https://doi.org/10.1016/j.jallcom.2023.171914.

[124] Obradović, M. D. *et al*. Electrochemical oxidation of ethanol on palladium-nickel nanocatalyst in alkaline media, Applied Catalysis B: Environmental. 189 (2016) 110–118, https://doi.org/10.1016/j.apcatb.2016.02.039.

[125] YANG, S. et al. High performance few-layer GaS photodetector and its unique photo-response in different gas environments, Nanoscale. 6 (2014) 2582–2587, https://doi.org/10.1039/C3NR05965K.

[126] Yao, Z. *et al*. Electrochemical-reduced graphene oxide-modified carbon fiber as Pt–Au nanoparticle support and its high efficient electrocatalytic activity for formic acid oxidation, Journal of Solid State Electrochemistry. 17 (2013) 2511–2519, https://doi.org/10.1007/s10008-013-2130-3.

[127] Yang, S. *et al*. Preparation and electrochemistry of graphene nanosheets–multiwalled carbon nanotubes hybrid nanomaterials as Pd electrocatalyst support for formic acid oxidation, Electrochimica Acta. 62 (2011) 242–249, https://doi.org/10.1016/j.electacta.2011.12.022.

[128] Garcia, A. B. S. *et al*. Effects of the Pechini's modified synthetic route on structural and photophysical properties of $Eu^{3+}$ or $Tb^{3+}$-doped $LaAlO_3$, Materials Research Bulletin. 143 (2021) 111462, 2021, https://doi.org/10.1016/j.materresbull.2021.111462.

[129] Kumar, D. A. *et al*. Effect of calcium doping on $LaCoO_3$ prepared by Pechini method, Powder Technology. 235 (2012) 140–147, https://doi.org/10.1016/j.powtec.2012.09.030.

[130] Li, S. *et al*. Comparison of amorphous, pseudohexagonal and orthorhombic $Nb_2O_5$ for high-rate lithium ion insertion, CrystEngComm. 18 (2016) 2532–2540, https://doi.org/10.1039/C5CE02069G.

[131] Pinto, B. A.; D'oliveira, A. S. C. M. Nb silicide coatings processed by double pack cementation: Formation mechanisms and stability, Surface and Coatings Technology. 409 (2021) 126913, https://doi.org/10.1016/j.surfcoat.2021.126913.

[132] Koch, M. *et al*. Towards hybrid one-pot/one-electrode Pd-NPs-based nanoreactors for modular biocatalysis, Biochemical Engineering Journal. 175 (2021) 108132–108132, https://doi.org/10.1016/j.bej.2021.108132.

[133] Kumari, N. *et al*. Dependence of photoactivity of niobium pentoxide ($Nb_2O_5$) on crystalline phase and electrokinetic potential of the hydrocolloid. 208 (2020) 110408–110408, https://doi.org/10.1016/j.solmat.2020.110408.

[134] Al-marri, A. H. *et al*. Enhanced photocatalytic properties of the $Nb_2O_5$/rGO for the degradation of methylene blue, Ionics. 29 (2023) 5505–5515, https://doi.org/10.1007/s11581-023-05255-w.

[135] Wolski, L. *et al*. Insight into methanol photooxidation over mono- (Au, Cu) and bimetallic (AuCu) catalysts supported on niobium pentoxide — An operando-IR study, Applied Catalysis B: Environmental. 258 (2019) 117978, https://doi.org/10.1016/j.apcatb.2019.117978.

[136] Batista, L. M. B. *et al*. Synthesis, characterization and evaluation of niobium catalysts in the flash pyrolysis of glycerol, Solid State Sciences. 97 (2019) 105977, https://doi.org/10.1016/j.solidstatesciences.2019.105977.

[137] Dias, D. T. *et al*. Photoacoustic Spectroscopy of Titanium Dioxide, Niobium Pentoxide, Titanium:Niobium, and Ruthenium-Modified Oxides Synthesized Using Sol–Gel

Methodology, Applied Spectroscopy. 78 (2024) 1028–1042, doi: 10.1177/00037028241268158.

[138] Shinohara, T. *et al*. Morphology Control of Energy-Gap-Engineered $Nb_2O_5$ Nanowires and the Regioselective Growth of CdS for Efficient Carrier Transfer Across an Oxide-Sulphide Nanointerface, Scientific Reports. 7 (2017), https://doi.org/10.1038/s41598-017-05292-2.

[139] Dai, Q. *et al*. A novel nano-fibriform C-modified niobium pentoxide by using cellulose templates with highly visible-light photocatalytic performance, Ceramics International. 46 (2020) 13210–13218, https://doi.org/10.1016/j.ceramint.2020.02.096.

[140] González-burciaga, L. A. *et al*. Characterization and Comparative Performance of $TiO_2$ Photocatalysts on 6-Mercaptopurine Degradation by Solar Heterogeneous Photocatalysis, Catalysts. 10 (2020) 118–118, https://doi.org/10.3390/catal10010118.

[141] Potdar, S. B. *et al*. Highly Photoactive Titanium Dioxide Supported Platinum Catalyst: Synthesis Using Cleaner Ultrasound Approach, Catalysts. 12 (2022) 78, https://doi.org/10.3390/catal12010078.